\documentclass[reprint,prx,aps,amsmath,amssymb,floatfix,superscriptaddress,longbibliography]{revtex4-2}

\usepackage{mathtools,amsmath}
\usepackage{graphicx} 
\usepackage[breaklinks=true,colorlinks,citecolor=teal,linkcolor=teal,urlcolor=teal]{hyperref}
\usepackage{physics}
\usepackage{siunitx}
\usepackage{bbold}
\usepackage{soul}
\usepackage{bm}
\usepackage[normalem]{ulem}
\usepackage{times}
\usepackage{enumerate}  
\usepackage{float}
\usepackage{amssymb}
\usepackage{xcolor} 
\usepackage{booktabs}

\begin{document}
\title{Artificially intelligent Maxwell's demon for optimal control of open quantum systems}

\author{Paolo A. Erdman}
\thanks{These authors contributed equally; Correspondence should be addressed to: \href{mailto:p.erdman@fu-berlin.de}{p.erdman@fu-berlin.de}}
\affiliation{Freie Universit{\" a}t Berlin, Department of Mathematics and Computer Science, Arnimallee 6, 14195 Berlin, Germany}

\author{Robert Czupryniak}
\thanks{These authors contributed equally; Correspondence should be addressed to: \href{mailto:p.erdman@fu-berlin.de}{p.erdman@fu-berlin.de}}
\affiliation{Department of Physics and Astronomy, University of Rochester, Rochester, NY 14627, USA}
\affiliation{Institute for Quantum Studies, Chapman University, Orange, CA 92866, USA}

\author{Bibek Bhandari}
\affiliation{Institute for Quantum Studies, Chapman University, Orange, CA 92866, USA}

\author{Andrew N. Jordan}
\affiliation{Institute for Quantum Studies, Chapman University, Orange, CA 92866, USA}
\affiliation{Department of Physics and Astronomy, University of Rochester, Rochester, NY 14627, USA}

\affiliation{The Kennedy Chair in Physics, Chapman University, Orange, CA 92866, USA}

\author{Frank No{\'e}}
\affiliation{Microsoft Research AI4Science, Karl-Liebknecht Str. 32, 10178 Berlin, Germany}
\affiliation{Freie Universit{\" a}t Berlin, Department of Mathematics and Computer Science, Arnimallee 6, 14195 Berlin, Germany}

\author{Jens Eisert}
\affiliation{Dahlem Center for Complex Quantum Systems, Freie Universit{\" a}t Berlin, 14195 Berlin, Germany}
\affiliation{Freie Universit{\" a}t Berlin, Department of Physics, Arnimallee 6, 14195 Berlin, Germany}
\affiliation{Rice University, Department of Chemistry, Houston, TX 77005, USA}

\author{Giacomo Guarnieri}
\affiliation{Department of Physics and INFN - Sezione di Pavia, University of Pavia, Via Bassi 6, 27100, Pavia, Italy}
\affiliation{Dahlem Center for Complex Quantum Systems, Freie Universit{\" a}t Berlin, 14195 Berlin, Germany}

\begin{abstract}
Feedback control of open quantum systems is of fundamental importance for practical applications in various contexts, ranging from quantum computation to quantum error correction and quantum metrology. Its use in the context of thermodynamics further enables the study of the interplay between information and energy. However, deriving optimal feedback control strategies is highly challenging, as it involves the optimal control of open quantum systems, the stochastic nature of quantum measurement, and the inclusion of policies that maximize a long-term time- and trajectory-averaged goal. In this work, we employ a reinforcement learning approach to automate and capture the role of a quantum Maxwell's demon: the agent takes the literal role of discovering optimal feedback control strategies in qubit-based systems that maximize a trade-off between measurement-powered cooling and measurement efficiency. Considering weak or projective quantum measurements, we explore different regimes based on the ordering between the thermalization, the measurement, and the unitary feedback timescales, finding different and highly non-intuitive, yet interpretable, strategies. In the thermalization-dominated regime, we find strategies with elaborate finite-time thermalization protocols conditioned on measurement outcomes. In the measurement-dominated regime, we find that optimal strategies involve adaptively measuring different qubit observables reflecting the acquired information, and repeating multiple weak measurements until the quantum state is ``sufficiently pure'', leading to random walks in state space. Finally, we study the case when all timescales are comparable, finding new feedback control strategies that considerably outperform more intuitive ones. We discuss a two-qubit example where we explore the role of entanglement and conclude discussing the scaling of our results to quantum many-body systems.
\end{abstract}

\maketitle

\section{Introduction}
Recent years have seen remarkable progress in the realization and control of various quantum technological platforms, ranging from superconducting qubits \cite{murch2013observing,kjaergaard2020superconducting,vijay2012stabilizing,weber2014mapping} to ultra-cold trapped ion \cite{haffner2008quantum,schneider2012experimental} and quantum dots \cite{josefsson2018quantum,prete2019thermoelectric,sothmann2014thermoelectric}. 
A key ingredient in achieving this feat is \textit{feedback control}, whereby a set of operations is performed depending on the information acquired through measurements about the controlled system~\cite{doyle2013feedback,aastrom2021feedback,jordan2024quantum}.
Crucially, as demonstrated by Maxwell's demon thought-experiment~\cite{plenio2001physics,maruyama2009colloquium} and Szilard's engine \cite{szilard1929}, feedback control represents the arch-stone bridging information theory, statistical mechanics and thermodynamics.
Indeed, a Maxwell's demon is an agent that can extract work from a thermalized system solely by acquiring information about it, leading to an apparent violation of the second law of thermodynamics.
\begin{figure}[!tb]
    \centering
    \includegraphics[width=0.99\columnwidth]{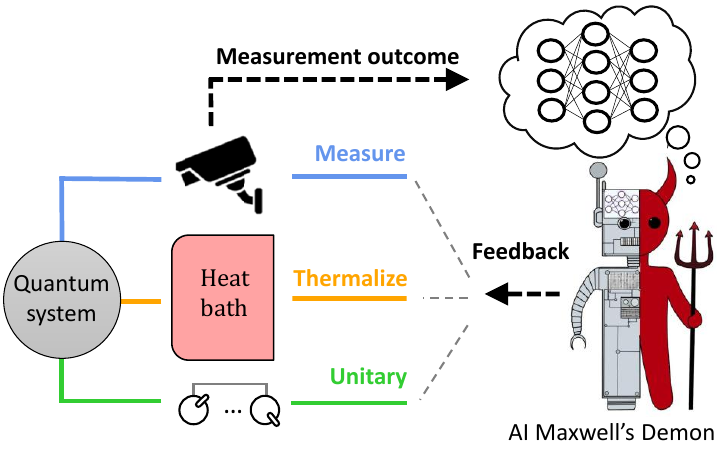}
    \caption{Schematic representation of a quantum Maxwell's demon. Based on previous measurement outcomes, the demon can decide whether to (partially) thermalize the quantum system, perform further measurements, or perform unitary feedback. The goal is to optimize the trade-off between cooling power and the cost of measuring the system. In this manuscript, we consider a reinforcement learning agent an actual Maxwell's demon.}
    \label{fig:demon}
\end{figure}
This apparent contradiction was later solved by Landauer~\cite{landauer1957} and Bennett \cite{bennett1982}, who accounted for the minimum cost of erasing the classical information acquired about the system, given by Landauer's limit \cite{landauer1961irreversibility}. For a random bit of information, this minimum work cost amounts on average to $k_{\rm B} T \ln 2$, $k_B$ being Boltzmann's constant, and $T$ the temperature. 
Although of fundamental importance, this limit notoriously represents a very loose estimate since many-body or macroscopic systems dissipate several orders of magnitude more.
However, Maxwell's demon has recently attracted great renewed interest due to technological leaps in miniaturization and control. 
Several state of the art experiments have successfully managed to probe Landauer's limit on classical platforms, including colloidal particles~\cite{berut2012experimental,berut2015information,jun2014high,gavrilov2017erasure}, 
nano-magnets ~\cite{hong2016experimental,martini2016experimental,gaudenzi2018quantum}, superconducting flux logic cells~\cite{saira2020nonequilibrium} and underdamped micro-mechanical oscillators ~\cite{dago2021information,dago2022dynamics}. 
This has opened the door to the 
investigation of thermodynamics of nano-scale quantum thermal machines~\cite{dubi2011colloquium,blickle2012realization,martinez2016brownian,rossnagel2016,josefsson2018quantum,maslennikov2019quantum,peterson2019experimental,lindenfels2019,onishchenko2022probing,erdman2019_prb,bhandari2021minimal}, and in particular to the impact of quantum information and feedback control on them~\cite{elouard2017extracting,ding2018measurement,elouard2018efficient,bresque2021two,buffoni2019quantum,landi2022informational,belenchia2022informational,PRXQuantum.2.030310}. 
Understanding this cost is hence crucial to optimize current and next-generation quantum devices~\cite{barato2015, gingrich2016dissipation, aamir2023, Pietzonka2017a, Dechant2018, horowitz2017proof, BaratoNJP2018, Holubec2018PRL, proesmans2017discrete, VanVu2020, Guarnieri2021PRL, Guarnieri2021PRE, falasco2020unifying, Hasegawa2023, VuSaito2023, dechant2020fluctuation, MacIeszczak2018, dechant2020fluctuation, Guar19c, SegalAgarwalla, GerryAgarwalla, Miller2020Landauer, Ray2023, jordan2006, Complexity} 
and ultimately minimizing their energetic footprint~\cite{auffeves2022quantum}.

The role of information is even more fundamental in quantum systems, as it becomes inseparable from the measurement procedure necessary to acquire and manipulate it. In contrast to classical machines, measurements intrinsically and necessarily 
alter the state and the energy content of quantum systems~\cite{elouard2017extracting,ding2018measurement,elouard2018efficient,bhandari2022con,bhandari2023meas,jordan2024quantum}.
All these factors paint a very rich and complex landscape, where determining the control strategies that give the optimal trade-off between power and efficiency becomes a difficult task, even for very simple models.

The effect of measurement and feedback in the context of quantum systems and the idea of a quantum Maxwell's demon have been studied in Refs.~\cite{lloyd1997,barato2013, koski2014_pnas,koski2015, camati2016, elouard2017extracting,chida2017,campisi2017,cottet2017,ding2018measurement,elouard2018efficient,potts2019thermodynamics,andersson2022,bhandari2022con,bhandari2023meas}.
However, their finite-time analysis and systematic optimization, crucial for actual quantum technologies, has yet to be carried out. Finite-time thermodynamic optimization of quantum systems is a notoriously complex problem, even in the absence of measurements and feedback, due to the out-of-equilibrium quantum dynamics of the system, the large optimization space given by all possible drivings, and the existence of trade-offs, 
e.g., between power and efficiency ~\cite{espositodot2010, abah2012single,janine2013dot,campo2014,campisi2015nonequilibrium, campisi2016, villazon2019, miller2019work,tobalina2019vanishing,erdman2019maximum, abiuso2019nonmarkov,abiuso2020_prl,bhandari2020geo,abiuso2020geometric,pancotti2020speed,dann2020quantum, wang2011performance, cavina2021, pablo2022geo,nettersheim2022power,das2023,carrega2024, razzoli2024}.
Whenever quantum measurements and feedback are taken into account, the problem becomes even richer, as the goal shifts from finding the best (fixed) protocol to finding the \textit{best feedback control strategy}, i.e., the policy which, based on the information gathered about the system, selects a specific optimal sequence of operations that maximize a target figure of merit. Notice that this optimization is further complicated by the stochastic nature of quantum measurement, which gives rise to stochastic trajectories, and the optimization goal is the long-term time- and trajectory-average of the figure of merit.
Crucially, if we choose as figure of merit the cooling power from a single thermal reservoir, the optimization goal evidently coincides with the prototypical behavior a Maxwell's demon, i.e., an agent that acquires information about a system to lower its temperature, and its thermodynamic cost is  characterized by the above-mentioned Landauer's limit. 

Changing gear for a moment, recent years have witnessed the rise of \emph{machine learning} methods applied across all scientific fields \cite{jordan2015}, including quantum mechanics \cite{carrasquilla2020,krenn2023,dawid2023}. Among these, \emph{reinforcement learning} (RL) has emerged as a powerful tool to prepare quantum states \cite{bukov2018,zhang2019,dalgaard2020,mackeprang2020,brown2021,porotti2022,metz2023}, including feedback control \cite{fosel2018,reuer2023}, to discover efficient implementations of quantum gates \cite{an2019,niu2019}, and for quantum error correction \cite{sweke2020}.
Recently, it has been applied in the context of quantum thermodynamics to design optimal time-dependent cycles and protocols \cite{sgroi2021,erdman2022,erdman2022_arxiv,erdman2023_pnas,erdman2023_prr,deng2024}. However, the full potential of RL as a tool to discover optimal feedback control strategies for quantum thermodynamic applications has not yet been explored.

In this work, we take a radical step in bringing together all the above ideas -- quantum feedback control, methods of reinforcement learning, and quantum thermodynamics -- 
in an entirely fresh fashion: in our approach, we
take the RL agent literally as a Maxwell's demon-like entity that, at every time-step, can decide to acquire information about a quantum system and to act on it accordingly. In this way, we give the agent in RL an ontological meaning and a thermodynamic interpretation.
Concretely, we employ RL to perform a systematic and comprehensive study to discover optimal feedback control strategies to maximize the long-term time-average and trajectory-averaged heat extracted from a single thermal bath, thus finding the ``best
artificially intelligent quantum Maxwell's demon'' without any prior knowledge about quantum mechanics, nor of thermodynamics.

As depicted in Fig.~\ref{fig:demon} the agent, acting effectively as a \textit{quantum Maxwell's demon}, has to choose at every time step whether it should perform a (partial) thermalization \cite{Thermalization} of the system with a heat bath (orange line), measure the quantum system (blue line), or apply unitary feedback (green line). Additionally, it must learn the value of some continuous time-dependent control governing the dynamics of the system. The aim is to maximize the long-term average extracted power, or equivalently, the cooling power, i.e., the heat extraction from the thermal bath, exploiting invasive quantum measurements and feedback while also minimizing the thermodynamic cost of measuring the system. Notice that, without measurements, it would not be thermodynamically possible to cool a single thermal bath. 

There are three timescales that are relevant in our analysis: (1) the \textit{thermalization timescale} $\tau_\text{t}>0$, 
representing the typical timescale for the quantum system to thermalize with the thermal bath,
(2) the \textit{measurement timescale} $\tau_\text{m}>0$, 
representing the typical timescale required to perform an approximately projective measurement, and (3) the \textit{unitary timescale} $\tau_\text{u}>0$, representing the typical timescale required to implement a rotation on the quantum system while decoupled from the thermal bath. Depending on the ordering between these timescales, we can be in completely different regimes that describe different physics and are thus characterized by profoundly different control strategies. In general, this regime will be dictated by the experimental details of the device. 

In this work, we wish to strike a balance between the complexity of a realistic description of the dynamics of the system and the interpretability of our results. We thus proceed incrementally: first, we consider the limiting case where only one timescale among $\tau_\text{t}$, $\tau_\text{m}$ and $\tau_\text{u}$ is relevant, the other two being much faster. This allows us to obtain interesting yet interpretable results from which we can learn general strategies. We then showcase the effectiveness of our method applying it to more complicated scenarios where multiple timescales are finite. In such cases, we find complicated control strategies that provably outperform intuitive strategies. For concreteness, we will consider one and two-qubit systems coupled to a thermal bath.

In Sec.~\ref{sec:model}, we introduce the model and the optimization objectives. In Sec.~\ref{sec:method}, we frame the optimization of the quantum Maxwell's demon performance as an RL task. In the following sections, we apply our method to different physical regimes, namely:

\begin{itemize}
\item 
\textit{Thermalization dominated regime}. In Sec.~\ref{sec:tau_t}, we consider a thermalization timescale, modeled with Lindblad dynamics, that is much slower than the other ones, i.e., $\tau_\text{t} \gg \tau_\text{m}, \tau_\text{u}$. While the strategy is simple when we are only interested in maximizing the cooling power, we find elaborate yet intuitive finite-time control strategies when we are interested in a trade-off between cooling power and measurement efficiency. The duration of the control sequence becomes gradually slower and slower as we are more interested in efficiency. We also compare our results with a more intuitive strategy inspired by the RL results, finding that the RL control performs better than a standard Maxwell's demon.

\item
\textit{Measurement dominated}. In Sec.~\ref{sec:tau_m}, we consider the measurement timescale to be much slower than the other ones, i.e., $\tau_\text{m} \gg \tau_\text{t}, \tau_\text{u}$. We model the finiteness of $\tau_\text{m}$ considering weak measurements, both discrete and continuous, and we consider both a fixed and learnable measurement basis. Here, we find elaborate yet interpretable control strategies, characterized by multiple repeated measurements, that vary abruptly depending on the measurement strength, shedding light on the relation between the generation of coherence during the measurement process, the measurement strength, the measured observable, and the performance of the Maxwell's demon. We further find that adaptively changing the measured observable outperforms a fixed observable. 
\item 
\textit{Measurement and thermalization dominated regime}. In Sec.~\ref{sec:tau_t_tau_m}, we consider comparable thermalization and measurement timescales that are much slower than the unitary timescale, i.e., $\tau_\text{m
} \sim  \tau_\text{t} \gg \tau_\text{u}$. The RL agent learns to perform thermalization strokes whose duration and line-shape adaptively depend on the level of quantum purity reached during the previous quantum measurements.
\item 
\textit{Two qubit setup: all timescales are finite}.
In Sec.\ref{sec:all_finite}, we consider a two-qubit setup where all timescales are finite and comparable, i.e., $\tau_\text{t} \sim \tau_\text{m} \sim \tau_\text{u}$. Here, we model finite-time thermalization and unitary dynamics using the Lindblad equation. For the measurement, we couple the main qubit to an additional auxiliary qubit that acts as a measurement probe that is then projectively measured. This can be considered as an explicit and finite-time modelization of a \emph{positive-operator-valued measure} (POVM). We consider both the case with and without counter-rotating terms in the qubit-qubit interaction. Here, we rigorously show that the strategies learned by the RL agent outperform intuitive control strategies, and we analyze the entanglement generated between the two qubits.
\end{itemize}

\section{Quantum Maxwell's demon}
\label{sec:model}
To realize a quantum Maxwell's demon \textit{gedankenexperiment}, we consider the setup sketched in Fig.~\ref{fig:demon}. It consists of a quantum system, a thermal bath, a measurement probe, and time-dependent controls on the quantum system. We discretize time in steps $t_i = i\Delta t$, where $\Delta t$ is the duration of each time step. In each time step, we assume that the quantum system can either undergo (partial) thermalization with the thermal bath, a quantum measurement, or unitary feedback at discretion of the Maxwell's demon.
Because of this, strictly speaking all three steps can be considered feedback; however rotations and Hamiltonian dynamics that pertain to unitary feedback are kept distinct for consistency.

\textit{Unitary feedback}: The system undergoes evolution in the absence of the bath and the measurement probe. We consider two different cases of unitary evolution: 1) unitary rotation of a qubit around the Bloch sphere given by 
\begin{equation}
U_{\vec{\phi}}=e^{-i\vec{\phi}\cdot \vec{\sigma}/2},
\end{equation}
where $\vec{\sigma} = \{\sigma_{\rm x},\sigma_{\rm y},\sigma_{\rm z}\}$ are Pauli matrices and $\vec{\phi}=\{\phi_{\rm x},\phi_{\rm y},\phi_{\rm z}\}$ are a set of tunable angles, and 2) Hamiltonian  dynamics of the system given by
\begin{equation}
	\dot{\rho}(t) = -\frac{i}{\hbar} [H_{\rm U}[u(t)], \rho(t)],
\end{equation}
where $H_{\rm U}[u(t)]$ is the family of Hamiltonians of the quantum system that depends on the time-dependent controls $t \mapsto u(t)$ (henceforth denoted by $u(t)$ for simplicity), $\rho(t)$ is the reduced density matrix of the quantum system, and $\hbar$ is Planck's reduced constant. The former case describes the regime where unitary dynamics is the fastest timescale, whereas the latter describes finite-time unitary dynamics induced by Hamiltonian dynamics.

\textit{Thermalization}: When the system is coupled to a thermal bath with inverse temperature $\beta=1/(k_BT)$, we describe its dynamics by the Lindblad master equation \cite{gorini1976,lindblad1976}
\begin{equation} \label{eq:master_eqn}
	\dot{\rho}(t) = -\frac{i}{\hbar} [H_{\rm T}[u(t)], \rho(t)] + \mathcal{D}[\rho(t)],
\end{equation}
where  $H_{\rm T}[u(t)]$ describes the
family of Hamiltonians of the system during thermalization (that, in principle, can be different from the family of Hamiltonians $H_{\rm U}[u(t)]$ describing the unitary evolution), and $\mathcal{D}[\cdot]$ is 
the dissipator describing the coupling of the system to the thermal bath. It can be expressed as
\begin{align}
	\mathcal{D}[\rho] &= \sum_k \gamma_{k,u(t)}\nonumber \\
	&\times \left( A_{k,u(t)}\rho A^\dagger_{k,u(t)} - \frac{1}{2}\left\{\rho,A^\dagger_{k,u(t)}A_{k,u(t)}\right\} \right),
 \label{eq:dissipator}
\end{align}
where $A_{k,u(t)}$ are the jump operators (possibly $u(t)$-dependent), and $\gamma_{k,u(t)}$ are the corresponding transition rates. We employ the global master equation to guarantee thermodynamic consistency \cite{hofer2017quantum}. Further, we consider a bosonic heat bath with flat spectral density.

{\it Measurement}: We consider both discrete and continuous, as well as strong and weak quantum measurements. All such scenarios can be encompassed in the general framework of POVMs, 
i.e., a set of Kraus operators $M_k$ -- one for each measurement outcome -- satisfying $\sum_k M_k^\dagger M_k = I$ for discrete measurements, and $\int dk M_k^\dagger M_k = I$ for continuous measurements \cite{zhang2017quantum,yanik2022thermodynamics}. The probability (probability density in the continuous case) of measuring outcome $k$ at time $t$ is given 
by
\begin{equation}
    p_k(t) = \Tr[\rho(t) M_k^\dagger M_k].
    \label{eq:pk_def}
\end{equation}
The post-measurement state $\rho_{k}(t+\Delta t)$, conditioned on the observation $k$ is given by
\begin{equation}
    \rho_{k}(t+\Delta t) = \frac{M_k\rho(t) M_k^\dagger}{\Tr[\rho(t) M_k^\dagger M_k]},
    \label{eq:state_meas}
\end{equation}
where we account for finite measurement time $\Delta t$.
When studying a qubit system, we consider quantum measurements of the observable $\sigma_\theta=\cos\theta \sigma_{\rm z} + \sin\theta \sigma_{\rm x}$, where $\theta \in [0,2\pi)$. Weak discrete quantum measurement are then described by the two Kraus operators $\{M_{+}, M_{-}\}$ given by \cite{yanik2022thermodynamics}
\begin{equation} \label{eq:discrete_measurement_operators}
M_{\pm} = \frac{1}{2}\left(\sqrt{\kappa}+\sqrt{1-\kappa}\right)I_2
\pm \frac{1}{2}\left(\sqrt{\kappa}-\sqrt{1-\kappa}\right)
\sigma_\theta,
\end{equation}
where $I_2$ is the $ 2\times 2$  identity, and $\kappa\in[1/2, 1]$ is an indicator of the strength of the discrete measurement \cite{jacobs2003,wiseman2009,jordan2024quantum}. The $\kappa\to 1$ limits describe strong (projective) measurements, where the demon acquires maximum information about the system, whereas $\kappa\to 1/2$ describes the opposite limit, where no information is acquired. Intermediate values of $\kappa$ describe the transition from strong to weak measurements.

In the case of continuous measurements, we have a continuum of Kraus operators $\{{M}_{k}\}_{k}$ where $k$ is now a continuous measurement outcome. They are given by \cite{jordan2024quantum} 
\begin{equation}
\begin{aligned}
&{M}_{k}=\bigg(\frac{\Delta t}{2\pi\tau_\text{m}}\bigg)^{\frac{1}{4}}\exp{-\frac{\Delta t\left(k I_2 -\sigma_{\theta}\right)^2}{4\tau_\text{m}}},
\end{aligned}
\label{eq:m_continuous}
\end{equation}
for the measurement of $\sigma_\theta$. 
Crucially $\tau_\text{m}$, which is the characteristic measurement timescale, can be seen as the inverse of the measurement strength \cite{dressel2017,yanik2022thermodynamics,jordan2024quantum} and is the time required
to achieve unit measurement signal to noise ratio \cite{dressel2017}.
When $\Delta t/\tau_\text{m}$ is large (small), the measurement is referred to as strong (weak).
Following Eq.~(\ref{eq:m_continuous}), the measurement readout is described by two Gaussian distributions with variance $\tau_\text{m}/\Delta t$ and mean $+1$ (associated with the $\sigma_\theta=+1$ measurement outcome) and $-1$  (associated with the $\sigma_\theta=-1$ measurement outcome). 
\subsection{Demon's optimization goals}
Let $P(t)$ be the instantaneous cooling power, i.e., the heat flux flowing out of the thermal bath. It will be zero when we measure or evolve unitarily. When thermalizing, it is given by \cite{alicki1979,binder2019}
\begin{equation}
	P(t) = \Tr[\mathcal{D}[\rho(t)] H_T[u(t)]].
\end{equation}Notice that such quantity depends on the specific trajectory, i.e., on the stochasticity of the measurement outcome and (eventually) on the stochasticity of the choice of the feedback. Since we are interested in the average long-term performance of the demon, we define the ``average long-term cooling power'' $\ev{P}$ as
\begin{equation}
	\ev*{P} := E\left[ \frac{1}{\mathfrak{T}} \int_{t_0}^\infty e^{-(t-t_0)/\mathfrak{T}}  P(t) \, dt\right],
\end{equation}
where $t_0$ is an arbitrary initial time, $E[\cdot]$ represents the average over all trajectories, and the time integral is an exponentially weighted time average with typical timescale $\mathfrak{T}$. Notice that, by taking such an average over all trajectories, $\ev*{O}$ does not depend on $t_0$. Ideally, we are interested in the limit $\mathfrak{T}\to \infty$, i.e., $\mathfrak{T} \gg \tau_\text{u},\tau_\text{t},\tau_\text{m}$.

To quantify the ``cost of measuring'', 
we consider the solution proposed by Bennett to the Maxwell's demon's paradox. When the demon measures the quantum system, it stores the measurement outcome as classical information. Landauer showed that the  minimum work necessary to erase such information is given by
\begin{equation} \label{eq:Landauer}
	W[\{p_k(t)\}_k] = \frac{1}{\beta} S[\{p_k(t)\}_k],
\end{equation}
where $S[\{p_k\}_k]$ is the Shannon entropy of the probability distribution $p_k$, and $p_k(t)$ is the distribution of measurement outcomes if we measure at time $t$, given by Eq.~(\ref{eq:pk_def}). Since the measurement outcomes are stored in a classical memory by the demon, the entropy of the classical memory will correspond to the entropy of the $p_k(t)$ distribution. We thus define as ``instantaneous dissipation rate'', the quantity
\begin{equation}
	D(t) = \sum_i  W[\{p_k(\tilde{t}_i)\}_k]\cdot \delta(t-\tilde{t}_i),
\end{equation}
where $\tilde{t}_i$ are the (stochastic) times at which we perform a measurement. As we did for the cooling power, we define the long-term average dissipation as
\begin{equation}
	\ev*{D} := E\left[\frac{1}{\mathfrak{T}} \int_{t_0}^\infty e^{-(t-t_0)/\mathfrak{T}}  D(t)\,dt\right] .
\end{equation}

Applying the second law of thermodynamics to the thermal bath, the quantum system, and the heat dissipated by Landauer erasure of the classical memories and averaging over time and trajectories, we find
\begin{equation}
    - \beta \mathfrak{T}\ev{P} + \beta \mathfrak{T}\ev{D} + \ev{\Delta S} \geq 0, 
\end{equation}
where the first term represents the entropy change of the thermal bath,
the second term is the (minimum) entropy dissipated by Landauer erasure, and the third term is the entropy change of the quantum system. Notice that, thanks to the long-time average and the average over trajectories, the entropy change 
of the quantum system will be zero in expectation. This leads to
\begin{equation}
	\eta := \frac{\ev*{P}}{\ev*{D}} \leq 1,
\end{equation}
where we define $\eta>0$ as the \textit{measurement efficiency}. Indeed, if $\eta=1$, we are converting into cooling power all of the heat that we dissipate by Landauer erasure. In this sense, we are making optimal use of all the classical information acquired about the system; if instead $\eta < 1$, we have irreversibilities.

Interestingly, one cannot simultaneously optimize the average cooling power $\ev*{P}$ and dissipation $\ev*{D}$. Indeed, zero dissipation implies a vanishingly small cooling power (since the overall transformation must be reversible), while a high cooling power can be achieved through many measurements and high dissipation. We thus perform a \textit{multi-objective optimization} searching for the Pareto-optimal trade-off between the two quantities. These are feedback control schemes where one objective ($\ev*{D}$ or $\ev*{P}$) cannot be further improved without sacrificing the other one \cite{erdman2023_prr}. The collection of $(\ev*{D},\ev*{P})$ points delivered by all Pareto-optimal feedback control schemes is known as the Pareto front. If the Pareto front is convex, it can be found by maximizing the figure of merit
\begin{equation}
	\ev*{F_c} := c \ev*{P} - (1-c)\ev*{D}
	\label{eq:f}
\end{equation}
with respect to all possible feedback control schemes and repeating such an optimization for all values of $c\in[0,1]$. Solutions found for $c=1$ correspond to maximum cooling power, whereas solutions for $c=0$ correspond to minimum dissipation (notice the minus sign in front of $\ev*{D}$). For intermediate values of $c$, we will find different trade-offs between the two objectives. Using the linearity of the expectation value, notice that we can interpret Eq.~(\ref{eq:f}) as the trajectory and time average of the instantaneous quantity
\begin{equation}
	F_c(t) := c\,P(t) - (1-c) D(t).
	\label{eq:ft}
\end{equation}

\section{Reinforcement Learning Formulation}
\label{sec:method}
\begin{figure}[!tb]
	\centering
	\includegraphics[width=0.99\columnwidth]{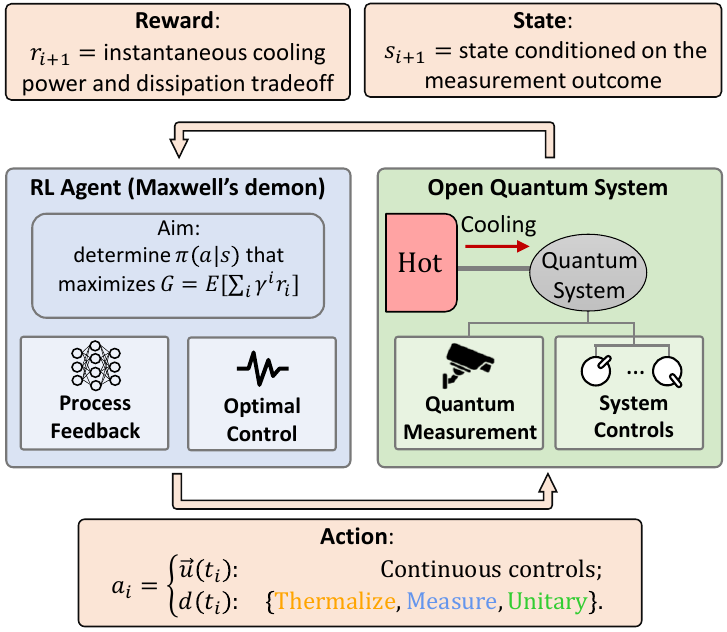}
	\caption{Schematic representation of the RL method learning to act as an optimal quantum Maxwell's demon. At every small time-step $t_i=i\Delta t$, the RL agent (blue box) interacts with an open quantum system (green box) performing actions $a_i$ (lower orange box) consisting of a discrete choice (whether to thermalize, measure, or evolve unitarily), and a continuous action (representing some time-dependent control). After the quantum state has evolved for a time-step $\Delta t$, the agent receives as feedback the state of the environment $s_{i+1}$, given by the density matrix of the quantum system conditioned on the measurement outcome, and a reward $r_{i+1}$ representing the optimization goal (upper orange boxes). The computer agent must learn an optimal policy $\pi^*(a|s)$ that maximizes the long-term and trajectory averaged trade-off between cooling power and cost of measurement. Through the trial and error attempt, the computer agent learns a gradually better and better policy until convergence. }
	\label{fig:setup}
\end{figure}

Reinforcement learning is a general optimization framework based on Markov decision processes \cite{sutton2018}, which consists of learning how to maximize a long-term goal by repeated interactions between an agent and an environment. As shown in Fig.~\ref{fig:setup}, interactions between the agent (blue box) and the environment (green box) occur at every time-step $t_i$. We choose as the environment a quantum system that can be coupled to a thermal bath, measured, and controlled by unitary evolution (green box). At each time-step $t_i = i\Delta t$, the agent chooses an action $a_i$ to perform on the environment (lower orange box) according to the \textit{policy function} $\pi(a_i|s_i)$, which represents the probability of choosing action $a_i$, given that the environment is in state $s_i$.
We choose as $a_i= \{d_i, u_i\}$ a combination of a discrete $d_i$ and continuous $u_i$ action. As discrete action, we choose the three possibilities $d_i \in \{ \text{Unitary, Thermalize, Measure} \}$ described in Sec.~\ref{sec:model}, corresponding respectively to performing unitary feedback, a (partial) thermalization, or a measurement. As continuous action $u_i$ we choose the value of the control $u(t_i)$, assumed to be constant in each time-interval $[t_i, t_{i+1}]$, corresponding either to the value of a Hamiltonian parameter or to a rotation angle of the qubit state. The environment then returns as feedback what is the state $s_{i+1}$ at the next time-step $t_{i+1}$, and a scalar quantity $r_{i+1}$, known as the reward.

Crucially, we choose as a state of the environment $s_i = \rho(t_i)$, i.e., the reduced density matrix of the quantum system \textit{conditioned on the measurement outcomes}. Since $\rho(t_i)$ encodes all information previously acquired by measuring the state, and since actions are chosen based on the state of the environment, this allows the agent to learn feedback control strategies $\pi(a_i|\rho(t_i))$ that make use of all previous measurement outcomes. 

The goal of RL in the so-called \textit{continuing task} is to identify an optimal policy function $\pi^*(a|\rho)$ such that, at every time-step $t_i$, the expectation value of the exponentially weighted sum of future rewards is maximized, i.e.,
\begin{equation}
	\pi^*(a|\rho) = \underset{\pi}{\text{argmax}}\,E\left[ r_{i+1} + \gamma  r_{i+2} + \gamma^2  r_{i+3} + \dots \right].
	\label{eq:pi_star}
\end{equation}
The reward $r_{i+1}$ is a (possibly stochastic) scalar quantity that only depends on the last state and the last action chosen, i.e., on $\rho(t_i)$ and $a_i$, whereas $\gamma \in [0,1)$ is the \textit{discount factor} which determines the timescale over which we optimize future rewards. The expectation value $E[\cdot]$, as previously, represents an average over all trajectories. 
Let us choose as the reward function
\begin{equation}
	r_{i+1} := \int_{t_i}^{t_{i+1}} F_c(t)\,dt.
	\label{eq:r}
\end{equation}
Plugging Eq.~(\ref{eq:r}) into Eq.~(\ref{eq:pi_star}), and using Eq.~(\ref{eq:ft}), 
we see that, in the limit $\Delta t\to 0$,
\begin{equation}
	\pi^*(a|\rho) = \underset{\pi}{\text{argmax}}\,\ev{F_c},
	\label{eq:pi_star_f}
\end{equation} 
where we choose
\begin{equation}
	\gamma := e^{-1/\mathfrak{T}}.
\end{equation}
As we can see, the discount factor $\gamma$ plays the role of the time-average timescale, with $\gamma \to 1$ corresponding to $\mathfrak{T}\to \infty$. 
Identifying the optimal policy $\pi^*(a|\rho)$ defined in Eq.~(\ref{eq:pi_star_f}) is a 
formidable task. Indeed, in $\ev*{F_c}$, we have an expectation value over time, over all possible stochastic measurement outcomes and stochastic actions, and the space of feedback schemes increases exponentially with the number of time steps.

Notice that with this choice of environment, state, and reward, the agent is behaving as an actual Maxwell's demon, and identifying the optimal policy precisely corresponds to discovering the best performing Maxwell's demon.
In this work, we employ a modification of the 
\emph{soft actor-critic} (SAC) algorithm \cite{haarnoja2018_pmlr,haarnoja2018_arxiv_sac, haarnoja2018_arxiv_walk} to solve the optimization problem stated in Eq.~(\ref{eq:pi_star_f}) with rewards given by Eq.~(\ref{eq:r}). The method, implemented in PyTorch \cite{paszke2017}, is detailed in App.~\ref{app:rl} (see also the Code and Data Availability Statement). In practice, we start from a random policy $\pi(a|\rho)$, and by interacting many times with the quantum system, we iteratively improve the policy until we find a (quasi) optimal policy. We either consider the case of maximum power ($c=1$) or repeat the optimization process for many values of $c$, and for each one, we compute the corresponding values of the dissipation $\ev*{D}$ and the cooling power $\ev*{P}$. Notably, the RL method naturally optimizes the sum of the reward averaging 
over time (with a timescale determined by $\gamma$) and over the trajectories, i.e., over the stochasticity of the measurement outcome and the policy. See App. \ref{app:rl} for details on the RL method and implementation.

\section{Thermalization dominated regime} 
\label{sec:tau_t}
We start by studying the regime where the thermalization timescale $\tau_\text{t}$ is the slowest one, i.e., $\tau_\text{t} \gg \tau_\text{m}, \tau_\text{u}$.
We show that the main impact of the thermalization timescale is the emergence of strategies where a single measurement is performed, followed by a carefully designed and finite-time thermalization protocol, conditioned on the last measurement outcome, whose duration depends on the power-efficiency trade-off (the more we are interested in the cooling power, the shorter the thermalization and the higher the dissipation).
This regime can be a realistic scenario in those setups
where the coupling to the thermal baths is engineered to be slow compared to the other timescales, as in Ref.~\cite{maillet2019}. We begin for simplicity with a single qubit, and later we generalize it to a larger system.
Specifically, we consider a single qubit system described by the Hamiltonian
\begin{equation} \label{eq:qubit_with_controlled_gap}
    H_\text{U}[u(t)] = H_\text{T}[u(t)] = u(t)\frac{E_0}{2} \sigma_{\rm z},
\end{equation}
where $E_0$ is a constant that sets the typical energy gap of the qubit, and $u(t)$
is a continuous time-dependent control. In the experiment described in Ref.~\cite{maillet2019}, this would correspond to a time-dependent gate voltage.

We describe finite-time thermalization using the Lindblad master equation with characteristic thermalization rate $\Gamma$. We have the following rates and jump operators for $k=\pm$ [see Eq.~(\ref{eq:dissipator})]
\begin{align}
	A_{\pm,u(t)} = \sigma_\pm,~~~~~~~~  \gamma_{\pm,u(t)} = \Gamma \, |n(\pm \beta u(t))|.
\end{align}
Here $\sigma_{\pm} =(\sigma_{\rm x} \pm i\sigma_{\rm y})/2$, $\Gamma$ is a constant that depends on the qubit-bath coupling strength, and $n(x)=(\exp(x)-1)^{-1}$ is the Bose-Einstein distribution function. We then choose a value of $\Delta t$ that is much smaller than the thermalization timescale of the qubit. To enforce that $\tau_\text{t} \gg \tau_\text{m}$, we consider projective measurements in the $\sigma_{\rm z}$ basis that occur in a single time-step $\Delta t$. 

Since the only control is on $\sigma_{\rm z}$ and we only perform measurements of $\sigma_{\rm z}$, the Lindblad dynamics of the system is fully characterized by the probability $p(t)=\Tr[\rho(t)\ket{1}\bra{1}]$ of being in the higher eigenstate $\ket{1}$ of $\sigma_{\rm z}$, and all dependence on the coherence drops. 
Mathematically, this is equivalent to a classical two-level system, which can be regarded as having no unitary dynamics timescale ($\tau_\text{u}=0$). Notice that by changing the sign of $u(t)$, we can exchange the ground and the excited state of the system; this will be exploited by the agent to apply conditional feedback to the system.

\begin{figure}[!tb]
	\centering
	\includegraphics[width=0.99\columnwidth]{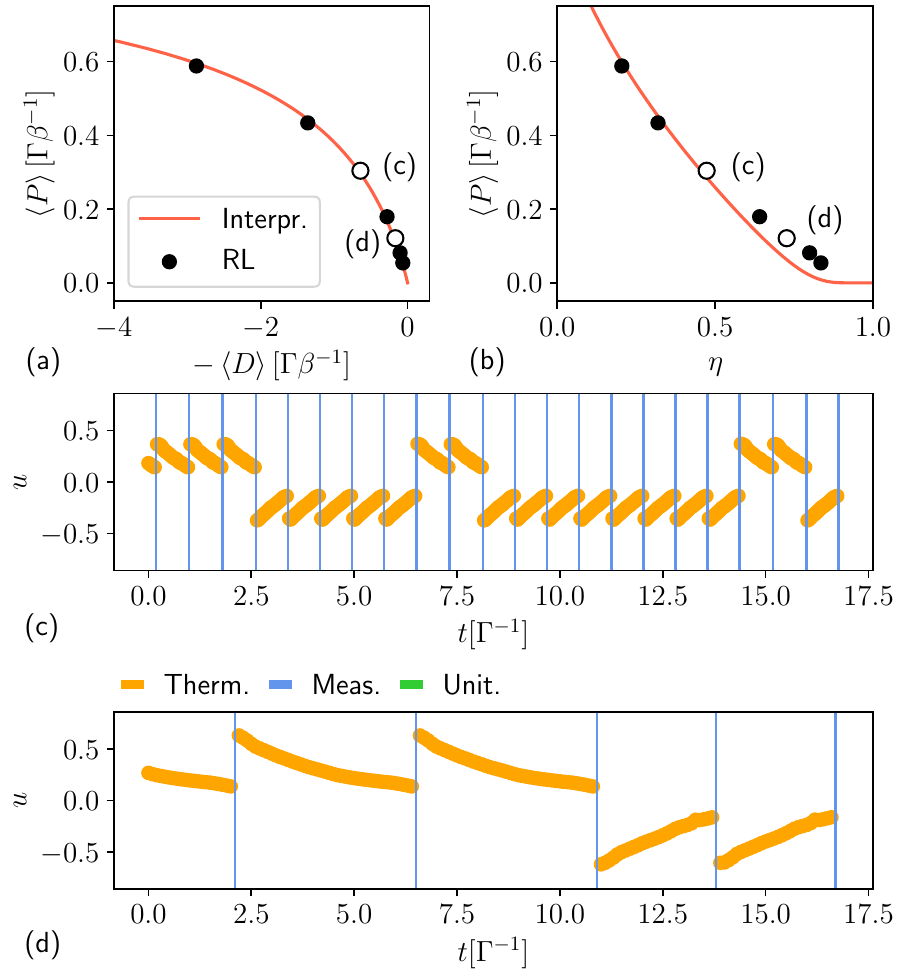}
	\caption{Maxwell's demon performance in the thermalization-dominated regime. Pareto front between the long-term average of the cooling power $\ev{P}$ and of the negative measurement dissipation $-\ev{D}$ (a), and between the cooling power $\ev{P}$ and the measurement efficiency $\eta$ (b). Each black dot corresponds to a separate RL optimization for different values of $c$, whereas the red line corresponds to the Pareto front of the interpretable policy described in Sec.~\ref{sec:tau_t}. Example of actions chosen by the agent along an arbitrary trajectory, as a function of time, in the $c=0.8$ case (c) and $c=0.65$ case (d). The corresponding points on the Pareto front are shown in (a,b) as empty black circles. The color corresponds to the discrete action (see legend), and the value shown on the $y$-axis corresponds to the value of $u(t)$ during the thermalization step. Measurements are shown as vertical lines. Parameters: $E_0=5\,\beta^{-1}$, $\Gamma=1\,(\beta\hbar)^{-1}$, $u(t)\in [-0.8, 0.8]$, and $\Delta t = \{0.003, 0.02, 0.03, 0.07, 0.1, 0.2, 0.3\} \,\Gamma^{-1}$ for $c=\{0.95,0.9, 0.8,0.7,0.65, 0.6, 0.58\}$ since the thermalization time increases as $c$ decreases.}
	\label{fig:therm_pareto}
\end{figure}
In Fig.~\ref{fig:therm_pareto}(a), we plot the Pareto front between the long-term average of the cooling power $\ev*{P}$ and of the negative dissipation $-\ev*{D}$. Each black dot corresponds to a separate optimization performed with a different value of $c$. In Fig.~\ref{fig:therm_pareto}(b), we plot the same points as in Fig.~\ref{fig:therm_pareto}(a), but plotting the Pareto front between the average cooling power and the measurement efficiency $\eta$. As we can see, we find a class of policies that trade between high power and high efficiency.

To visualize the type of policies that we find, in Figs.~\ref{fig:therm_pareto}(c,d), we plot an example of the chosen actions along an arbitrary trajectory as a function of time, respectively, for $c=0.8$ and $c=0.65$. The corresponding value on the Pareto front is shown in Figs.~\ref{fig:therm_pareto}(c,d) as an empty black circle.
In both cases, the agent 
learns the following policy:
\begin{itemize}
    \item Perform a projective measurement (blue vertical line).
    \item Perform feedback: if the system is in the ground state, do nothing; if it is in the excited state, change the sign of $u(t)$ as to have the system in the ground state.
    \item Perform a finite-time thermalization of the qubit (orange dots) while ramping the value of $u(t)$ (plotted on the $y$-axis). The qubit, being in the ground state, absorbs heat from the bath, thus cooling it.
\end{itemize}
Feedback conditioned on the last measurement outcome is clearly visible in Fig.~\ref{fig:therm_pareto}(c,d) as changes of sign of the orange thermalization ramps. Furthermore, as we would intuitively expect, the agent automatically learns to increase the duration of the thermalization as we shift our interest from high power to high efficiency. Indeed, a slow thermalization with a slow ramping of the control allows for the maximization of the heat extracted per measurement at the cost of taking a longer time, thus decreasing the cooling power but increasing the measurement efficiency. 

In the $c\to 1$ limit, in which we are only interested in power, we find that the optimal policy consists of performing measurements followed by infinitesimally short thermalization. Interestingly, this implies that the optimal duration of the thermalization [given by the duration of the orange ramps in Figs.~\ref{fig:therm_pareto}(c,d)], is only determined by the value $c$, i.e., by how much we are interested in trading 
power and efficiency. We also notice that the agent learns to never perform unitary actions: indeed, in this system, a unitary evolution would not change the state $p(t)$, nor would it exchange any heat with the thermal bath, making it always a suboptimal choice. The slight asymmetry in the thermalization ramps in Fig.~\ref{fig:therm_pareto}(d) between positive and negative values of the control $u(t)$ is likely due to numerical imperfections in the RL optimization and hints at the fact that the performance only weakly depends on the exact shape of the thermalization ramp. We now confirm this substantially more quantitatively.

Thanks to the intuition gained from analyzing the behavior of the RL agent, we define an ``interpretable policy'' that behaves as in the bullets described above, but thermalizes the qubit while keeping the control $u(t)$ fixed at some value $\bar{u}$ for some time $\bar{\tau}$. Therefore, we do not perform the smooth ramping of $u(t)$ shown in Figs.~\ref{fig:therm_pareto}(c,d), but we keep it constant. We numerically optimize its performance for many values of $c$, with respect to the two parameters $\bar{u}$ and $\bar{\tau}$, and plot the corresponding performance in the Pareto fronts in Figs.~\ref{fig:therm_pareto}(a,b) as a red thin line.

Notably, the interpretable policy captures the main features of the Pareto front and the increase of $\tau$ from $0$ (when we are only interested in power) to a larger and larger value as we shift our interest from power to efficiency. However, the non-trivial ramping of the control $u(t)$ discovered by the RL agent slightly outperforms the interpretable policy in the high-efficiency regime [see the black dots in Fig.~\ref{fig:therm_pareto}(b) that lie above the red curve for $\eta>0.5$]. This is due to the further reduction of irreversibilities during the thermalization process thanks to a smooth ramping of $u(t)$, a phenomenon that was proven analytically in the slow-driving regime and observed experimentally \cite{scandi2022}.

\section{Measurement dominated regime}
\label{sec:tau_m}
We now study the regime where the measurement timescale $\tau_\text{m}$ is the slowest one, i.e., $\tau_\text{m} \gg \tau_\text{t}, \tau_\text{u}$. 
We show that optimal feedback control strategies involve adaptively measuring the qubit in different bases according to the acquired information. Furthermore, the main impact of the measurement timescale is generally the emergence of weak measurements that are repeated until the state is sufficiently pure, leading to a \textit{random walk} in state space, followed by short unitary and thermalization steps conditioned on the previous outcomes. The use of continuous or discrete measurements leads to qualitatively similar feedback control strategies, with the former exploring a larger portion of the state space of the qubit because of the continuous measurement outcomes.
This regime is experimentally relevant since in most superconducting qubit platforms, the time to perform, e.g., a $\pi$ pulse, i.e., a unitary rotation, is much faster than the timescale for a measurement readout \cite{gunyho2024single,nguyen2024programmable}. 

Specifically, we consider a single qubit system whose Hamiltonian is given by Eq.~(\ref{eq:qubit_with_controlled_gap}). We then proceed incrementally: in Subsec.~\ref{sec:meas_dis}, we consider discrete weak measurements of a fixed observable that may or may not generate coherence in the Hamiltonian basis. In Subsec.~\ref{subsec:learn_basis}, we allow the agent to learn and adaptively change the observable to measure. In Subsec.~\ref{sec:cont_x_z}, we consider continuous weak measurements.

\subsection{Discrete, fixed measurements}
\label{sec:meas_dis}
We start by considering a fixed measurement basis.
In this section, we model the three discrete actions in the following way. During thermalization, we keep the qubit's energy gap constant at $E_0=\beta^{-1}/2$, and we consider a qubit-bath coupling strength of $\Gamma\Delta t=0.8$, which brings the qubit state close to a thermal state in a single time-step $\Delta t$.
We consider discrete yet non-projective measurements of $\sigma_{\rm x(z)}$ with measurement strength $\kappa$ as described by Eq.\  (\ref{eq:discrete_measurement_operators}). When performing the unitary dynamics, in a single time-step $\Delta t$ we rotate the state of qubit about the $y$-axis of an angle $\phi_{\rm y} \in [0,2\pi]$ that the RL agent can choose.

\begin{figure}[!tb]
	\centering
	\includegraphics[width=0.99\columnwidth]{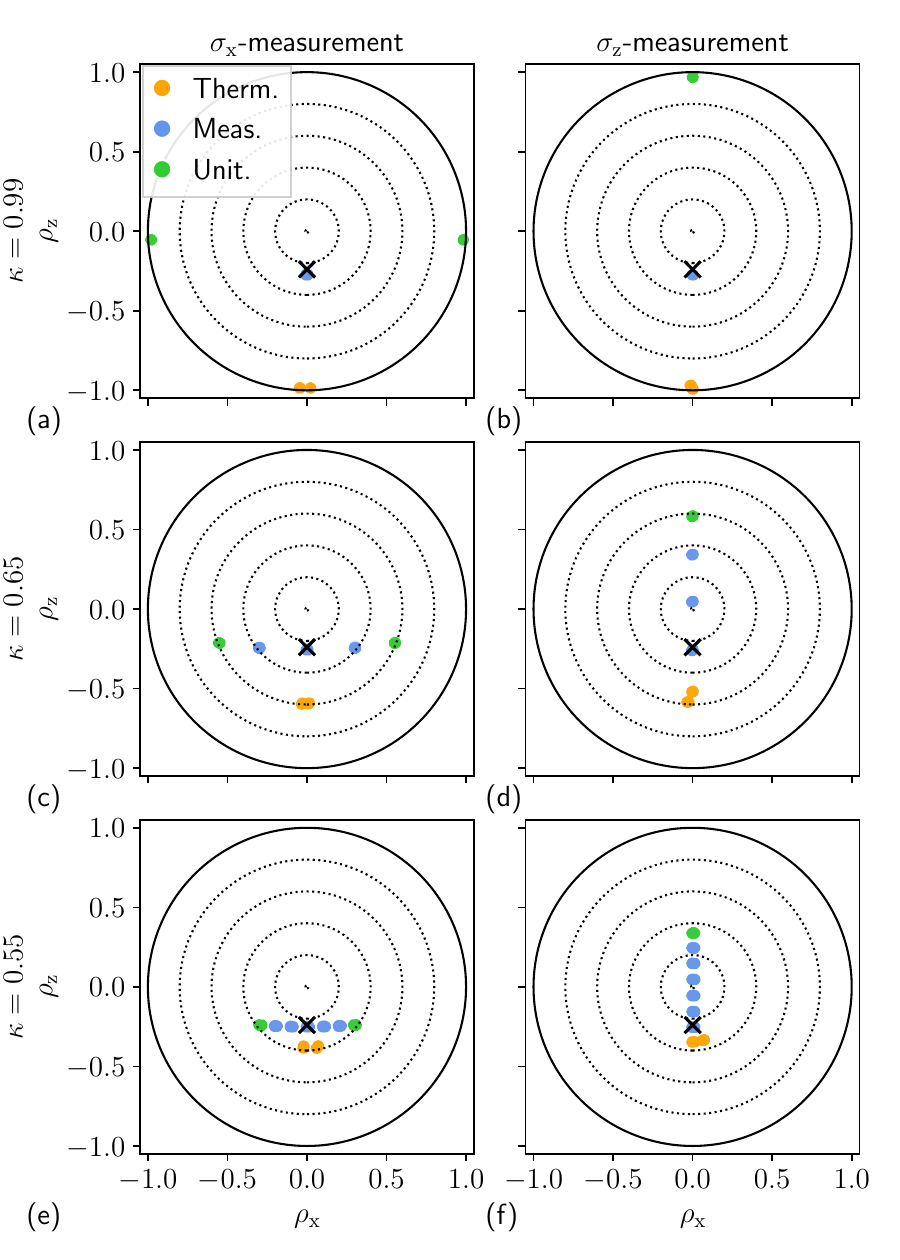}
	\caption{Actions chosen by the RL agent as a function of qubits state represented as a point on the Bloch sphere in the measurement-dominated regime with discrete fixed measurements. The type of action is represented by the color of a point, and the black cross represents the thermal state. We allow the RL agent to cool the thermal bath either by $\sigma_{\rm x}$ measurement (left column) or $\sigma_{\rm z}$ measurement (right column). $\rho_{\rm x}$ and $\rho_{\rm z}$ denote qubit's $x$ and $z$ coordinates on the Bloch sphere. Each row represents a different measurement strength, respectively $\kappa=\{0.99, 0.65, 0.55\}$. Each plot represents a trajectory using the RL policy for 10,000 time steps with parameters $\Gamma=0.8 \,(\beta\hbar)^{-1}$, $\Delta t=0.8\,\Gamma^{-1}$. $E_0=0.5\,\beta^{-1}$.  }
	\label{fig:Bloch_discr_meas}
\end{figure}

In Fig.\ \ref{fig:Bloch_discr_meas} we show the results of the RL agent in various scenarios: each row represents a different value of the measurement strength $\kappa$, and each column the type of measurement that the agent can perform (either $\sigma_{\rm x}$, left, or $\sigma_{\rm z}$, right).
The position of the dots in Fig.~\ref{fig:Bloch_discr_meas}
represent the states of the qubit visited while interacting with the RL agent (their $\rho_{\rm z}=\Tr[\rho\sigma_{\rm z}]$ and $\rho_{\rm x}=\Tr[\rho\sigma_{\rm x}]$ components are plotted, since $\rho_{\rm y}=\Tr[\rho\sigma_{\rm y}]=0$), and the color of the dots represents the corresponding action chosen by the agent after visiting such states.

Interestingly, aside from important differences for different values of $\kappa$ commented below, the following general and interpretable strategy emerges from all cases.
\begin{enumerate}
    \item Let us assume that the qubit is in a thermal state (black cross in Fig.~\ref{fig:Bloch_discr_meas}).
    \item Perform a measurement of the qubit (blue dots). 
    \item If the purity of the state reaches a threshold value (green dots), go to step 4; otherwise, go back to step 2 (blue dots).
    \item Perform a unitary rotation of the qubit to the negative $z$-axis (orange dots).
    \item Thermalize the qubit, leading us back to step 1.
\end{enumerate}
This strategy can be easily interpreted. While unitary control cannot change the purity of the state (i.e., the radius of the state on the Bloch sphere), the measurement process gives us information about the state of the system and thus tends to purify the state of the qubit. The RL agent learns to repeat measurements until the state is pure enough and then rotates the state such that we are nearest to the ground state of the qubit (corresponding to $\rho_{\rm z}=-1$). Then, the qubit is allowed to thermalize, absorbing heat from the bath and consequently decreasing its purity. The process is then repeated. 

As we could expect, for equal measurement strength $\kappa$, measuring along $\sigma_{\rm z}$ and $\sigma_{\rm x}$ leads to similar control strategies, with the state either moving along the $y$ or $x$ axis (see blue and green dots in Fig.~\ref{fig:Bloch_discr_meas}). Measuring along $\sigma_{\rm z}$ displays a small advantage since we can skip the unitary rotation if the measurement process naturally drives the state near the ground state. On the contrary, if we measure $\sigma_{\rm x}$, we always have to rotate the state since the Hamiltonian and the measurement operator do not commute.

\begin{figure}[!tb]
	\centering
	\includegraphics[width=0.99\columnwidth]{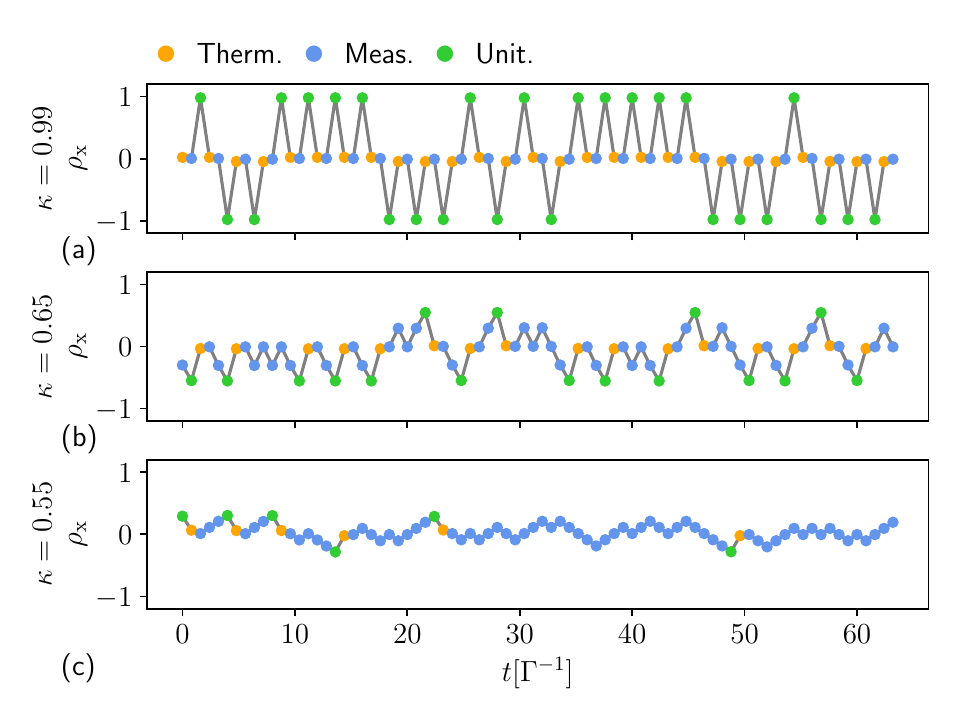}
	\caption{Plots of the state component $\rho_{\rm x}$, as a function of time, corresponding to the results shown in Fig.~\ref{fig:Bloch_discr_meas}(a,c,e) relative to the $\sigma_{\rm x}$ measurement case. An arbitrary trajectory is shown. The colors of the dots indicate the actions chosen by the RL agent at a given moment.}
	\label{fig:x_logs}
\end{figure}

Notably, the optimal control strategy strongly depends on the measurement strength $\kappa$. For nearly projective measurements ($\kappa=0.99$), shown in Fig.~\ref{fig:Bloch_discr_meas}(a,b), the RL agent naturally discovers a known control strategy described in Ref.~\cite{yanik2022thermodynamics}. This corresponds to the limit where, after a single measurement, we are guaranteed to have a pure state. Therefore, assuming we are in a thermal state, it is always sufficient to perform one measurement step (single blue dot), then one unitary step (green dots), followed by a thermalization step (orange dot). As discussed above, the unitary step can be skipped in the $\sigma_{\rm z}$-measurement case if the post-measurement state is already the ground state of the qubit [orange dot in panel (b)].

However, as we move to weaker measurements [see Fig.~\ref{fig:Bloch_discr_meas}(c-f)], the agent starts repeating multiple measurements (blue dots) before performing a unitary rotation (green dots). In this regime, the state of the qubit experiences a \textit{random walk} in state space until it reaches a target purity (which is automatically learned by the RL agent). 
This is clearly visible in Fig.~\ref{fig:x_logs}, where a trajectory of visited states (represented by $\rho_{\rm x}$) and corresponding actions (colors) are plotted as a function of time for the $\sigma_{\rm x}$-measurement case. As in Fig.~\ref{fig:Bloch_discr_meas}, every row corresponds to a different value of $\kappa$ (qualitatively similar results are found in the $\sigma_{\rm z}$ case, see App.~\ref{sec:app_discr_sigma_z}).
As we can see, the weaker the measurement, the longer the random walk and the number of states visited while measuring (compare $\kappa=0.65$ with $\kappa=0.55$ in Figs.~\ref{fig:Bloch_discr_meas} and \ref{fig:x_logs}).

Another interesting feature is the change of target purity (measured, e.g., as the norm of the Bloch vector) before rotating the state as $\kappa$ varies. This can be seen noticing how the green dots move towards the center in Fig.~\ref{fig:Bloch_discr_meas} and towards $\rho_{\rm x}=0$ in Fig.~\ref{fig:x_logs} as we decrease the measurement strength. Since reaching a target purity requires more and more time as the measurement gets weaker, the RL agent learns to ``save time'' by choosing a lower and lower target purity at which the state is rotated and then thermalized.

\subsection{Discrete, adaptive measurements}
\label{subsec:learn_basis}
In this section, we consider the same setup and model as in the previous subsection. However, 
we explore the possibility of learning the best observable to measure instead of fixing it to either $\sigma_{\rm z}$ or $\sigma_{\rm x}$. As we will see, the optimal feedback control strategy involves adaptively changing the measurement basis according to the acquired information. Therefore, during each measurement step, we consider a measurement of $\sigma_\theta$ [see Eq.~(\ref{eq:discrete_measurement_operators})] and allow the agent to choose any value of $\theta \in [0,\pi]$. Furthermore, since the agent chose to rotate the qubit along the negative $z$-axis in all previous results, we here fix the unitary step to be a rotation to the negative $z$-axis and remove the freedom of choosing the rotation angle. The thermalization step is the same as in Sec.~\ref{sec:meas_dis}.

\begin{figure}[!tb]
	\centering
\includegraphics[width=0.99\columnwidth]{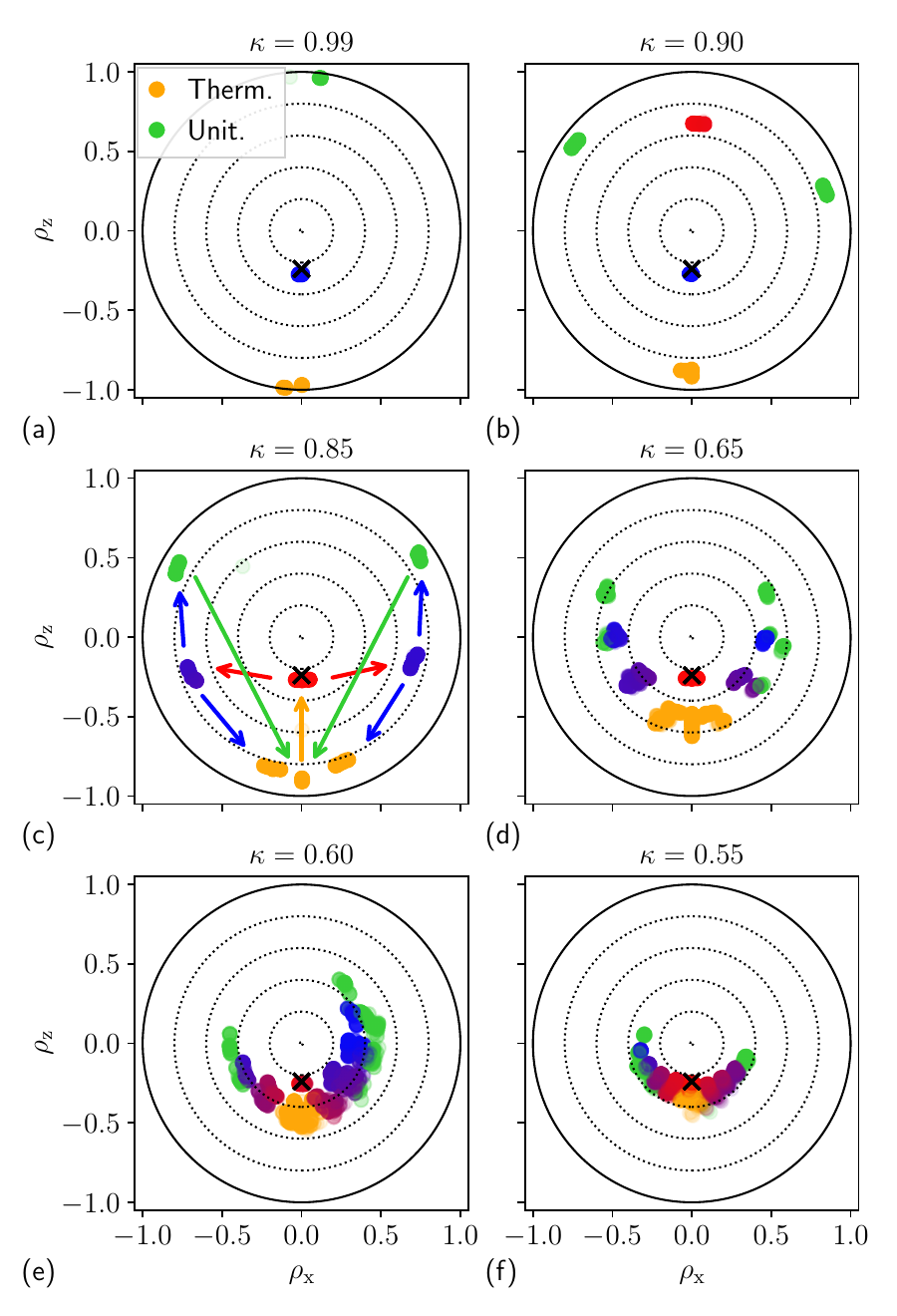}
\includegraphics[width=0.99\columnwidth]{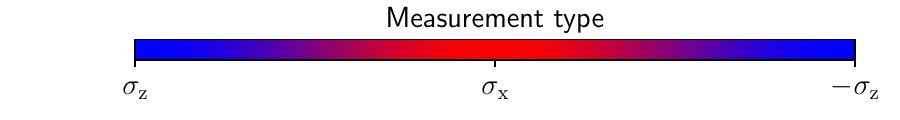}
	\caption{Actions chosen by the RL agent as a function of qubits state represented as a point on the Bloch sphere in the measurement-dominated regime with discrete, adaptive measurements. The action is represented by the color of a point (see legend). When a measurement is performed, we allow the RL agent to choose the angle of discrete measurement (see color bar). The black cross is the thermal state. Each plot corresponds to a different measurement strength and corresponds to a trajectory of 10,000 time-steps interacting with the RL agent. The same parameters were used as in Fig.~\ref{fig:Bloch_discr_meas}.}
	\label{fig:bloch_angle}
\end{figure}
In Fig.\ \ref{fig:bloch_angle}, we plot the results we find in the same style as Fig.~\ref{fig:Bloch_discr_meas}, but now the measurement action is not plotted as a blue dot, but as a color between blue and red depending on the learned angle (see color bar). 
For $\kappa=0.99$ [Fig.~\ref{fig:bloch_angle}(a)], the RL agent only chooses $\sigma_{\rm z}$ measurements, and thus finds the same cycle as in Fig.\ \ref{fig:Bloch_discr_meas}(b). 

\begin{figure}[!tb]
	\centering
	\includegraphics[width=0.99\columnwidth]{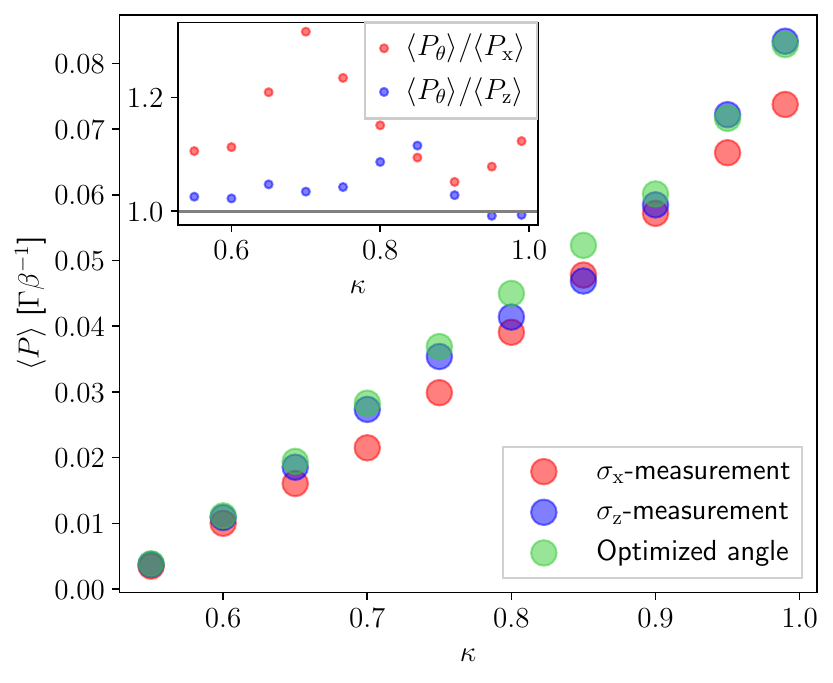}
	\caption{Comparison of the cooling power for fixed measurements of $\sigma_{\rm x}$ (red dots), fixed measurements of $\sigma_{\rm z}$ (blue dots), and for learnable measurements of $\sigma_\theta$ (green dots). The inset contains the ratio between the cooling power $\ev{P_\theta}$ with a learnable observable, and the cooling power $\ev{P_{\rm x}}$ ($\ev{P_{\rm z}}$) with fixed measurement of $\sigma_{\rm x}$ ($\sigma_{\rm z}$). Every dot is computed averaging over 100,000 time steps. The agent was trained for the same set of parameters as in Fig.~\ref{fig:Bloch_discr_meas} and  Fig.~\ref{fig:bloch_angle}. }
	\label{fig:policy_comparison}
\end{figure}

However, as we decrease the measurement strength, we see the emergence of points whose color interpolates between blue and red. This implies that the agent is performing different measurements based on the conditional state of the qubit.
It is thus not optimal to fix a single measurement basis, but we can increase the demon's performance by varying it adaptively. To prove that this is indeed the case, in Fig.~\ref{fig:policy_comparison}, we plot the cooling power as a function of $\kappa$, fixing $\sigma_{\rm x}$ measurements (red dots), $\sigma_{\rm z}$ measurements (blue dots), and with a tunable observable (green dots). In the inset, we further plot the performance improvement ratio
$\ev*{P_\theta}/\ev*{P_{\rm z}}$ ($\ev*{P_\theta}/\ev*{P_{\rm x}}$) i.e., the ratio between the power $\ev*{P_\theta}$ delivered by the agent with learnable angles, and the power $\ev*{P_{\rm z}}$ ($\ev*{P_{\rm x}}$) fixing the measurement basis to $\sigma_{\rm z}$ ($\sigma_{\rm x}$). Indeed, the ratio is greater or equal to $1$ in the majority of the cases, reaching up to $30\%$ improvements at $\kappa=0.7$ with respect to measurements of $\sigma_{\rm x}$. Interestingly, in all cases, the power decreases roughly linearly with the measurement strength parameter $\kappa$.

We notice that the same general feedback control strategy, detailed as bullet points in Sec.~\ref{sec:meas_dis}, also emerges in this case. This consists of a series of measurements until a target purity is reached, followed by a unitary rotation and a thermalization step. Also in this case, as the measurement strength decreases (i.e., as $\kappa$ goes from $1$ to $1/2$), the target purity reached when performing a unitary rotation (green dots) is progressively smaller and smaller in order to limit the duration of the measurements. 

However, the novelty here is in the choice of the measurement basis. As we can see from Fig.~\ref{fig:bloch_angle}(c-f), the agent chooses a measurement angle $\theta$ (blue to red dots) roughly perpendicular to the direction of the conditional qubit's state on the Bloch sphere. Indeed, when the state only has a $\rho_{\rm z}$ component, we see red dots corresponding to measurements along $\sigma_{\rm x}$. As the state acquires a $\rho_{\rm x}$ component, also the measurement angle changes (purple and blue dots).
\begin{figure}[!tb]
	\centering
	\includegraphics[width=0.99\columnwidth]{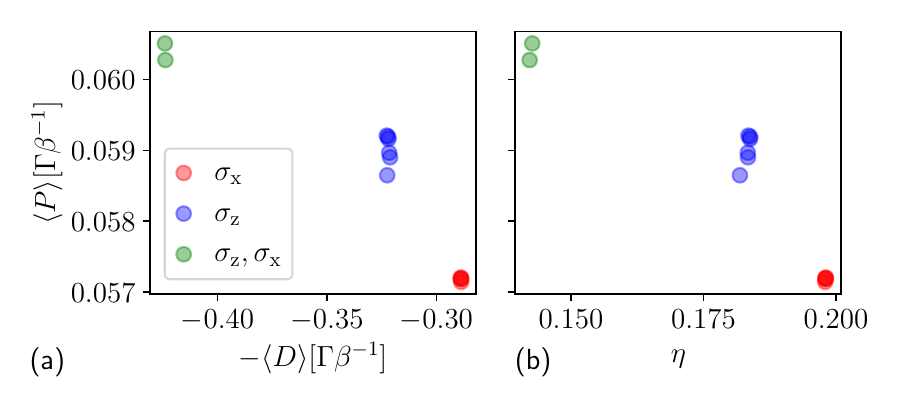}
	\caption{
 Maxwell's demon performance in the measurement-dominated regime with discrete, adaptive measurements. Pareto front between the average cooling power $\ev{P}$ and negative dissipation $-\ev{D}$ (a), and between $\ev{P}$ and the measurement efficiency $\eta$ (b). Each dot corresponds to a separate RL optimization for different values of $c$.  The red (blue) dots correspond to RL strategies that turn out to only measure $\sigma_{\rm x}$ ($\sigma_{\rm z}$), and the green dots that include both $\sigma_{\rm x}$ and $\sigma_{\rm z}$ measurements. We fix $\kappa=0.9$, and the other system parameters are as in Fig.~\ref{fig:Bloch_discr_meas}. }
	\label{fig:power_vs_penalty}
\end{figure}

This effect can be interpreted in the following way: as shown in App.~\ref{sec:app_second_measurement_angle}, and as discussed in Refs.~\cite{jacobs2003,jordan2006}, the average post-measurement purity $\bar{\gamma}$ of the qubit's state is given by
\begin{equation}
    \bar{\gamma} = 
     1 -\frac{1}{2} \frac{(1-l^2)(1-|{\rho}|^2)}{1-l^2 |{\rho}|^2\cos^2\alpha},
\end{equation}
where $l=2\kappa-1$, $|{\rho}|^2=\rho_{\rm x}^2 + \rho_{\rm z}^2$, and $\alpha$ is the angle between the state on the Bloch sphere the measurement axis. As can be easily seen, the average purity $\bar{\gamma}$ is maximized for $\alpha=\pm \pi/2$, which is indeed the measurement perpendicular to the qubit's state. Interestingly, the RL agent automatically learns to choose adaptively the measurement strategy that allows to maximize the purification of the state and thus the cooling power. To visualize this, let us consider Fig.~\ref{fig:bloch_angle}(c). When we are in the thermal state (black cross), the agent chooses a red action corresponding to a measurement of $\sigma_{\rm x}$, which is perpendicular to the qubit state. The state then stochastically evolves along the red arrows. Then, a blue measurement is performed, corresponding to a  $\sigma_\theta$ measurement, which is again roughly perpendicular to the qubit state. The state then evolves stochastically along the blue arrows. Now, the target purity is high enough, so the agent either decides to thermalize the state (orange dots and arrows) directly, or to rotate it (green dots and arrows) and then thermalize it (orange arrow). For lower values of the measurement strength, we see that the variation of the state after each measurement becomes smaller and smaller, and more measurements are performed. These measurements are roughly perpendicular to the qubit state, although we find small deviations that actually deliver marginally higher performance; see App.~\ref{sec:app_second_measurement_angle} for details.
The slight left-right asymmetries observed for some values of $\kappa$ are attributed to imperfect training of the RL agent.

So far, we have studied the maximization of the power. We conclude this section by assessing the impact of optimizing the trade-off between cooling power and measurement efficiency. In Fig.~\ref{fig:power_vs_penalty}, we show the Pareto front describing the trade-off between the average cooling power and measurement dissipation [panel (a)] and between power and efficiency [panel (b)] fixing $\kappa=0.90$. 
The points tend to cluster in three different regions. 
It indicates that the RL agent abruptly changes the control strategy as we shift our interest between power and efficiency; these strategies are explicitly shown in App.~\ref{sec:app_penalty}.
The highest power solutions, corresponding to the green dots, perform both $\sigma_{\rm z}$ and $\sigma_{\rm x}$ measurements, while solutions which sacrifice power for efficiency only perform $\sigma_{\rm z}$ (blue dots) or $\sigma_{\rm x}$ measurements (red dots). Therefore, as we shift our interest from power to efficiency, we observe changes both in the feedback control strategy and in the measurement basis.

\subsection{Continuous measurements}
\label{sec:cont_x_z}
\begin{figure}[!tb]
	\centering
	\includegraphics[width=0.99\columnwidth]{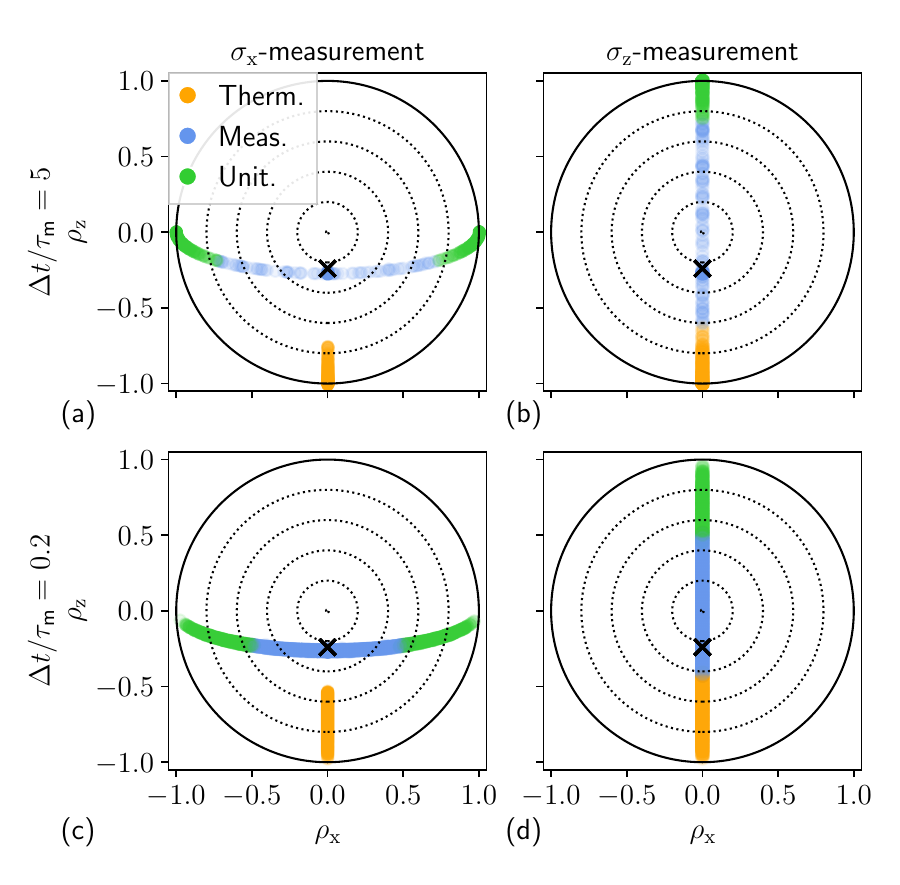}
	\caption{Actions chosen by the RL agent as a function of qubit's state represented as a point on the Bloch sphere in the measurement-dominated regime with continuous measurements. The type of action is represented by the color of a point. The black cross denotes the thermal state. The measurement is fixed and given by $\sigma_{\rm x}$ (left column) or $\sigma_{\rm z}$ (right column), and the measurement strength $\Delta t/\tau_\text{m}$ is reported on the plot. Each plot is generated by a trajectory with 10,000 time steps. The optimization was performed with parameters $\Gamma = 0.8\, (\beta\hbar)^{-1}$, $\Delta t=0.8\,\Gamma^{-1}$, $E_0=0.5\,\beta^{-1}$.}
	\label{fig:x_z_cont}
\end{figure}
We now consider the influence of continuous measurements instead of discrete measurements. As we will see, we find similar results as in the discrete case, but because of the continuous measurement outcomes, many more qubit states are visited, resulting in a ``smoothed'' version of the discrete measurement case.
Specifically, we model the thermalization and unitary parts as in the previous section (Sec.~\ref{subsec:learn_basis}), and we model the measurements using the continuous measurement formalism described in Eq.\  (\ref{eq:m_continuous}). 

As we did in Fig.~\ref{fig:Bloch_discr_meas} for discrete measurements, in Fig.~\ref{fig:x_z_cont}, we plot our results for the continuous measurement case using the same plotting style. In particular, the states and actions chosen by the agent are plotted for different values of the measurement strength $\Delta t/\tau_\text{m}$ by row (larger $\Delta t/\tau_\text{m}$ corresponds to stronger measurements), and for the $\sigma_{\rm x}$ and $\sigma_{\rm z}$ cases by column. In App.~\ref{app:cont_meas}, we consider the case where the RL agent can choose an arbitrary measurement angle.

Interestingly, we find feedback control strategies similar to the discrete measurement case, which can thus be understood and interpreted in a similar way. However, because of the continuous measurement outcomes, the qubit visits a much larger range of states. This results in qualitatively similar plots comparing Fig.~\ref{fig:x_z_cont} to \ref{fig:Bloch_discr_meas}, but the points appear now to be ``smeared out''.

\begin{figure}[!tb]
	\centering
	\includegraphics[width=0.99\columnwidth]{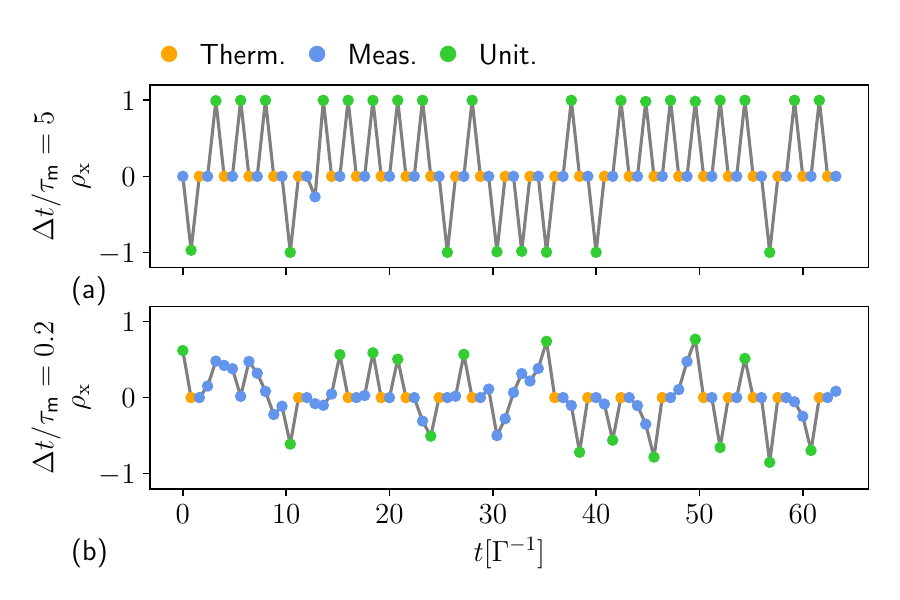}
	\caption{Plots of the $x$-coordinate of the qubit, as a function of time, along an arbitrary trajectory considering continuous $\sigma_{\rm x}$ measurements studied in Fig.~\ref{fig:x_z_cont}(a,c). The colors of the dots indicate the discrete action chosen by the RL agent (see legend). Each row corresponds to a different measurement strength $\Delta t/\tau_\text{m}$.}
	\label{fig:cont_x_logs}
\end{figure}
To better visualize the random walk performed by the state during the measurement process, in Fig.\ \ref{fig:cont_x_logs}, we plot the $\rho_{\rm x}$ component of the state, as a function of time, for the $\sigma_{\rm x}$ measurement case corresponding to Fig.~\ref{fig:x_z_cont}(a,c) (see App.~\ref{app:cont_meas} for the $\sigma_{\rm z}$ measurement case, where qualitatively similar results are found). For strong measurements [Fig.~\ref{fig:cont_x_logs}(a)], one single measurement step is typically sufficient, although there are some rare cases in which a second consecutive measurement is necessary to achieve a sufficiently high purity [see, e.g., the two consecutive blue points in Fig.~\ref{fig:cont_x_logs}(a) at $t=12\Gamma^{-1}$]. For weaker measurements [Fig.~\ref{fig:cont_x_logs}(b)], an average of almost $3$ consecutive measurements is observed before performing a unitary operation and thermalization, resulting in the random-walk behavior that we previously observed in the discrete measurement case.

\section{Measurement and thermalization dominated regime}
\label{sec:tau_t_tau_m}

Here we study the regime where both the measurement and thermalization timescales are comparable and much slower than the unitary dynamics, i.e., $\tau_\text{t}\sim \tau_\text{m} \gg \tau_\text{u}$. 
We combine the modelization of finite-time thermalization from Sec.~\ref{sec:tau_t} with the continuous measurement studied in Sec.~\ref{sec:cont_x_z}. As we show, the RL agent naturally finds solutions that display features from these two distinct regimes, i.e., optimized thermalization strokes and multiple measurements, leading to a random walk in state space. However, the line-shape and the duration of the thermalization strokes are now conditioned on the (stochastic) purity reached after various consecutive measurements.

We consider the Hamiltonian of Eq.~(\ref{eq:qubit_with_controlled_gap}) and describe the thermalization of the qubit using the same master equation as in Sec.~\ref{sec:tau_t}. We recall that the thermalization speed is described by the rate $\Gamma$ and that the agent can arbitrarily tune the gap of the qubit $u(t)$ during the thermalization. We consider continuous measurements of $\sigma_{\rm z}$, as described in Sec.~\ref{sec:cont_x_z}; we recall that the measurement strength is determined by the ratio $\Delta t/\tau_\text{m}$. The unitary part consists of evolving the system according to Eq.~(\ref{eq:qubit_with_controlled_gap}) but, as in Sec.~\ref{sec:tau_t}, the coherence of the qubit is not relevant. Therefore, the unitary evolution can be considered to be the fastest timescale.

\begin{figure}[!tb]
	\centering
	\includegraphics[width=0.99\columnwidth]{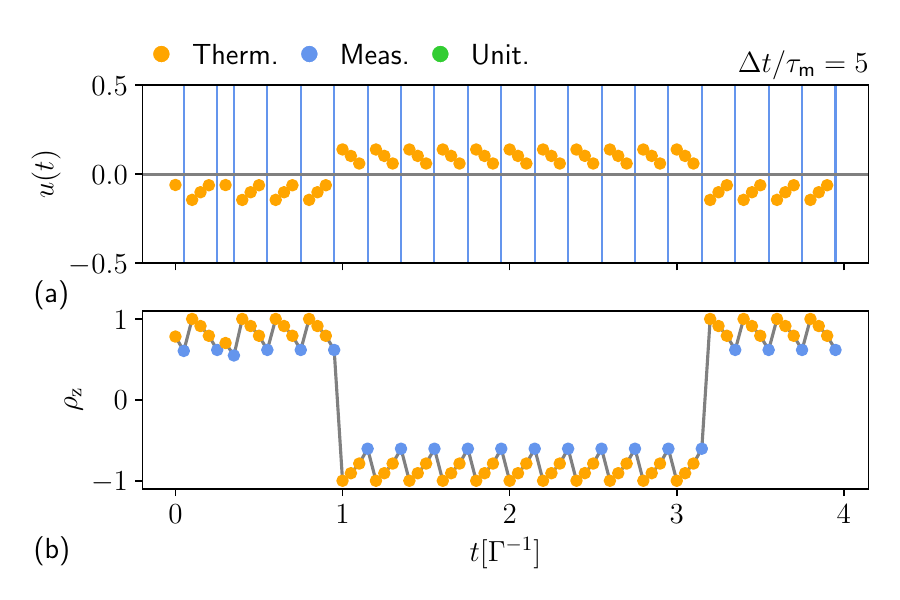}
	\includegraphics[width=0.99\columnwidth]{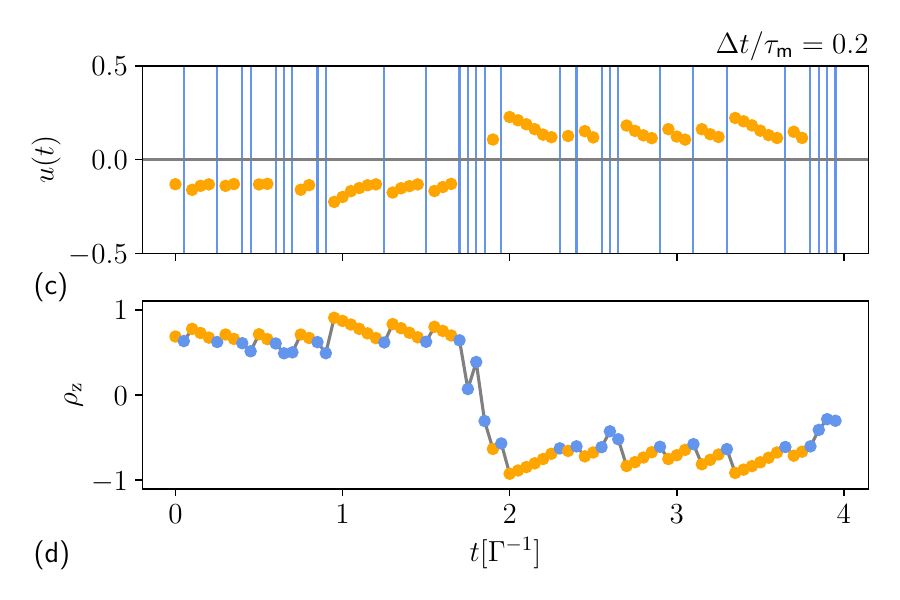} 
	\caption{ Qubit gap $u(t)$ during thermalization (a,c) and qubits $z$ coordinate (b,d) as a function of time along an arbitrary trajectory in the measurement and thermalization dominated regime. The RL agent is allowed to thermalize the qubit and perform continuous measurement of $\sigma_{\rm z}$ with strength quantified by the ratio $\Delta t/\tau_\text{m}$ reported on the plots. 
    Parameters used to generate the policy: $\Gamma= (\beta\hbar)^{-1}$, $\Delta t=0.05\, \Gamma^{-1}$,  $E_0=5\,\beta^{-1}$, $u(t)\in[-0.8,0.8]$.}
	\label{fig:gap_cont_action}
\end{figure}

In Fig.\ \ref{fig:gap_cont_action}(a,b), we plot the RL  results for $\Delta t/\tau_\text{m}=5$, i.e., for rather strong measurements, whereas in Fig.~\ref{fig:gap_cont_action}(c,d) we consider $\Delta t/\tau_\text{m}=1/5$ corresponding to a weak measurement. In both cases, we plot an arbitrary trajectory showing the control $u(t)$ as a function of time in panels (a,c), and the state $\rho_{\rm z}$, as a function of time, in panels (b,d). 

As expected, the control strategy shown in Fig.~\ref{fig:gap_cont_action}(a), corresponding to strong measurements, resembles the results in Fig.~\ref{fig:therm_pareto}(c), where the measurement was projective. Indeed, the feedback control strategy consists of performing a single measurement, followed by thermalizing the qubit while continuously ramping the gap of the qubit and choosing the correct sign of $u(t)$ based on the measurement outcome. However, here we witness rare cases in which a single measurement outcome is inconclusive, i.e., it does not result in a purification of the state (see, e.g., the two nearly consecutive blue dots at $t=0.3\Gamma^{-1}$). In such cases, the agent decides to perform a second measurement after a single thermalization time-step, since the measurement did not sufficiently purify the state as in the other measurements shown.

As we reduce the measurement strength, we observe multiple repeated measurements, producing a random walks in state space, that tend to purify the state [blue lines and dots in Fig.~\ref{fig:gap_cont_action}(c,d)]. This is followed by smooth thermalization strokes [orange dots in Fig.~\ref{fig:gap_cont_action}(c,d)], whose duration and line-shape depend on the purity reached during the previous measurements.
 On average, such measurements are repeated consecutively $2.5$ times. Interestingly, there are also cases in which the measurement decreases the purity of the state (see, e.g., at $t=1.8\Gamma^{-1}$, where $|\rho_{\rm z}|$ decreases after a blue dot, i.e., after a measurement).

\section{Two qubits setup: all timescales are comparable}
\label{sec:all_finite}
Here we show that our method can also be applied to the considerably harder problem when all three timescales are comparable ($\tau_\text{t}\sim \tau_\text{m}\sim \tau_\text{u}$) in a two-qubit setting, allowing to find interpretable strategies that considerably outperform more intuitive feedback control strategies. 
By extending our quantum system from a one- to two-qubits setup, we allow the RL agent to explore novel physical effects that can emerge in larger systems. In particular, the RL agent can now generate entanglement between the qubits and exploit it to optimize its performance.
We analyze the relation between the Maxwell's demon performance, and the generation of entanglement between the qubits considering both the absence and presence of counter-rotating terms in the qubit-qubit interaction.
To this end, we adopt the following description.

\begin{figure}[!tb]
	\centering
\includegraphics[width=0.99\columnwidth]{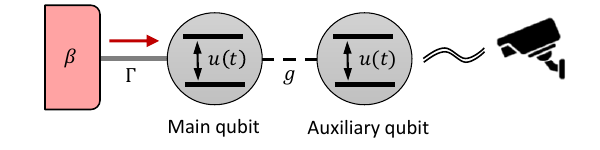} 
	\caption{Schematic representation of the Maxwell's demon setup studied in Sec.~\ref{sec:all_finite} when all timescales $\tau_\text{t}\sim \tau_\text{m}\sim \tau_\text{u}$ are comparable. When thermalizing, the main qubit is decoupled from the auxiliary qubit and put in contact with a thermal bath with inverse temperature $\beta$. During the unitary evolution, the two qubits are coupled together and interact. Projective measurements are performed only on the auxiliary qubit thus only partial information about the system is acquired.}
	\label{fig:2_qubits}
\end{figure}

As in the previous sections, we consider a main (m) qubit that can be coupled to a thermal bath (see Fig.~\ref{fig:2_qubits}). However, we model finite-time measurements by considering a second auxiliary (a) qubit that can interact with the main qubit through finite-time Hamiltonian dynamics and can then be measured projectively. This can be seen as a finite-time modelization of a non-projective measurement, where correlations are established between the qubits through unitary dynamics, and then information about the main qubit is acquired by projectively measuring the auxiliary qubit. 

We model the three discrete actions in the following way: unitary dynamics is described by the Schr{\"o}dinger equation with the following time-dependent Hamiltonian
\begin{equation}
	H_U[u(t)] = u(t) \frac{E_0}{2} \left(\sigma^{\text{(m)}}_{\rm z} +  \sigma^{\text{(a)}}_{\rm z}\right) + g H_\text{int},
\end{equation}
which describes two resonant interacting qubits~\cite{bresque2021two}. Here $u(t)$ is the time-dependent control that we allow the agent to optimize in a continuous interval $[u_\text{min},u_\text{max}]$, and $g$ represents the interaction strength. We consider two possible interactions
\begin{align}
    H_\text{int}^{\text{(no-counter)}} &= \left( \sigma^{\text{(m)}}_+ \sigma^{\text{(a)}}_- +  \sigma^{\text{(m)}}_- \sigma^{\text{(a)}}_+\right), \\
    H_\text{int}^{\text{(counter)}} &= \sigma_{\rm x}^{\text{(m)}}\sigma_{\rm x}^{\text{(a)}}.
    \label{eq:h_int_count}
\end{align}
This corresponds, respectively, to neglecting ($H_\text{int}^{\text{(no-counter)}}$) and considering ($H_\text{int}^{\text{(counter)}}$) the counter-rotating terms $\sigma^{\text{(m)}}_+\sigma^{\text{(a)}}_+ + \sigma^{\text{(m)}}_-\sigma^{\text{(a)}}_-$. The coupling  $H_\text{int}^{\text{(counter)}}$, for example, can be obtained by inductively coupling two superconducting qubits \cite{PhysRevApplied.15.034065}. Further, as shown in Ref.~\cite{nguyen2024programmable}, the interaction between the two qubits can be suitably programmed through the parametric driving of the qubits. During the thermalization stroke, we decouple the two qubits, and we describe the thermalization of the main qubit using the same model as in Sec.~\ref{sec:tau_t}, where we recall that the thermalization rate is given by $\Gamma$. The measurement stroke is also described as in Sec.~\ref{sec:tau_t}, i.e., performing a projective measurement along the $z$-axis, but here the measurement is only performed on the auxiliary qubit.

\begin{figure}[!tb]
	\centering	\includegraphics[width=0.99\columnwidth]{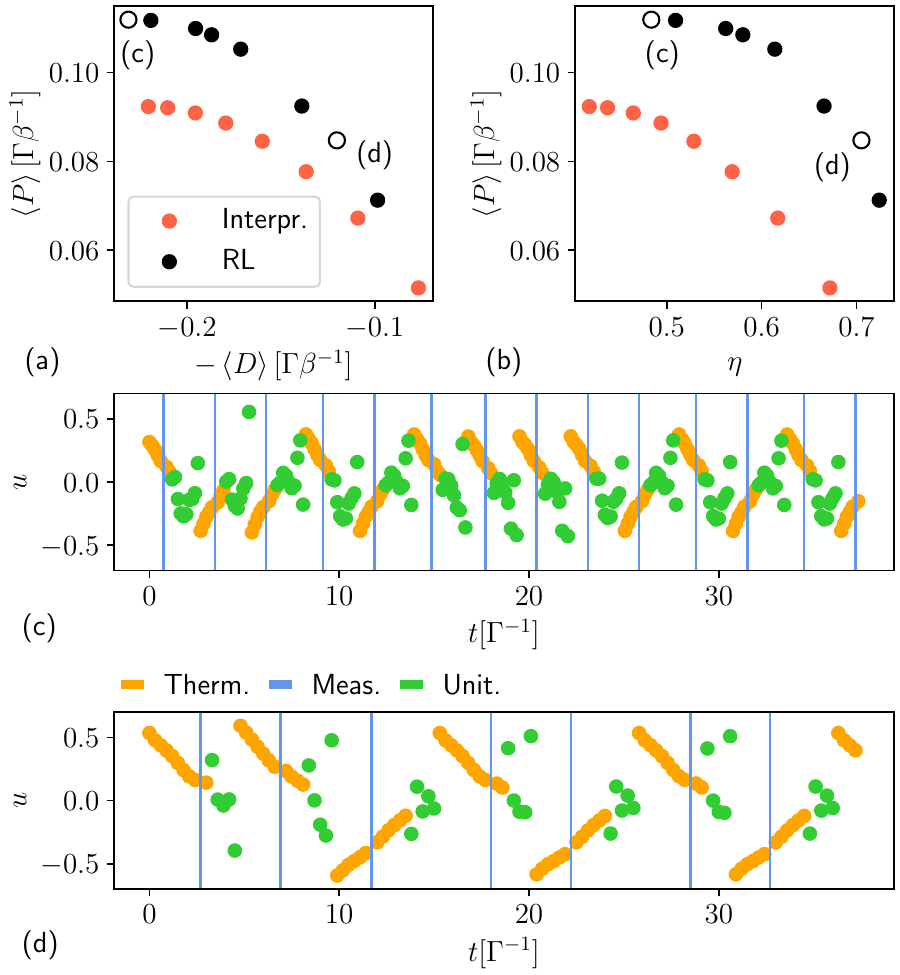}
	\caption{Performance of the Maxwell's demon model described in Sec.~\ref{sec:all_finite}, when all timescales are finite, and the interaction between the two qubits is described by $H_\text{int}^{\text{(no-counter)}}$. In panels (a,b), the black dots correspond to the RL results, while the red one corresponds to the interpretable strategy described in Sec.~\ref{sec:all_finite}. The empty dots in panels (a and b) correspond to the two control strategies plotted respectively in panel (c) (corresponding to $c=1$) and panel (d) (corresponding to $c=0.7$). The colors in (c,d) correspond to the three discrete actions (see legend). Parameters: $E_0=5\beta^{-1}$, $g=\beta^{-1}$, $\Gamma=1(\beta\hbar)^{-1}$, $u(t)\in [-0.8, 0.8]$, and $\Delta t = 0.15\Gamma^{-1}$ for $c=\{1, 0.95, 0.9, 0.85, 0.8\}$ and $\Delta t = 0.3\Gamma^{-1}$ for $c=\{0.75, 0.7, 0.65\}$.}
\label{fig:all_finite_res_no_counter}
\end{figure}
Before showing the results using the RL optimization, we devise an interpretable strategy inspired by the RL results that we use to benchmark the performance of the RL method. The Hamiltonian $H_U[u(t)]$, with $H_\text{int} = H_\text{int}^{\text{(no-counter)}}$, has an interesting property: let us assume that the state of the joint system is given by
\begin{equation}
 	\rho = \rho^{\text{(m)}}\otimes \ket*{\psi^{\text{(a)}}}\bra*{\psi^{\text{(a)}}},
 	\label{eq:rho_swap}
\end{equation}
with $\ket*{\psi^{\text{(a)}}}= \ket*{0}$ or $\ket*{ \psi^{\text{(a)}}}= \ket*{1}$. It can be shown analytically that, after a time
\begin{equation}
	\tau_{\text{swap}} = \frac{\pi}{2g},
\end{equation}
the state of the two qubits exactly swaps up to a global phase. We, therefore, consider the following ``interpretable feedback control strategy'':
\begin{itemize}
    \item Perform a measurement of the auxiliary  qubit.
    \item If it is in the ground state, do nothing.
    \item If it is in the excited state, change the sign of $u(t)$, such that now it is in the ground state.
    \item Since the post-measurement state is of the form of Eq.~(\ref{eq:rho_swap}), we perform a unitary stroke of duration $\tau_{\text{swap}}$ keeping the control $u(t)$ constant at some value $\bar{u}$.
    \item The main qubit will be in the ground state because of the swap. We thus let it thermalize with the bath at the same value of the control $\bar{u}$, for time $\bar{\tau}$, resulting in heat extraction from the thermal bath.
\end{itemize}
This interpretable strategy is then optimized numerically with respect to $\bar{u}$ and $\bar{\tau}$.

\subsection{Without counter-rotating terms}
In Fig.~\ref{fig:all_finite_res_no_counter}, we compare the performance of the RL agent against the interpretable strategy choosing $H_\text{int} = H_\text{int}^{\text{(no-counter)}}$. From Fig.~\ref{fig:all_finite_res_no_counter}(a,b), we see that the Pareto front of the RL strategy (black dots) is visibly better than that of the ``intuitive strategy'' (red dots), both in the high power and high-efficiency regime. 

In Fig.~\ref{fig:all_finite_res_no_counter}(c,d), we show a trajectory of the feedback control strategy as a function of time, whose performance is shown in Fig.~\ref{fig:all_finite_res_no_counter}(a,b) as white dots. We notice that it is similar in spirit to the ``intuitive strategy'' since we verified that the duration of the unitary stroke (green dots) is approximately $\tau_\text{swap}$. However, the RL agent learns again to modulate the gap of the qubit while thermalizing, and this gives us an advantage that is substantially larger than the advantage observed in Sec.~\ref{sec:tau_t} where we considered the thermalization-dominated regime. As expected, the thermalization strokes slow down as we shift our interest from high power [Fig.~\ref{fig:all_finite_res_no_counter}(c)] to high efficiency [Fig.~\ref{fig:all_finite_res_no_counter}(d)].
 
\begin{figure}[!tb]
	\centering
	\includegraphics[width=0.99\columnwidth]{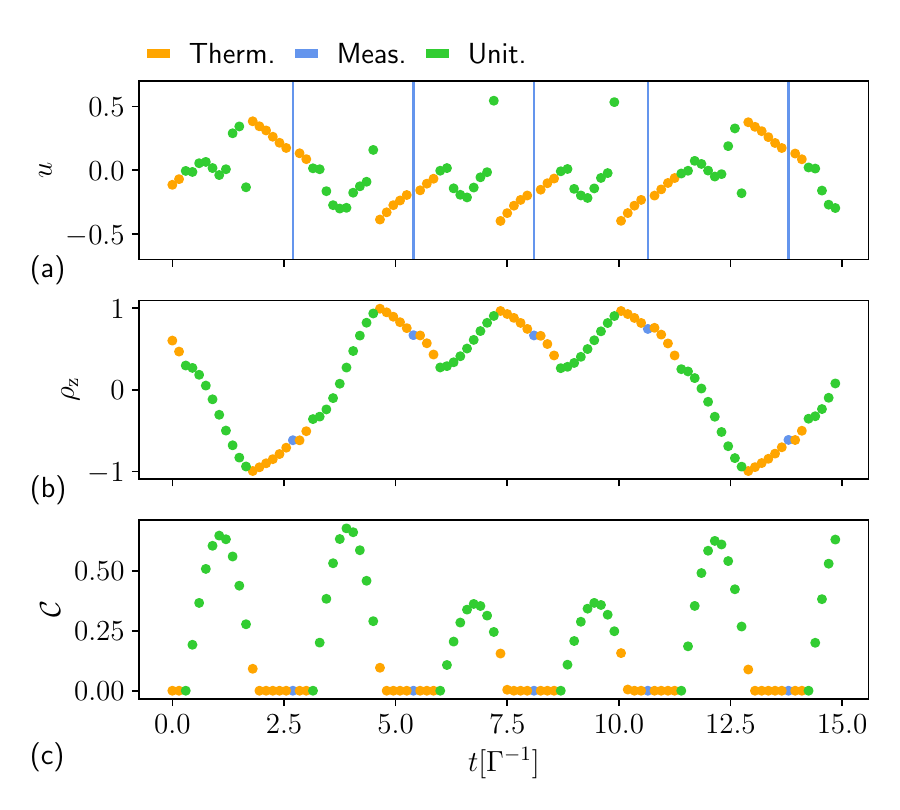}
	\caption{Analysis of the control strategy shown in Fig.~\ref{fig:all_finite_res_no_counter}(c). The control $u(t)$, the only non-null component $\rho_{\rm z}$ of the reduced density matrix of the main qubit, and the concurrence $\mathcal{C}$ between the qubits are respectively plotted in panels (a,b,c) as a function of time considering an arbitrary trajectory. The color corresponds to the discrete action shown in the legend.
 }
 \label{fig:all_finite_no_counter_prot}
\end{figure}
To further interpret the RL results, in Fig.~\ref{fig:all_finite_no_counter_prot}, we have plotted, 
respectively, the control $u(t)$ (a), the only non-null component $\rho_{\rm z}$ of the Bloch vector of the main qubit tracing out the auxiliary qubit (b), and the concurrence $\mathcal{C}$ (c) between the two qubits in the maximum power case ($c=1$) shown in Fig.~\ref{fig:all_finite_res_no_counter}(c).
This plot sheds light on the cooling mechanism and allows us to interpret what the RL agent is doing. As we see in panel (b), during the unitary strokes (green dots), the purity of the qubit's state $1$ increases, almost reaching $1$. During this process, entanglement is developed between the qubits, as seen in panel (c). Once the purity is high enough, the agent thermalizes the qubit, cooling the heat bath and destroying entanglement between the qubits. The auxiliary qubit is then measured again, and the process is repeated. Interestingly, based on the outcome of the measurement process, the control $u(t)$ may change sign, leading to a different trajectory also of the state, and to a different amount of entanglement between the qubits.

\subsection{Including counter-rotating terms}
\label{subsec:2qubits_res_counter}
\begin{figure}[!tb]
	\centering
\includegraphics[width=0.99\columnwidth]{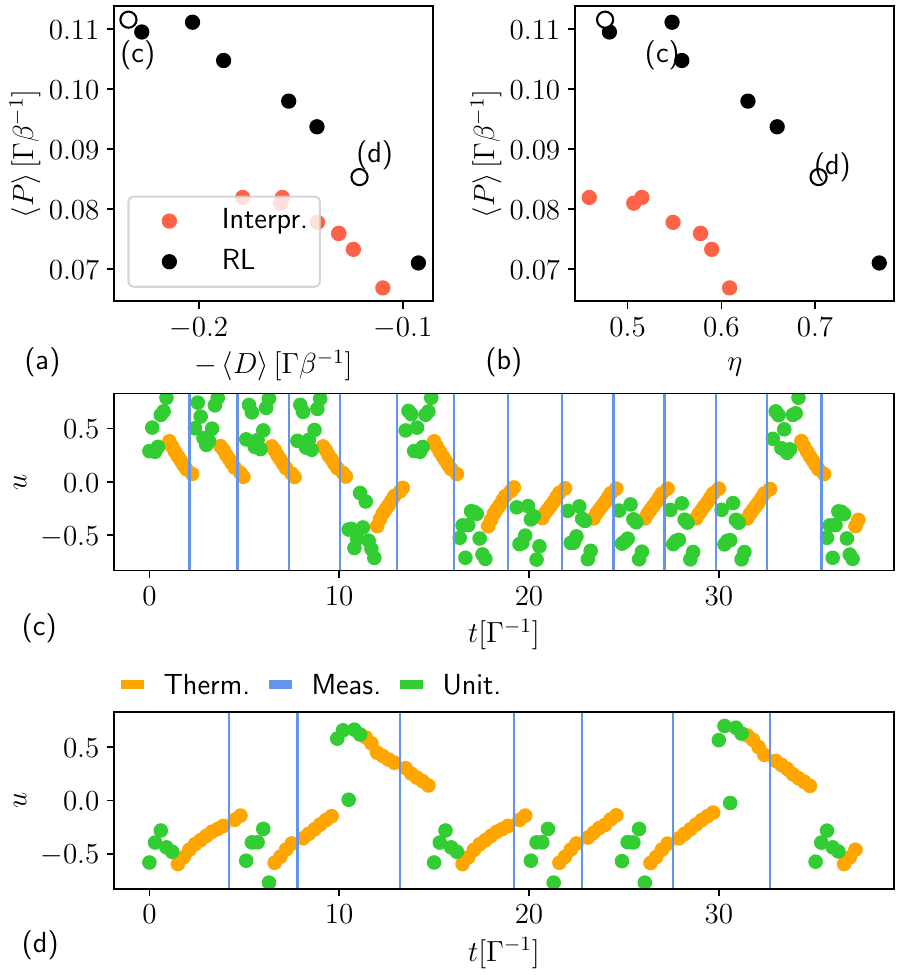}
	\caption{Performance of the Maxwell's demon model considered in Fig.~\ref{fig:all_finite_res_no_counter}, but including the counter-rotating terms in the interactions, i.e., using $H_\text{int}^\text{(count)}$ instead of $H_\text{int}^\text{(no-count)}$. Plotting style and parameters are the same as in Fig.~\ref{fig:all_finite_res_no_counter}, except for the choice of $\Delta t$ give by $\Delta t=0.15 \Gamma^{-1}$ for $c=\{1, 0.95, 0.9\}$; $\Delta t=0.2\Gamma^{-1}$ for $c=0.85$; $\Delta t= 0.3\Gamma^{-1}$ for $c=\{0.8, 0.75, 0.7\}$; and $\Delta t=0.4\Gamma^{-1}$ for $c=0.65$.}
\label{fig:all_finite_res_counter}
\end{figure}
We now consider the effect of the presence of the counter-rotating terms in the coupling between the two qubits, i.e., we consider the interacting Hamiltonian given by $H_\text{int}^{\text{(counter)}}$ in Eq.~(\ref{eq:h_int_count}).
In Fig.~\ref{fig:all_finite_res_counter}, we plot the results for this case in the same style as Fig.~\ref{fig:all_finite_res_no_counter}. Because of the counter-rotating terms, the interaction between 
the qubits will not perfectly swap the qubit states, making this a harder optimization problem. As we can see from Fig.~\ref{fig:all_finite_res_counter}(a,b), the RL strategy produces a substantially higher power and higher efficiency than the interpretable strategy. As expected, the thermalization timescale increases as we shift our interest from high power [\ref{fig:all_finite_res_counter}(c)] to high efficiency [\ref{fig:all_finite_res_counter}(d)], and the RL agent learns a non-intuitive unitary stroke to purify the state of the main qubit (green dots).

Interestingly, comparing Fig.~\ref{fig:all_finite_res_counter}(b) 
with Fig.~\ref{fig:all_finite_res_no_counter}(b), we see that the best performance curve of the demon using RL, i.e., its Pareto front, is roughly the same with and without counter-rotating terms. Conversely, the performance of the interpretable strategy is decreased by the presence of the counter-rotating terms since they do not allow a perfect swap of the qubit states.

\begin{figure}[!tb]
	\centering
\includegraphics[width=0.99\columnwidth]{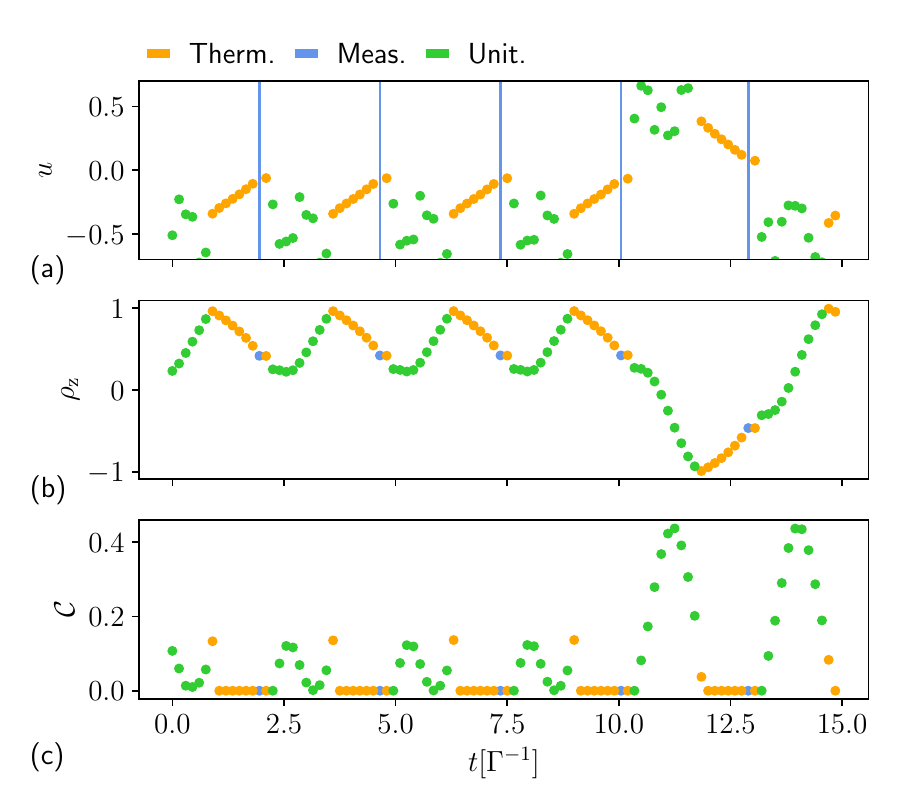}
	\caption{Analysis of the control strategy shown in Fig.~\ref{fig:all_finite_res_counter}(c) plotted in the same style as Fig.~\ref{fig:all_finite_no_counter_prot}.
 }
\label{fig:all_finite_counter_prot}
\end{figure}
To conclude, we analyze the effect of the RL control strategy in the maximum power case ($c=1$) in Fig.~\ref{fig:all_finite_counter_prot}, as we did in Fig.~\ref{fig:all_finite_no_counter_prot} for the case without counter-rotating terms.
Interestingly, the cooling mechanism seems quite similar regardless of the presence of counter-rotating terms. Indeed, after measuring the auxiliary qubit, the agent purifies the state [green dots in \ref{fig:all_finite_counter_prot}(b)], and entanglement is generated during this process [green dots in \ref{fig:all_finite_counter_prot}(c)]. Once the state is pure enough, the agent thermalizes the main qubit, extracting heat from the bath and making sure that the main qubit is in the ground state (by choosing appropriately the sign of $u$ based on the measurement outcome). However, based on the measurement outcome, the amount of entanglement generated during the unitary feedback can be lower than in the case without counter-rotating terms [compare the green dots in Fig.~\ref{fig:all_finite_no_counter_prot}(c) and Fig.~\ref{fig:all_finite_counter_prot}(c)].

\section{Conclusions and outlook}
In this work, we have introduced a RL-based framework to discover optimal feedback control strategies in open quantum systems, such that the RL agent operates as an actual Maxwell's demon. 

This compelling feature, on the one hand, allows us to literally regard the
agent in RL as a Maxwell's demon that acquires thermodynamically relevant information about the system, given it an ontological interpretation, providing an appealing picture.

On the other hand, this approach serves as a tool to come up with strategies that seem hardly possible to 
identify with conventional tools. The approach taken
allows us to systematically maximize the multi-objective performance of a finite-time quantum Maxwell's demon, which aims to maximize the information-induced average cooling power while minimizing the measurement cost. 
We study different physical regimes characterized by a different ordering between the measurement timescale, the unitary feedback timescale, and the thermalization timescale. Our framework optimizes the system's long-term time-averaged and trajectory-averaged performance.
We find feedback control strategies that are interpretable yet outperform more intuitive strategies, shedding light on important aspects that should be considered when designing a quantum Maxwell's demon.

When the thermalization timescale is the slowest, the RL optimization results in non-trivial control strategies that outperform more intuitive ones. Moreover, we show that the duration of the thermalization stroke is dictated by the trade-off between cooling power and measurement post.

When the measurement timescale is the slowest, we found that the RL method first purifies the state through repeated measurements, leading to a random walk in the qubit's state space. In this process, we consider both discrete and continuous weak measurements. This is followed by unitary feedback and thermalization. Notably, we show that adaptively choosing different measurement observables leads to a performance enhancement with respect to fixing the measurement. We further showed significant changes in the optimal measurement observable as we shifted our interest from high cooling power to low measurement cost.

We then consider the case when thermalization and measurement timescales are comparable and slower than the unitary feedback timescale. After performing multiple measurements, the RL agent learns carefully modulated thermalization ramps whose line-shape and duration depends on the purity reached during the previous measurements.

Finally, we have demonstrated that our method can be applied to a considerably more complex problem of two qubits when all three timescales are comparable, and that we can substantially outperforms more intuitive strategies. To this end, we considered two interacting qubits: one acting as the main qubit that can thermalize with the bath but not be measured, and the other one acting as an auxiliary qubit that we can measure and that indirectly provides us with information about the main qubit through finite-time interactions. 
We find intriguing 
and highly counter-intuitive feedback control strategies, where entanglement between the qubits is generated and then destroyed by measurements and thermalizations. 

The introduction of RL methods to discover feedback control strategies in quantum thermodynamics paves the way for numerous future research directions. The impact of measurements and feedback on the performance of driven
quantum heat engines and refrigerators operating between two different temperatures can be rigorously assessed, potentially revealing performance enhancements induced by quantum measurements. The impact of such feedback control schemes on power fluctuations can also be assessed and compared to recent results on the thermodynamic uncertainty relations \cite{pietzonka2018universal}. Building on our two-qubit example, many-body interacting systems could also be optimized using this approach, revealing the impact that local or global measurements on many-body systems could have on the thermodynamic performance of such systems.

Finally, the use of advanced neural network architectures, such as the one proposed in Ref.~\cite{erdman2023_pnas}, could allow the RL agent to learn how to act as an optimal quantum Maxwell's demon by interacting directly with an experimental device without even knowing the exact model describing the dynamics of the system. This relies on using the time-series of chosen actions, rather than the quantum state, as state for the RL agent. While the size of the Hilbert space scales exponentially with the system size, this approach only scales with the number of controls, therefore scaling favorably in the system size, enabling also the theoretical optimization of substantially larger systems.

\section*{Code and Data Availability Statement}
The reinforcement learning code, together with the generated data that is presented in this manuscript, will be made publicly available respectively on a GitHub repository and on Figshare upon publication of the manuscript on a peer-reviewed journal.

\section*{Acknowledgements}
PAE gratefully acknowledges funding by the Berlin Mathematics Center MATH+ (AA2-18).
JE has been supported by the DFG (FOR 2724, CRC 183), the BMBF (QSolid), and the ERC (DebuQC). RC, BB and ANJ acknowledge the support of Chapman University, U. S. Army Research Office under grant\ W911NF-22-1-0258 and the U.S. Department of Energy (DOE), Office of Science, Basic Energy Sciences (BES), under Award No.\ DESC0017890. 
J.E. acknowledges funding by the DFG (FOR 2724, for which this is an inter-node collaboration reaching an important milestone, and CRC 183), the FQXI, the Quantum Flagship (Millenion, for which is again the result of an inter-node collaboration), the BMBF (DAQC), and the ERC (DebuQC). FN gratefully acknowledges funding by the BMBF (Berlin Institute for the Foundations of Learning and Data—BIFOLD), the European Research Commission (ERC CoG 772230) and the Berlin Mathematics Center MATH+ (AA1-6, AA2-8). Numerical calculations have been performed using PyTorch \cite{paszke2017} and the QuTiP2 toolbox \cite{johansson2013}. We thank Ketevan Kotorashvili for helping with the figure design.

\newpage

\appendix

\section{Measurement dominated regime: Discrete $\sigma_{\rm z}$ measurements} \label{sec:app_discr_sigma_z}
\begin{figure}[!b]
    \centering
    \includegraphics[width=0.99\columnwidth]{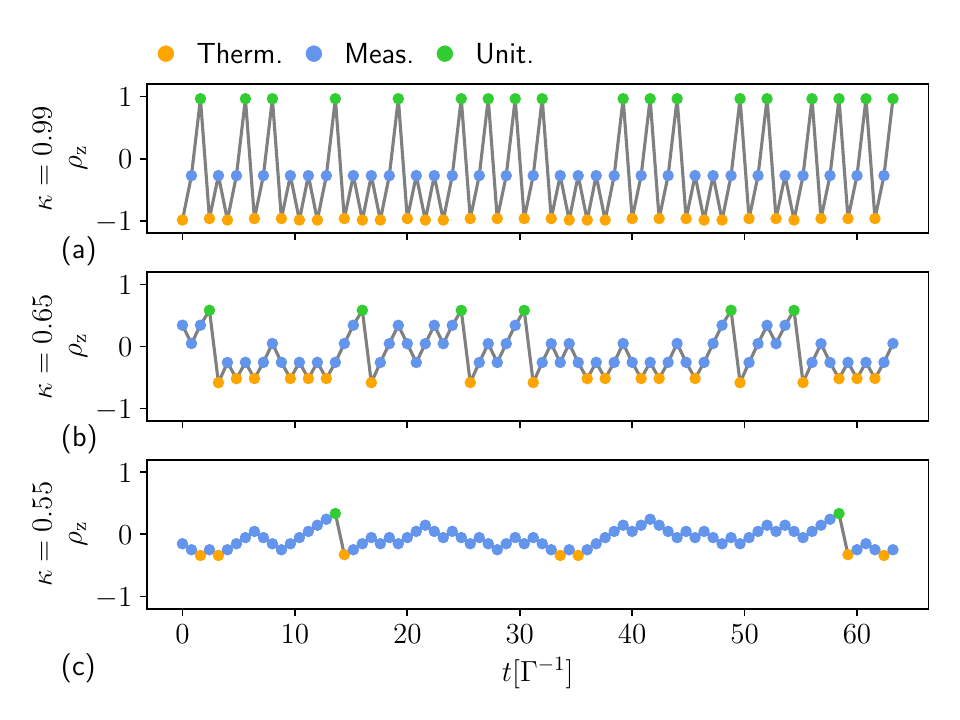}
    \caption{Plots of the state component $\rho_{\rm z}$, as a function of time, corresponding to the results shown in Fig.~\ref{fig:Bloch_discr_meas}(b,d,f) relative to the discrete $\sigma_{\rm z}$ measurement case in the measurement dominated regime. Each row corresponds to a different value of the discrete measurement strength $\kappa$. An arbitrary trajectory is shown. The colors of the dots indicate the actions chosen by the RL agent at a given moment.}
    \label{fig:z_logs}
\end{figure}

In this appendix we show additional results for the model and regime studied in Sec.~\ref{sec:meas_dis}, i.e., the measurement dominated regime where discrete measurements in a fixed basis are performed. While Fig.~\ref{fig:x_logs} in the main text shows a trajectory of visited states and corresponding actions as a function of time for the $\sigma_{\rm x}$-measurement case, in Fig.~\ref{fig:z_logs} we show the same plot, but for the $\sigma_{\rm z}$ measurement case. Each row corresponds to a different value of $\kappa$, and here the state is represented by the $\rho_{\rm z}$ component of the qubit state. As 
we can see, the $\sigma_{\rm x}$- and $\sigma_{\rm z}$-measurement cases display qualitatively similar features.

\section{Measurement dominated regime: Discrete adaptive measurements. Role of the measurement basis.} 
\label{sec:app_second_measurement_angle}
In this appendix, we first show that a measurement perpendicular to the state on the Bloch sphere is the one that maximizes the average increase in purity of the state. We then show how, in fact, the RL agent chooses a measurement angle that is not exactly perpendicular to the qubit's state. While this may seem like an imperfection of the RL method, we show that this actually leads to a (very small) improvements in the performance. 

We start by proving that the measurements that maximize the average purity of the state are perpendicular to the qubit state on the Bloch sphere. Let us consider an arbitrary qubit state of the form
\begin{equation}
    \rho = \frac{1}{2}\left( I_2 + \rho_{\rm x} \sigma_{\rm x} + \rho_{\rm z}\sigma_{\rm z}\right),
\end{equation}
since the measurements and dynamics considered in this manuscript never drives it onto $\sigma_{\rm y}$. 

Let us consider a weak measurement with measurement angle $\theta$ described by the Kraus operators $M_\pm$ defined in Eq.~(\ref{eq:discrete_measurement_operators}). For later convenience, let us define the vectors
\begin{equation}
\begin{aligned}
    \vec{\rho} &= \rho_{\rm x} \hat{x} + \rho_{\rm z} \hat{z}, \\
    \hat{n} &= \sin\theta\hat{x} + \cos\theta\hat{z},
\end{aligned}
\end{equation}
where $\theta$ is the angle defined below Eq.~(\ref{eq:state_meas}), and $\hat{n}$ describes the axis along which we are measuring.

Using Eq.~(\ref{eq:pk_def}) and the properties of the Pauli matrices, the probability of measuring outcomes $\pm$ can be written as
\begin{equation}
    p_\pm = \Tr\left[ \rho M_\pm^\dagger M_\pm \right] = \frac{1}{2}\left[ 1 \pm (2\kappa-1)\vec{\rho}\cdot\hat{n} \right].
\end{equation}
The post-measurement purity of the state, conditioned on the measurement outcome, is given by
\begin{equation}
    \Tr\left[ \rho_\pm^2 \right] = \frac{1}{p_\pm^2} \Tr\left[ \left( \rho M_\pm^\dagger M_\pm \right)^2 \right],
\end{equation}
where we have used Eq.~(\ref{eq:state_meas}) to compute the post-measurement state and the cyclic property of the trace. Inserting the explicit expressions for $\rho$ and $M_\pm$, we can rewrite it as
\begin{multline}
    \Tr\left[ \rho_\pm^2 \right] =   \frac{1}{4p_\pm^2} \Big[ 
 \text{Tr}\left[\rho^2\right] 
 + (2\kappa-1)^2 \text{Tr}\left[(\rho \sigma_\theta)^2\right] \\
 \pm 2(2\kappa-1)\text{Tr}\left[\rho^2 \sigma_\theta\right]
 \Big].
 \label{eq:app_s1}
\end{multline}
Using the properties of the Pauli matrices, we can rewrite the three traces in Eq.~(\ref{eq:app_s1}) as
\begin{equation}
    \text{Tr}\left[\rho^2\right] =
    \frac{1}{2} \left( 1 + |\vec{\rho}|^2 \right),
\end{equation}
\begin{equation}
    \text{Tr}\left[ (\rho\sigma_\theta)^2 \right] =
   \frac{1}{2}\left[
   1+ (\vec{\rho}\cdot\hat{n})^2
    -|\vec{\rho}\times \hat{n}|^2
   \right],
\end{equation}
\begin{equation}
    \text{Tr}\left[\rho^2\sigma_\theta \right] = 
    {\vec{\rho}\cdot\hat{n}}.
\end{equation}
Plugging these expressions back into Eq.~(\ref{eq:app_s1}) leads to
\begin{multline}
    \Tr\left[ \rho_\pm^2 \right] =  
    \frac{1}{8p_\pm^2} \Big[ 
    1 + |\vec{\rho}|^2
 + \\ (2\kappa-1)^2 \left[
   1+ (\vec{\rho}\cdot\hat{n})^2
    -|\vec{\rho}\times \hat{n}|^2
   \right]
 \pm 4(2\kappa-1)\vec{\rho}\cdot\hat{n}
 \Big].
\end{multline}
This leads to the final expression for the average purity $\bar{\gamma}$ given by
\begin{multline}
    \bar{\gamma} = p_+\Tr\left[ \rho_+^2 \right] + p_-\Tr\left[ \rho_-^2 \right] = \\
     1 -\frac{1}{2} \frac{(1-l^2)(1-|\vec{\rho}|^2)}{1-l^2 |\vec{\rho}|^2\cos^2\alpha},
     \label{eq:avg_purity}
\end{multline}
where we defined $l=2\kappa-1$, and where $\alpha$ is the angle between the state on the Bloch sphere and the measurement axis, explicitly defined by 
\begin{equation}
    \vec{\rho}\cdot\hat{n} = |\vec{\rho}|\cos\alpha.    
\end{equation}
Since $|l|, |\vec{\rho}| \leq 1$, $\bar{\gamma}$ is trivially maximized by minimizing $\cos^2\alpha$, i.e., when $\cos\alpha=0$ corresponding to
\begin{equation}
    \alpha = \pm\frac{\pi}{2}.
\end{equation}
This proves that perpendicular measurements maximize the average post-measurement purity.

We now show that the RL agent actually decides to perform measurements that are not exactly perpendicular to the qubit's state on the Bloch sphere, and we show that this indeed leads to a slight performance enhancement.

Let's consider the policy shown in Fig.~\ref{fig:bloch_angle}(c) that can be described by the following procedure: 
\begin{enumerate}
    \item Initialize the qubit in the thermal state.
    \item Perform $\sigma_{\rm x}$ measurement of the qubit. Determine the post-measurement state coordinate $\rho_{\rm x}$.
    \item If $\rho_{\rm x}>0$, perform the second measurement of $\sigma_\theta$ with $\theta=\theta_+$, where $\theta$ is the measurement angle that enters the measurement operator in 
    Eq.\  (\ref{eq:discrete_measurement_operators}). If $\rho_{\rm x}<0$, perform the second measurement of $\sigma_\theta$ with $\theta=\theta_-$. The angles $\theta=\theta_\pm$ can be determined from the analysis of the RL policies. If the measurement increased the qubit's coordinate $\rho_{\rm z}$, proceed to step 4. Otherwise, skip step 4.
    \item Perform the unitary feedback, i.e., rotate the qubit to the negative $z$-axis.
    \item Thermalize the qubit, leading us back to step 2. 
\end{enumerate}

Interestingly, the RL method finds the same exact procedure also for $\kappa=0.7, 0.75, 0.8,$ and $0.85$. 
We thus implement numerically the policy above, comparing the resulting expected cooling power
when the angles $\theta_\pm$ are chosen to maximize the post-measurement purity, i.e. perpendicular to the state on the Bloch sphere, and when the angles are chosen according to the RL results. Out of symmetry considerations, we assume $ |\theta_+ - \pi/2| = |\theta_- - \pi/2|$. 

The measurement angles $\theta_\pm^{\rm (RL)}$ in the RL-case are chosen according to
\begin{equation}
    |\theta_+^{\rm (RL)} - \pi/2| = {E} \left[|\theta_j - \pi/2| \right], 
\end{equation}
where the angles $\theta_j$ are collected from the simulations performed by trained RL agents.

\begin{figure}[!tb]
	\centering
	\includegraphics[width=0.99\columnwidth]{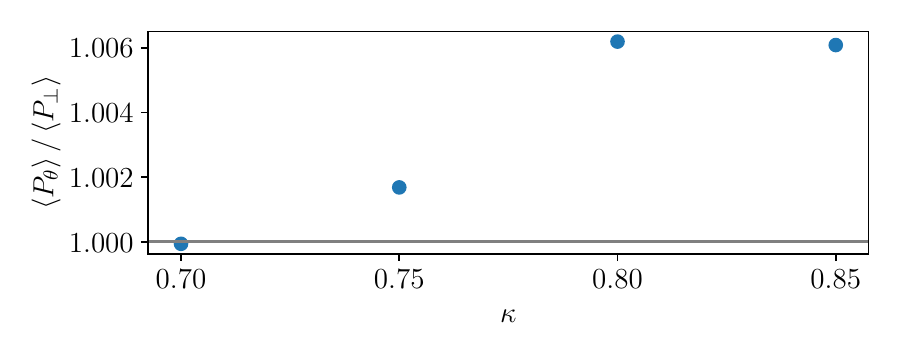}
	\caption{Ratio of the cooling powers for the policy given in App.~\ref{sec:app_second_measurement_angle} when the measurement angle is based on RL ($\ev*{P_\theta}$), and when it is chosen to be perpendicular to the state on the Block sphere ($\ev*{P_\perp}$), i.e. to maximize the average post-measurement purity. The ratio larger than one indicates a slight advantage of the strategy based on RL optimization. All system parameters are the same as Fig.~\ref{fig:bloch_angle}, except for considering a full thermalization during the thermalization action (corresponding to the $\Gamma\Delta t \gg 1$ limit). }
	\label{fig:cooling_power_ratio}
\end{figure}

In Fig.~\ref{fig:cooling_power_ratio}, 
we plot the ratio between the power $\ev*{P_\theta}$ using the RL-chosen measurement angle, and the power $\ev*{P_\perp}$ found choosing the angle that maximizes the average post measurement purity, for $\kappa=0.7,0.75,0.8,0.85$. As we can see, the RL-chosen angles provide a small ($<1\%$) yet measurable improvement of the cooling power. This indicates that the choice of measurement angle in step 3 is mainly, but not exclusively, dictated by the intent of maximizing the average post-measurement purity of the qubit. 

\section{Measurement dominated regime: Discrete adaptive measurements. Power-efficiency trade-off}
\label{sec:app_penalty}
\begin{figure}[!tb]
	\centering
	\includegraphics[width=0.99\columnwidth]{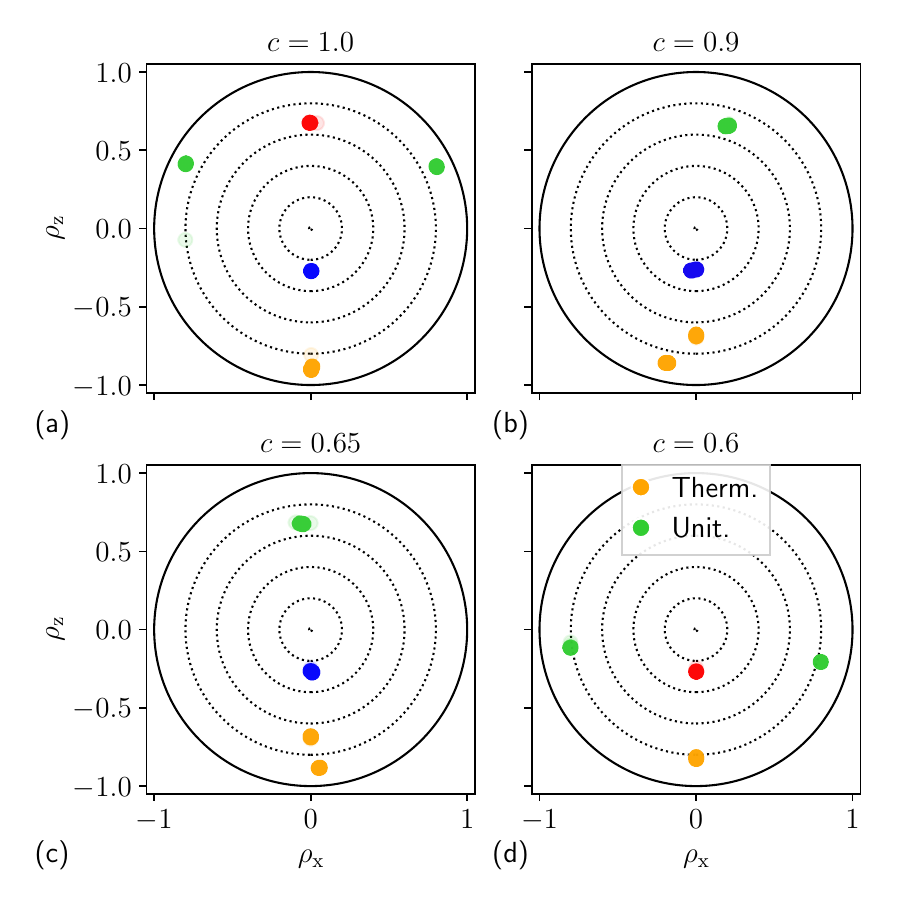}
	\includegraphics[width=0.99\columnwidth]{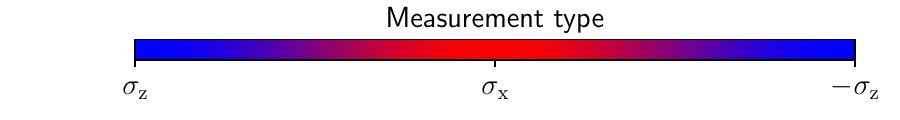} 
	\caption{
 Actions chosen by the RL agent as a function of the qubit's state represented as a point on the Bloch sphere in the sames style as Fig.~\ref{fig:bloch_angle}. Each panel corresponds to a representative point along the Pareto front in Fig.~\ref{fig:power_vs_penalty} computed with a different value of $c$ shown on the plot.}
	\label{fig:penalty_bloch}
\end{figure}
At the end of Sec.~\ref{subsec:learn_basis}, we asses the impact of power-efficiency trade-offs in the measurement dominated regime when the RL agent can adaptively choose the measurement basis. In particular, in Fig.~\ref{fig:power_vs_penalty} we showed a Pareto-front displaying different power-efficiency trade-offs, and we claim that differently colored points correspond to different measurement strategies.

In Fig.~\ref{fig:penalty_bloch} we explicitly show how the optimal strategy changes as we change the trade-off between power and efficiency, i.e., as we change the parameter $c$. Indeed, in Fig.~\ref{fig:penalty_bloch}(a) we see that, for $c=1$, the agent performs both $\sigma_{\rm x}$ (red dot) and $\sigma_{\rm z}$ (blue dot) measurements. This corresponds to the green points along the Pareto front of Fig.~\ref{fig:power_vs_penalty}. In Figs.~\ref{fig:penalty_bloch}(b,c), corresponding to $c=0.9$ and $c=0.65$, we see that the agent only performs $\sigma_{\rm z}$ measurements. This corresponds to the blue points along the Pareto front of Fig.~\ref{fig:power_vs_penalty}. In Fig.~\ref{fig:penalty_bloch}(d), corresponding to $c=0.6$, we see that the RL agent only performs $\sigma_{\rm x}$ measurements, corresponding to the red points along the Pareto front of Fig.~\ref{fig:power_vs_penalty}.

\section{Measurement dominated regime: continuous measurements}
\label{app:cont_meas}
In this appendix we show additional results for the model and regime studied in Sec.~\ref{sec:cont_x_z}, i.e., the measurement dominated regime where continuous measurements are performed in a fixed measurement basis. In particular, we consider the same model as in Sec.~\ref{sec:cont_x_z}, but we additionally discuss the case where the measurement basis is chosen by the RL agent, and we show additional plots for the $\sigma_{\rm z}$-measurement case.

\begin{figure}[!tb]
	\centering
	\includegraphics[width=0.99\columnwidth]{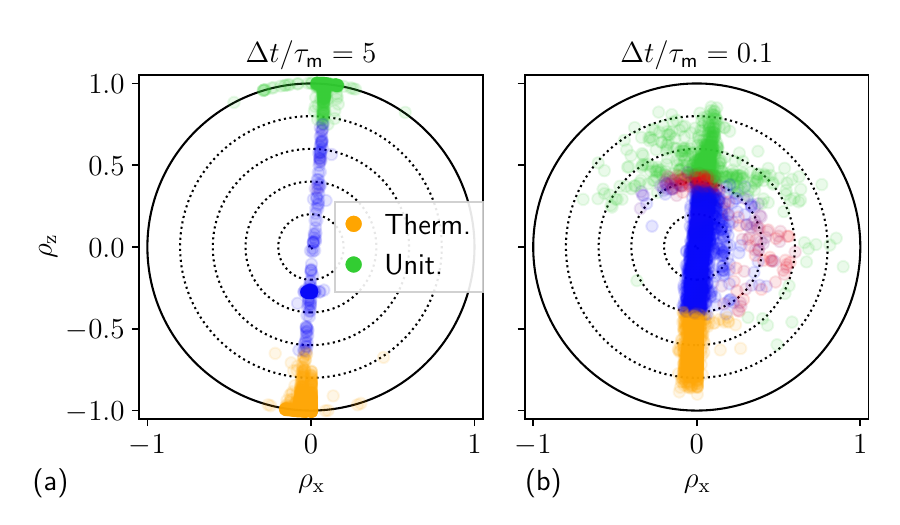}
	\includegraphics[width=0.99\columnwidth]{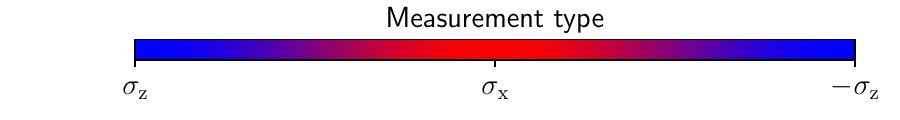}
	\caption{Actions chosen by the RL agent, as a function of qubit's state represented as a point on the Bloch sphere, plotted in the same style and with the same model as Fig.~\ref{fig:x_z_cont}, but here the RL agent can choose the measurement basis (blue to red dots). 
Each panel represents a different value of the continuous measurement strength $\Delta t/\tau_\text{m}$ shown on the plot.}
	\label{fig:cont_bloch_angle}
\end{figure}
In Fig.~\ref{fig:cont_bloch_angle} we show the results allowing the RL agent the freedom of choosing the measurement angle, plotted in the same style as Fig.~\ref{fig:x_z_cont}. 
We observe that the strong measurement case [Fig.~\ref{fig:cont_bloch_angle}(a)] corresponds to the cooling scheme powered by continuous $\sigma_{\rm z}$ measurement (blue points). Reducing the measurement strength resulted in the emergence of qubit states for which the $\sigma_{\rm x}$ measurement is performed, marked by the red points in Fig.~\ref{fig:cont_bloch_angle}(b).

\begin{figure}[!tb]
	\centering
	\includegraphics[width=0.99\columnwidth]{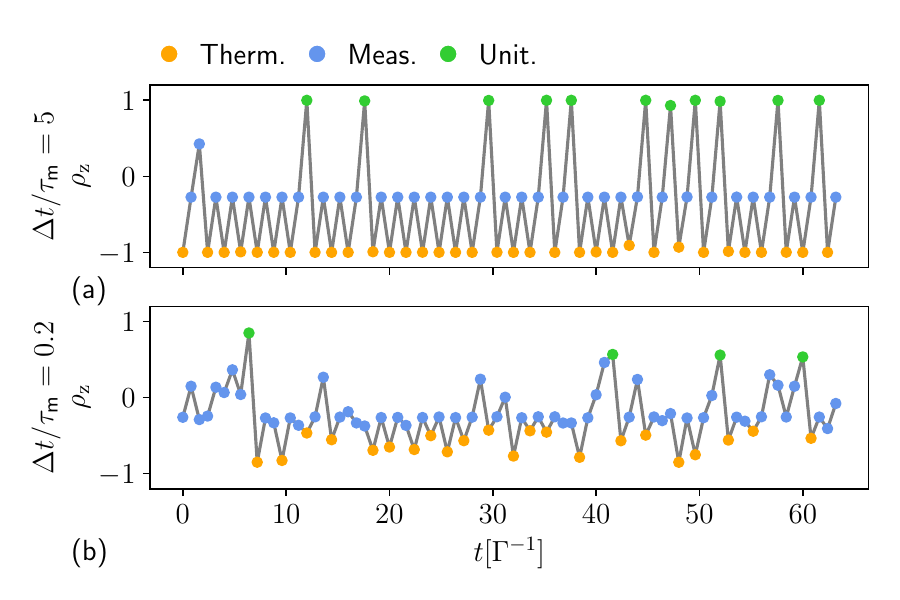}
	\caption{Plots of the state component $\rho_{\rm z}$, as a function of time, corresponding to the results shown in Fig.~\ref{fig:x_z_cont}(b,d) relative to the $\sigma_{\rm z}$ continuous measurement case. Each row corresponds to a different value of the continuous measurement strength $\Delta t/\tau_\text{m}$. An arbitrary trajectory is shown. The colors of the dots indicate the actions chosen by the RL agent at a given moment. }
	\label{fig:cont_z_logs}
\end{figure}
While in Fig.~\ref{fig:cont_x_logs} we showed a trajectory in the $\sigma_{\rm x}$ measurement case, in Fig.~\ref{fig:cont_z_logs} we additionally show a trajectory of the $\rho_{\rm z}$ component of the state, as a function of time, corresponding to Fig.~\ref{fig:x_z_cont}(b,d). As we can see, the results are qualitatively similar to the $\sigma_{\rm x}$ measurement case.

\section{Reinforcement learning algorithm}
\label{app:rl}
In this appendix we provide specific information on the reinforcement learning algorithm that was used, and details of the optimization carried out in the manuscript. 

We employ a generalization of the soft actor-critic method, first introduced in \cite{haarnoja2018_pmlr, haarnoja2018_arxiv_sac, haarnoja2018_arxiv_walk} for continuous actions only, to a combination of discrete and continuous actions \cite{christodoulou2019,delalleau2019,erdman2022, erdman2023_pnas}. In particular, we use the method and code that was implemented in Ref.~\cite{erdman2023_pnas} but, aside from minor implementation details that will be described below, we only change the neural network architectures that describe the policy function and the value function. Therefore, in the following we provide an overview of the method, we describe the differences compared to Ref.~\cite{erdman2023_pnas}, but we refer to \cite{erdman2023_pnas} for a more detailed explanation of the method.
Our code is implemented in PyTorch \cite{paszke2017} and is based on numerous modifications of the
SAC implementation provided by Spinning Up from OpenAI \cite{spinningup2018}. For each of the different physical cases discussed in this manuscript, we implemented an RL environment that computes the state evolution based on the chosen action, and computes the rewards. These calculations were performed using the QuTiP2 toolbox \cite{johansson2013}.
For code and data to reproduce our results, see the Code and Data Availability Statement.

\subsection{Soft actor-critic algorithm}
As described in Eq.~(\ref{eq:pi_star}) of the main text, the goal of RL is to determine the policy function $\pi(a|s)$ that maximizes the expected sum of future rewards, i.e. 
\begin{equation}
	\pi^*(a|s) = \underset{\pi}{\text{argmax}}\,\mathop{\mathrm{E}_\pi}\limits_{ s\sim \mu_\pi} \left[ \sum_{k=0}^\infty \gamma^k\,r_{k+1}  \Big| s_0=s \right],
 \label{eq:pi_star_app}
\end{equation}
where $E_\pi$ denotes the expectation value over the state evolution and over the actions chosen according to the policy $\pi$, and $s_0=s$ is sampled from the steady-state distribution of states $\mu_\pi$ visited using $\pi$.
In the soft actor-critic method, Eq.~(\ref{eq:pi_star_app}) is solved using the idea of \textit{policy iteration} \cite{sutton2018}, i.e. iterating over two steps: the \textit{policy evaluation} and \textit{policy improvement} step. In the policy evaluation step, the quality of the current policy is evaluate by estimating its value function $Q^\pi(s,a)$ (the critic). In the policy improvement step, the value function is used to improve the policy function (the actor). 
In practice, we start from an initially random policy, apply it for some steps onto the environment, use the collected experience to perform policy evaluation and policy improvement, then we collect new observations using the new policy, and so on, until convergence.

The soft actor-critic method balances exploration and exploitation \cite{sutton2018} using an entropy regularized maximization objective \cite{haarnoja2018_pmlr, haarnoja2018_arxiv_sac,haarnoja2018_arxiv_walk}, which corresponds to maximizing a trade-off of the sum of future rewards, and the entropy of the policy. Since we are dealing both with discrete and continuous actions, we use the formulation detailed in Ref.~\cite{erdman2023_pnas}. 
We decompose the policy function $\pi(d,u|s)$, where $a=\{d,u\}$ is a combination of a discrete ($d$) and continuous ($u$) action, using the following identity
\begin{equation}
    \pi(d,u|s) = \pi_{\mathrm{D}}(d|s) \cdot \pi_{\mathrm{C}}(u|d,s).
    \label{eq:pi_decomp}
\end{equation}
Here $\pi_{\mathrm{D}}(d|s)$ is the marginal probability of choosing the discrete (D) action $d$, whereas $\pi_{\mathrm{C}}(u|d,s)$ is the conditional probability of choosing the continuous (C) action $u$, given that the discrete action $d$ was chosen.
Let us denote as
\begin{equation}
    H[P] = \mathop{\mathrm{E}}\limits_{x\sim P}[ -\log P(x) ]
\end{equation}
the entropy of a probability function (or probability density in the continuous case) $P(x)$.
Using the decomposition in Eq.~(\ref{eq:pi_decomp}), we can write
the entropy of the policy function as
\begin{equation}
    H[\pi(\cdot|s)] = H^{\pi}_\text{D}(s) + H^{\pi}_\text{C}(s),
    \label{eq:h_decomposition}
\end{equation}
where 
\begin{align}
    H^{\pi}_\text{D}(s) &= H[\pi_{\mathrm{D}}(\cdot|s)], &  
    H^{\pi}_\text{C}(s) &= \sum_d \pi_{\mathrm{D}}(d|s) H[\pi_{\mathrm{C}}(\cdot|d,s)],
\end{align}
represent, respectively, the entropy of the discrete actions and the average (differential) entropy of the continuous action.

Our aim is now to solve the following equation \cite{erdman2023_pnas}
\begin{multline}
    \pi^* = \mathrm{arg}\max_\pi\, \mathop{\mathrm{E}_\pi}\limits_{s\sim \mathcal{B}}\Big[ \sum_{k=0}^{\infty} \gamma^k \,\Big(r_{k+1} + \alpha_\text{D} H^\pi_\text{D}(s_k) \\ 
     + \alpha_\text{C}H^\pi_\text{C}(s_k)  \Big) \Big| s_0 =s  \Big],
    \label{eq:pi_star_final}
\end{multline}
where $\alpha_\text{C}, \alpha_\text{D} \geq 0$ are two ``temperature'' parameters that balance the trade-off between exploration and exploitation. Indeed, for $\alpha_\text{C}=\alpha_\text{D}=0$, Eq.~(\ref{eq:pi_star_final}) coincides with the standard RL optimization objective of Eq.~(\ref{eq:pi_star_app}). For $\alpha_\text{C},\alpha_\text{D}>0$, we are finding trade-offs between maximizing the rewards, and increasing respectively the entropy of the continuous and discrete policy. Since the entropy is an indicator of the randomness of a probability distribution, the larger are $\alpha_\text{C}$ and $\alpha_\text{D}$, the more we are incentivizing an explorative behavior of the agent. Notice that in Eq.~(\ref{eq:pi_star_final}) replaced the (unknown) state distribution $\mu_\pi$ with a first-in-first-out replay buffer $\mathcal{B}$ that is populated during training by storing the observed one-step transitions $(s_i,a_i, r_{i+1}, s_{i+1})$.

The parameters $\alpha_\text{C}$ and $\alpha_\text{D}$ are automatically tuned during training to reach, respectively, a target entropy $\bar{H}_\text{C}$ and $\bar{H}_\text{D}$ of the continuous and discrete policy functions. Then, during training, we decrease them to switch from a more explorative to a more deterministic behavior.

Another important function in actor-critic methods is the value function $Q^\pi(s,a)$ \cite{sutton2018}. As discussed in Ref.~\cite{erdman2023_pnas}, from Eq.~(\ref{eq:pi_star_final}) we define the value function as 
\begin{multline}
    Q^\pi(s,a) = 
    \text{E}_\pi  \Big[ r_{1} + 
    \sum_{k=1}^{\infty} \gamma^k \,\Big(r_{k+1} + \alpha_\text{D} H^\pi_\text{D}(s_k) \\
     + \alpha_\text{C}H^\pi_\text{C}(s_k)  \Big) \Big| s_0=s, a_0=a \Big],
    \label{eq:q_def}
\end{multline}
which not only represents the sum of future rewards that we would obtain starting from state $s$, performing action $a$, and then choosing all future actions according to $\pi$, but it also incorporates the entropy terms. This function plays the role of the critic, and gives us a measure of the expected quality of a given policy $\pi$.

In the soft actor-critic method we use neural networks to parameterize the policy function $\pi(d,u|s)$ and the value function $Q^\pi(s,a)$. 

The value function is represented by a neural network $Q_\phi(s,a)$ that depends on a set of trainable parameters $\phi$.
As done in Ref.~\cite{erdman2022}, we use a feed-forward neural network that takes $(s,u)$ as input, and outputs $3$ values corresponding to the value of $\{ Q(s,u,d) \}_d$, for $d\in \{\text{Measure}, \text{Thermalize}, \text{Unitary}\}$. In particular, we use the ReLU function as non-linearity, two hidden layers, and a linear output layer.
Since the state $s$ coincides with the density matrix of the system, we expand the density matrix onto the computation basis, separate the real and imaginary parts, and arrange all elements into a vector. Although not strictly necessary, we further concatenate the last chosen  continuous action $u$ to the state vector. 

The policy function is described by the two functions $\pi_{\text{D},\theta}(d|s)$ and $\pi_{\text{C},\theta}(u|d,s)$, where $\theta$ is a set of trainable parameters. In particular, we first choose a discrete action $d$ according to the discrete distribution $\pi_{\text{D},\theta}(d|s)$, and then we choose as $\pi_{\text{C},\theta}(u|d,s)$ a squashed Gaussian policy, i.e. the distribution of the variable
\begin{equation}
\begin{aligned}
 \tilde{u}(\xi|d,s) &= 
    u_\text{a} + \frac{u_\text{b} - u_\text{a}}{2}[1+ \tanh\left( \mu_\theta(d,s) + \sigma_\theta(d,s)\cdot \xi )  \right)],   
     \\  \xi &\sim \mathcal{N}(0,1),
\end{aligned}
\label{eq:u_tilda}
\end{equation}
where $\mu_\theta(d,s)$ and $\sigma_\theta(d,s)$ represent respectively the mean and standard deviation of the Gaussian distribution, $\mathcal{N}(0,1)$ is the normal distribution with zero mean and unit variance, and $u\in[u_\mathrm{a},u_\mathrm{b}]$. This is the so-called reparameterization trick. The functions $\pi_\text{D}(d|s)$, $\mu_\theta(d,s)$ and $\sigma_\theta(d,s)$ are computed from a trainable neural network in the following way. We use a single feed-forward neural network that takes the state as input (in the same way as described for the value function), and outputs $9$ parameters: $\{\pi_\text{D}(d|s)\}_d$, $\{\mu_d\}_d$ and $\{m_d\}_d$ for $d\in \{\text{Measure}, \text{Thermalize}, \text{Unitary}\}$. More specifically, we use two hidden layers, the ReLU function as non-linearity, and then a linear layer to output  $\{\mu_d\}_d$ and $\{m_d\}_d$, and a soft-max function to output $\{\pi_\text{D}(d|s)\}_d$. Then $\mu_\theta(d,s)=\mu_d$, and we compute
\begin{equation}
    \sigma_\theta(d,s) = \sqrt{ m_d^2 + \epsilon},
\end{equation}
where $\epsilon=10^{-8}$ is a numerical safety parameter, to ensure that the standard deviation is positive.

To determine the value function $Q_\phi(s,a)$, the soft actor-critic introduces two value functions $Q_{\phi_1}(s,a)$ and $Q_{\phi_2}(s,a)$, and their parameters $\phi_i$, for $i=1,2$, are determined minimizing the following loss function \cite{erdman2023_pnas}
\begin{equation}
    L_Q(\phi_i) = \mathop{\mathrm{E}}\limits_{(s,a,r,s^\prime)\sim \mathcal{B}} \left[ ( Q_{\phi_i}(s,a) - y(r,s^\prime))^2  \right],
    \label{eq:q_loss}
\end{equation}
where 
\begin{multline}
    y(r,s^\prime) = r +  \gamma \underset{a^\prime \sim \pi(\cdot|s^\prime)}{\text{E}} \Big[ \min_{j=1,2}Q_{\phi_{\text{targ},j}}(s^\prime,a^\prime) + \alpha_\text{D} H_\text{D}(s^\prime) \\
     + \alpha_\text{C}H_\text{C}(s^\prime) \Big],
    \label{eq:y_1}
\end{multline}
and where $\phi_{\mathrm{targ},j}$, for $j=1,2$, are parameters that are not updated when minimizing the loss function; instead, they are constant during backpropagation, and then they are updated according to Polyak averaging, i.e.
\begin{equation}
    \phi_{\mathrm{targ},i} \leftarrow \rho_\mathrm{polyak} \phi_{\mathrm{targ},i} + (1-\rho_\mathrm{polyak})\phi_{i},
\end{equation}
where $\rho_\mathrm{polyak}$ is a hyperparameter. 
The parameters defining the policy function $\pi_\theta(a|s)$ are determined minimizing the following loss function \cite{erdman2023_pnas}
\begin{equation}
\begin{aligned}
&
\begin{multlined}
    L_\pi(\theta) = 
    \underset{ \substack{s\sim \mathcal{B}  }}{\text{E}}\Big[\sum_d \pi_{\text{D},\theta}(d|s) \Big(  \alpha_\text{D} \log\pi_{\text{D},\theta}(d|s) +\\
     \alpha_\text{C} \log\pi_{{\text{C}},\theta}(\tilde{u}_{\theta}(\xi|d,s)|d,s) - 
    \min_{j=1,2} Q_{\phi_j}(s,\tilde{u}_{\theta}(\xi|d,s),d)  \Big) \Big],
    \end{multlined}
    \\ &\quad\quad \xi \sim \mathcal{N}(0,1).
\end{aligned}
    \label{eq:pi_loss}
\end{equation}
Finally, to prevent the temperature parameters from becoming negative, we express them as $\alpha_\text{D}(\beta_\text{D})=e^{\beta_\text{D}}$ and $\alpha_\text{C}(\beta_\text{C})=e^{\beta_\text{C}}$, and we determine the $\beta_\text{D}$, $\beta_\text{C}$ parameters minimizing the following loss functions \cite{erdman2023_pnas}
\begin{equation}
    L_\text{D}(\beta_\text{D}) = \alpha_\text{D}(\beta_\text{D})\underset{ \substack{s\sim \mathcal{B} }}{\text{E}}\left[ -\sum_d \pi_{\text{D}}(d|s) \log\pi_{\text{D}}(d|s) - \bar{H}_\text{D} \right],
    \label{eq:d_loss}
\end{equation}
and 
\begin{equation}
\begin{aligned}
&
\begin{multlined}
        L_\text{C}(\beta_\text{C}) = \alpha_\text{C}(\beta_\text{C}) \\
    \cdot\underset{ \substack{s\sim \mathcal{B} }}{\text{E}}\left[ -\sum_d \pi_{\text{D}}(d|s)[ \log\pi_{\text{C}}(\tilde{u}_{\theta}(\xi|d,s)|d,s) ] - \bar{H}_\text{C} \right],
    \end{multlined}
    \\ &\quad\quad \xi \sim \mathcal{N}(0,1).
\end{aligned}
    \label{eq:c_loss}
\end{equation}

\subsection{Training details}
We now provide additional details on the actual implementation and training. As mentioned previously, we store all one-step transitions  $(s_k,a_k, r_{k+1}, s_{k+1})$ in a first-in-first-out replay buffer $\mathcal{B}$ of fixed dimension from which batches one one-step transitions are sampled to estimate the loss functions. In order to start from an exploratory behavior, as in Ref.~\cite{erdman2023_pnas} we first choose actions randomly (according to a uniform distribution) for a fixed number of initial time-steps. Furthermore, for a different number of fixed initial steps, we do not perform any updates to the value function, the policy function, nor to the temperature parameters. This is to allow the replay buffer to have enough transitions before sampling transitions from it. After this initial phase, we repeat $n_\text{updates}$ optimization steps of all quantities, every $n_\text{updates}$ steps such that, overall, the number of updates and of time-steps coincide. In particular, we use the ADAM optimizer \cite{kingma2014} with default hyperparameters and learning rate reported in Tables~\ref{tab:hyperparams}, \ref{tab:hyperparams2} to minimize $L_Q(\phi_i)$ $L_\pi(\theta)$, and stochastic gradient descent with default hyperparameters and learning rate reported in Tables~\ref{tab:hyperparams},  \ref{tab:hyperparams2} to minimize $L_\text{D}(\beta_\text{D})$ and $L_\text{C}(\beta_\text{C})$. As mentioned in the previous section, we transition from an exploratory to a more deterministic behaviour during training by scheduling the target entropy terms $\bar{H}_\text{D}$ and $\bar{H}_\text{C}$ as follows 
\begin{multline}
    \bar{H}_a(n_\text{steps}) = \bar{H}_{a,\text{end}}  \\
    + (\bar{H}_{a,\text{start}}-\bar{H}_{a,\text{end}})\exp(-n_\text{steps}/\bar{H}_{a,\text{decay}}),
\end{multline}
where $a=\text{C},\text{D}$, $n_\text{steps}$ is the current step number, and $\bar{H}_{a,\text{start}}$, $\bar{H}_{a,\text{end}}$ and $\bar{H}_{a,\text{decay}}$ are hyperparameters. Furthermore, in order to have hyperparameters for the target entropy that do not depend on the size of the continuous action interval $[u_\text{min},u_\text{max}]$, when computing the entropy of the policy for the loss functions, we assume that the continuous actions always lie in the interval $[-1,1]$.

\begin{table*}[h]
\centering
\begin{tabular}{lccccccc}
\toprule
\begin{tabular}{c}Hyperparameter  \\ ~ \\ ~ \\ \end{tabular} ~ & ~ 
\begin{tabular}{c}Fig.~\ref{fig:therm_pareto}\\  $c=\{0.95, 0.9, $ \\ \multicolumn{1}{r}{$ 0.8, 0.7, 0.65\}$} \end{tabular} ~ &  ~ \begin{tabular}{c}Fig.~\ref{fig:therm_pareto}\\  $c=\{0.6,0.58\}$ \\ ~ \end{tabular} ~ &  ~ 
\begin{tabular}{c}Figs.~\ref{fig:all_finite_res_no_counter},\ref{fig:all_finite_no_counter_prot}\\  $c=\{1, 0.95, 0.9, $  \\ \multicolumn{1}{r}{$ 0.85, 0.8, 0.75, 0.7\}$}  \end{tabular} ~ &  ~ 
\begin{tabular}{c}Figs.~\ref{fig:all_finite_res_no_counter},\ref{fig:all_finite_no_counter_prot}\\  $c=0.65$ \\ ~ \end{tabular} ~ &  ~ \begin{tabular}{c}Figs.~\ref{fig:all_finite_res_counter},\ref{fig:all_finite_counter_prot}\\  $c=\{1,0.9,$ \\ \multicolumn{1}{r}{$0.8,0.75,0.7\}$} \end{tabular} ~ &  ~  
\begin{tabular}{c}Figs.~\ref{fig:all_finite_res_counter},\ref{fig:all_finite_counter_prot}\\  $c=\{0.95,$ \\ \multicolumn{1}{r}{$0.85,0.65\}$} \end{tabular} ~   
\\
\midrule
Batch size  & 256 & `` & ``& ``& ``& `` \\
Training steps & 160k & 320k & 160k & `` & 320k& `` \\
ADAM learning rate  & 0.001 & `` & ``& `` & 0.0003 & `` \\
SGD learning rate  & 0.003& `` & ``& `` & `` & ``\\
$\mathcal{B}$ size  & 80k & `` & ``  & 160k & 80k& 180k \\
$\rho_\text{polyak}$  & 0.995& ``& ``& `` & ``& ``\\
Hidden layers size  ~ & ~ (256, 128)& `` & ``& `` & (256, 256)& `` \\
Initial random steps  &  5k & `` & ``& `` & ``& ``\\
First update at step  & 1k & ``  & ``& `` & ``& ``\\
$n_\text{updates}$  & 50 & `` & ``& ``  & ``& ``\\
$\bar{H}_{\text{C},\text{start}}$  & 0.8 & `` & `` & 0.77 & 0.74& `` \\
$\bar{H}_{\text{C},\text{end}}$  & -3 & ``  & ``& `` & ``& ``\\
$\bar{H}_{\text{C},\text{decay}}$  & 60k & ``  & `` & 120k & 180k& `` \\
$\bar{H}_{\text{D},\text{start}}$ & $\ln 3$ & `` & ``& `` & `` & ``\\
$\bar{H}_{\text{D},\text{end}}$  & 0.01 & `` & ``& `` & ``& ``\\
$\bar{H}_{\text{D},\text{decay}}$ & 30k & `` & `` & 60k & 90k& 140k\\
$\gamma$ & 0.998 & `` & ``& ``& ``& ``\\
\bottomrule
\end{tabular}
\caption{Training hyperparameters used for each figure reported in the top row. The `` symbol means the same hyperparameter as to its left.}
\label{tab:hyperparams}
\end{table*}

\begin{table*}[h]
\centering
\begin{tabular}{lcccccc}
\toprule
\begin{tabular}{c}Hyperparameter  \\ ~ \\ ~ \\ \end{tabular} ~ & ~ 
\begin{tabular}{c}Figs.~\ref{fig:Bloch_discr_meas}, \ref{fig:x_logs}, \ref{fig:bloch_angle}, \\ \multicolumn{1}{r}{\ref{fig:policy_comparison}, \ref{fig:power_vs_penalty}, \ref{fig:z_logs}, \ref{fig:penalty_bloch}} \\ ~ \end{tabular} ~ &  ~ 
\begin{tabular}{c}Fig.~\ref{fig:x_z_cont}(a,c), \\ Fig.~\ref{fig:cont_x_logs} \\~  \end{tabular} ~ &  ~ 
\begin{tabular}{c}Fig.~\ref{fig:x_z_cont}(b,d), \\   \ref{fig:cont_z_logs} \\ ~ \end{tabular} ~ &  ~ 
\begin{tabular}{c}Fig.~\ref{fig:gap_cont_action} \\  ~ \\ ~ \end{tabular} ~ &  ~ 
\begin{tabular}{c}Fig.~\ref{fig:cont_bloch_angle}\\  ~ \\ ~ \end{tabular} ~ &  ~ 
\\
\midrule
Batch size  & 128 & 256 & ``& `` & `` \\
Training steps & 250k & `` & `` & 300k & 250k  \\
ADAM learning rate  & 0.0008 & 0.0003 & ``& `` & ``  \\
SGD learning rate  & 0.001& `` & ``& `` & `` \\
$\mathcal{B}$ size  & 80k & 100k & 80k & ``  & `` \\
$\rho_\text{polyak}$  & 0.995& ``& ``& `` & `` \\
Hidden layers size  ~ & ~ (128, 128)& `` & ``& `` & ``  \\
Initial random steps  &  5k & `` & ``& `` & `` \\
First update at step  & 1k & ``  & ``& `` & `` \\
$n_\text{updates}$  & 50 & `` & ``& `` & ``  & \\
$\bar{H}_{\text{C},\text{start}}$  & 0.8 & `` & ``& ``  & `` \\
$\bar{H}_{\text{C},\text{end}}$  & -3 & ``  & ``& `` & `` \\
$\bar{H}_{\text{C},\text{decay}}$  & 100k & ``  & 80k& ``  &  `` \\
$\bar{H}_{\text{D},\text{start}}$ & $\ln 3$ & `` & ``& `` & `` \\
$\bar{H}_{\text{D},\text{end}}$  & 0.01 & `` & ``& `` & `` \\
$\bar{H}_{\text{D},\text{decay}}$ & 100k & `` & 80k & `` & 100k  \\
$\gamma$ & 0.998 & `` & ``& `` & ``\\
\bottomrule
\end{tabular}
\caption{Training hyperparameters used for each figure reported in the top row. The `` symbol means the same hyperparameter as to its left.}
\label{tab:hyperparams2}
\end{table*}

All hyperparameters used during training are reported in Tables~\ref{tab:hyperparams} and \ref{tab:hyperparams2}. Most results were found on the first run, except for Figs.~\ref{fig:all_finite_res_counter}, \ref{fig:all_finite_counter_prot}, \ref{fig:cont_bloch_angle} where the optimization was sometimes repeated a few times to improve the result. The values of $c$ used for the multi-objective optimizations are all reported in Table~\ref{tab:hyperparams}, except for Fig.~\ref{fig:power_vs_penalty}, where we have used $c=\{1, 0.95, 0.9,\dots, 0.45, 0.4\}$, and multiplied the cooling power by a factor $35$ and the dissipation term by a factor $4$ during the optimization of $\ev*{F_c}$.
When the continuous control $u(t)\in[u_\text{min},u_\text{max}]$ corresponds to the measurement angle $\theta$, we found that the RL method converges to better results if we choose an interval $[u_\text{min},u_\text{max}]$ that is larger than $[0,\pi]$. In particular, we used $[-0.2,3.4]$ and $[-0.3,3.6]$ respectively when optimizing the discrete and continuous measurement case.


\begin{thebibliography}{155}%
\makeatletter
\providecommand \@ifxundefined [1]{%
 \@ifx{#1\undefined}
}%
\providecommand \@ifnum [1]{%
 \ifnum #1\expandafter \@firstoftwo
 \else \expandafter \@secondoftwo
 \fi
}%
\providecommand \@ifx [1]{%
 \ifx #1\expandafter \@firstoftwo
 \else \expandafter \@secondoftwo
 \fi
}%
\providecommand \natexlab [1]{#1}%
\providecommand \enquote  [1]{``#1''}%
\providecommand \bibnamefont  [1]{#1}%
\providecommand \bibfnamefont [1]{#1}%
\providecommand \citenamefont [1]{#1}%
\providecommand \href@noop [0]{\@secondoftwo}%
\providecommand \href [0]{\begingroup \@sanitize@url \@href}%
\providecommand \@href[1]{\@@startlink{#1}\@@href}%
\providecommand \@@href[1]{\endgroup#1\@@endlink}%
\providecommand \@sanitize@url [0]{\catcode `\\12\catcode `\$12\catcode
  `\&12\catcode `\#12\catcode `\^12\catcode `\_12\catcode `\%12\relax}%
\providecommand \@@startlink[1]{}%
\providecommand \@@endlink[0]{}%
\providecommand \url  [0]{\begingroup\@sanitize@url \@url }%
\providecommand \@url [1]{\endgroup\@href {#1}{\urlprefix }}%
\providecommand \urlprefix  [0]{URL }%
\providecommand \Eprint [0]{\href }%
\providecommand \doibase [0]{https://doi.org/}%
\providecommand \selectlanguage [0]{\@gobble}%
\providecommand \bibinfo  [0]{\@secondoftwo}%
\providecommand \bibfield  [0]{\@secondoftwo}%
\providecommand \translation [1]{[#1]}%
\providecommand \BibitemOpen [0]{}%
\providecommand \bibitemStop [0]{}%
\providecommand \bibitemNoStop [0]{.\EOS\space}%
\providecommand \EOS [0]{\spacefactor3000\relax}%
\providecommand \BibitemShut  [1]{\csname bibitem#1\endcsname}%
\let\auto@bib@innerbib\@empty
\bibitem [{\citenamefont {Murch}\ \emph {et~al.}(2013)\citenamefont {Murch},
  \citenamefont {Weber}, \citenamefont {Macklin},\ and\ \citenamefont
  {Siddiqi}}]{murch2013observing}%
  \BibitemOpen
  \bibfield  {author} {\bibinfo {author} {\bibfnamefont {K.}~\bibnamefont
  {Murch}}, \bibinfo {author} {\bibfnamefont {S.}~\bibnamefont {Weber}},
  \bibinfo {author} {\bibfnamefont {C.}~\bibnamefont {Macklin}},\ and\ \bibinfo
  {author} {\bibfnamefont {I.}~\bibnamefont {Siddiqi}},\ }\bibfield  {title}
  {\bibinfo {title} {Observing single quantum trajectories of a superconducting
  quantum bit},\ }\href {https://doi.org/10.1038/nature12539} {\bibfield
  {journal} {\bibinfo  {journal} {Nature}\ }\textbf {\bibinfo {volume} {502}},\
  \bibinfo {pages} {211} (\bibinfo {year} {2013})}\BibitemShut {NoStop}%
\bibitem [{\citenamefont {Kjaergaard}\ \emph {et~al.}(2020)\citenamefont
  {Kjaergaard}, \citenamefont {Schwartz}, \citenamefont {Braum{\"u}ller},
  \citenamefont {Krantz}, \citenamefont {Wang}, \citenamefont {Gustavsson},\
  and\ \citenamefont {Oliver}}]{kjaergaard2020superconducting}%
  \BibitemOpen
  \bibfield  {author} {\bibinfo {author} {\bibfnamefont {M.}~\bibnamefont
  {Kjaergaard}}, \bibinfo {author} {\bibfnamefont {M.~E.}\ \bibnamefont
  {Schwartz}}, \bibinfo {author} {\bibfnamefont {J.}~\bibnamefont
  {Braum{\"u}ller}}, \bibinfo {author} {\bibfnamefont {P.}~\bibnamefont
  {Krantz}}, \bibinfo {author} {\bibfnamefont {J.~I.-J.}\ \bibnamefont {Wang}},
  \bibinfo {author} {\bibfnamefont {S.}~\bibnamefont {Gustavsson}},\ and\
  \bibinfo {author} {\bibfnamefont {W.~D.}\ \bibnamefont {Oliver}},\ }\bibfield
   {title} {\bibinfo {title} {Superconducting qubits: Current state of play},\
  }\href {https://doi.org/10.1146/annurev-conmatphys-031119-050605} {\bibfield
  {journal} {\bibinfo  {journal} {Ann. Rev. Cond. Matt. Phys.}\ }\textbf
  {\bibinfo {volume} {11}},\ \bibinfo {pages} {369} (\bibinfo {year}
  {2020})}\BibitemShut {NoStop}%
\bibitem [{\citenamefont {Vijay}\ \emph {et~al.}(2012)\citenamefont {Vijay},
  \citenamefont {Macklin}, \citenamefont {Slichter}, \citenamefont {Weber},
  \citenamefont {Murch}, \citenamefont {Naik}, \citenamefont {Korotkov},\ and\
  \citenamefont {Siddiqi}}]{vijay2012stabilizing}%
  \BibitemOpen
  \bibfield  {author} {\bibinfo {author} {\bibfnamefont {R.}~\bibnamefont
  {Vijay}}, \bibinfo {author} {\bibfnamefont {C.}~\bibnamefont {Macklin}},
  \bibinfo {author} {\bibfnamefont {D.}~\bibnamefont {Slichter}}, \bibinfo
  {author} {\bibfnamefont {S.}~\bibnamefont {Weber}}, \bibinfo {author}
  {\bibfnamefont {K.}~\bibnamefont {Murch}}, \bibinfo {author} {\bibfnamefont
  {R.}~\bibnamefont {Naik}}, \bibinfo {author} {\bibfnamefont {A.~N.}\
  \bibnamefont {Korotkov}},\ and\ \bibinfo {author} {\bibfnamefont
  {I.}~\bibnamefont {Siddiqi}},\ }\bibfield  {title} {\bibinfo {title}
  {{Stabilizing Rabi oscillations in a superconducting qubit using quantum
  feedback}},\ }\href {https://doi.org/10.1038/nature11505} {\bibfield
  {journal} {\bibinfo  {journal} {Nature}\ }\textbf {\bibinfo {volume} {490}},\
  \bibinfo {pages} {77} (\bibinfo {year} {2012})}\BibitemShut {NoStop}%
\bibitem [{\citenamefont {Weber}\ \emph {et~al.}(2014)\citenamefont {Weber},
  \citenamefont {Chantasri}, \citenamefont {Dressel}, \citenamefont {Jordan},
  \citenamefont {Murch},\ and\ \citenamefont {Siddiqi}}]{weber2014mapping}%
  \BibitemOpen
  \bibfield  {author} {\bibinfo {author} {\bibfnamefont {S.}~\bibnamefont
  {Weber}}, \bibinfo {author} {\bibfnamefont {A.}~\bibnamefont {Chantasri}},
  \bibinfo {author} {\bibfnamefont {J.}~\bibnamefont {Dressel}}, \bibinfo
  {author} {\bibfnamefont {A.~N.}\ \bibnamefont {Jordan}}, \bibinfo {author}
  {\bibfnamefont {K.}~\bibnamefont {Murch}},\ and\ \bibinfo {author}
  {\bibfnamefont {I.}~\bibnamefont {Siddiqi}},\ }\bibfield  {title} {\bibinfo
  {title} {Mapping the optimal route between two quantum states},\ }\href
  {https://doi.org/10.1038/nature13559} {\bibfield  {journal} {\bibinfo
  {journal} {Nature}\ }\textbf {\bibinfo {volume} {511}},\ \bibinfo {pages}
  {570} (\bibinfo {year} {2014})}\BibitemShut {NoStop}%
\bibitem [{\citenamefont {H{\"a}ffner}\ \emph {et~al.}(2008)\citenamefont
  {H{\"a}ffner}, \citenamefont {Roos},\ and\ \citenamefont
  {Blatt}}]{haffner2008quantum}%
  \BibitemOpen
  \bibfield  {author} {\bibinfo {author} {\bibfnamefont {H.}~\bibnamefont
  {H{\"a}ffner}}, \bibinfo {author} {\bibfnamefont {C.~F.}\ \bibnamefont
  {Roos}},\ and\ \bibinfo {author} {\bibfnamefont {R.}~\bibnamefont {Blatt}},\
  }\bibfield  {title} {\bibinfo {title} {Quantum computing with trapped ions},\
  }\href {https://doi.org/10.1016/j.physrep.2008.09.003} {\bibfield  {journal}
  {\bibinfo  {journal} {Phys. Rep.}\ }\textbf {\bibinfo {volume} {469}},\
  \bibinfo {pages} {155} (\bibinfo {year} {2008})}\BibitemShut {NoStop}%
\bibitem [{\citenamefont {Schneider}\ \emph {et~al.}(2012)\citenamefont
  {Schneider}, \citenamefont {Porras},\ and\ \citenamefont
  {Schaetz}}]{schneider2012experimental}%
  \BibitemOpen
  \bibfield  {author} {\bibinfo {author} {\bibfnamefont {C.}~\bibnamefont
  {Schneider}}, \bibinfo {author} {\bibfnamefont {D.}~\bibnamefont {Porras}},\
  and\ \bibinfo {author} {\bibfnamefont {T.}~\bibnamefont {Schaetz}},\
  }\bibfield  {title} {\bibinfo {title} {Experimental quantum simulations of
  many-body physics with trapped ions},\ }\href
  {https://doi.org/10.1088/0034-4885/75/2/024401} {\bibfield  {journal}
  {\bibinfo  {journal} {Rep. Prog. Phys.}\ }\textbf {\bibinfo {volume} {75}},\
  \bibinfo {pages} {024401} (\bibinfo {year} {2012})}\BibitemShut {NoStop}%
\bibitem [{\citenamefont {Josefsson}\ \emph {et~al.}(2018)\citenamefont
  {Josefsson}, \citenamefont {Svilans}, \citenamefont {Burke}, \citenamefont
  {Hoffmann}, \citenamefont {Fahlvik}, \citenamefont {Thelander}, \citenamefont
  {Leijnse},\ and\ \citenamefont {Linke}}]{josefsson2018quantum}%
  \BibitemOpen
  \bibfield  {author} {\bibinfo {author} {\bibfnamefont {M.}~\bibnamefont
  {Josefsson}}, \bibinfo {author} {\bibfnamefont {A.}~\bibnamefont {Svilans}},
  \bibinfo {author} {\bibfnamefont {A.~M.}\ \bibnamefont {Burke}}, \bibinfo
  {author} {\bibfnamefont {E.~A.}\ \bibnamefont {Hoffmann}}, \bibinfo {author}
  {\bibfnamefont {S.}~\bibnamefont {Fahlvik}}, \bibinfo {author} {\bibfnamefont
  {C.}~\bibnamefont {Thelander}}, \bibinfo {author} {\bibfnamefont
  {M.}~\bibnamefont {Leijnse}},\ and\ \bibinfo {author} {\bibfnamefont
  {H.}~\bibnamefont {Linke}},\ }\bibfield  {title} {\bibinfo {title} {A
  quantum-dot heat engine operating close to the thermodynamic efficiency
  limits},\ }\href {https://doi.org/10.1038/s41565-018-0200-5} {\bibfield
  {journal} {\bibinfo  {journal} {Nature Nanotech.}\ }\textbf {\bibinfo
  {volume} {13}},\ \bibinfo {pages} {920} (\bibinfo {year} {2018})}\BibitemShut
  {NoStop}%
\bibitem [{\citenamefont {Prete}\ \emph {et~al.}(2019)\citenamefont {Prete},
  \citenamefont {Erdman}, \citenamefont {Demontis}, \citenamefont {Zannier},
  \citenamefont {Ercolani}, \citenamefont {Sorba}, \citenamefont {Beltram},
  \citenamefont {Rossella}, \citenamefont {Taddei},\ and\ \citenamefont
  {Roddaro}}]{prete2019thermoelectric}%
  \BibitemOpen
  \bibfield  {author} {\bibinfo {author} {\bibfnamefont {D.}~\bibnamefont
  {Prete}}, \bibinfo {author} {\bibfnamefont {P.~A.}\ \bibnamefont {Erdman}},
  \bibinfo {author} {\bibfnamefont {V.}~\bibnamefont {Demontis}}, \bibinfo
  {author} {\bibfnamefont {V.}~\bibnamefont {Zannier}}, \bibinfo {author}
  {\bibfnamefont {D.}~\bibnamefont {Ercolani}}, \bibinfo {author}
  {\bibfnamefont {L.}~\bibnamefont {Sorba}}, \bibinfo {author} {\bibfnamefont
  {F.}~\bibnamefont {Beltram}}, \bibinfo {author} {\bibfnamefont
  {F.}~\bibnamefont {Rossella}}, \bibinfo {author} {\bibfnamefont
  {F.}~\bibnamefont {Taddei}},\ and\ \bibinfo {author} {\bibfnamefont
  {S.}~\bibnamefont {Roddaro}},\ }\bibfield  {title} {\bibinfo {title}
  {Thermoelectric conversion at 30 k in inas/inp nanowire quantum dots},\
  }\href {https://doi.org/10.1021/acs.nanolett.9b00276} {\bibfield  {journal}
  {\bibinfo  {journal} {Nano Lett.}\ }\textbf {\bibinfo {volume} {19}},\
  \bibinfo {pages} {3033} (\bibinfo {year} {2019})}\BibitemShut {NoStop}%
\bibitem [{\citenamefont {Sothmann}\ \emph {et~al.}(2014)\citenamefont
  {Sothmann}, \citenamefont {S{\'a}nchez},\ and\ \citenamefont
  {Jordan}}]{sothmann2014thermoelectric}%
  \BibitemOpen
  \bibfield  {author} {\bibinfo {author} {\bibfnamefont {B.}~\bibnamefont
  {Sothmann}}, \bibinfo {author} {\bibfnamefont {R.}~\bibnamefont
  {S{\'a}nchez}},\ and\ \bibinfo {author} {\bibfnamefont {A.~N.}\ \bibnamefont
  {Jordan}},\ }\bibfield  {title} {\bibinfo {title} {Thermoelectric energy
  harvesting with quantum dots},\ }\href
  {https://doi.org/10.1088/0957-4484/26/3/032001} {\bibfield  {journal}
  {\bibinfo  {journal} {Nanotechnology}\ }\textbf {\bibinfo {volume} {26}},\
  \bibinfo {pages} {032001} (\bibinfo {year} {2014})}\BibitemShut {NoStop}%
\bibitem [{\citenamefont {Doyle}\ \emph {et~al.}(2013)\citenamefont {Doyle},
  \citenamefont {Francis},\ and\ \citenamefont
  {Tannenbaum}}]{doyle2013feedback}%
  \BibitemOpen
  \bibfield  {author} {\bibinfo {author} {\bibfnamefont {J.~C.}\ \bibnamefont
  {Doyle}}, \bibinfo {author} {\bibfnamefont {B.~A.}\ \bibnamefont {Francis}},\
  and\ \bibinfo {author} {\bibfnamefont {A.~R.}\ \bibnamefont {Tannenbaum}},\
  }\href {https://doi.org/10.1007/978-0-387-85460-1_1} {\emph {\bibinfo {title}
  {Feedback control theory}}}\ (\bibinfo  {publisher} {Dover Publications},\
  \bibinfo {address} {New York},\ \bibinfo {year} {2013})\BibitemShut {NoStop}%
\bibitem [{\citenamefont {{\AA}str{\"o}m}\ and\ \citenamefont
  {Murray}(2021)}]{aastrom2021feedback}%
  \BibitemOpen
  \bibfield  {author} {\bibinfo {author} {\bibfnamefont {K.~J.}\ \bibnamefont
  {{\AA}str{\"o}m}}\ and\ \bibinfo {author} {\bibfnamefont {R.}~\bibnamefont
  {Murray}},\ }\href {https://doi.org/10.1515/9781400828739} {\emph {\bibinfo
  {title} {Feedback systems: an introduction for scientists and engineers}}}\
  (\bibinfo  {publisher} {Princeton University Press},\ \bibinfo {year}
  {2021})\BibitemShut {NoStop}%
\bibitem [{\citenamefont {Jordan}\ and\ \citenamefont
  {Siddiqi}(2024)}]{jordan2024quantum}%
  \BibitemOpen
  \bibfield  {author} {\bibinfo {author} {\bibfnamefont {A.~N.}\ \bibnamefont
  {Jordan}}\ and\ \bibinfo {author} {\bibfnamefont {I.~A.}\ \bibnamefont
  {Siddiqi}},\ }\href
  {https://www.cambridge.org/us/universitypress/subjects/physics/quantum-physics-quantum-information-and-quantum-computation/quantum-measurement-theory-and-practice?format=HB&isbn=9781009100069}
  {\emph {\bibinfo {title} {Quantum measurement: Theory and practice}}}\
  (\bibinfo  {publisher} {Cambridge University Press},\ \bibinfo {year}
  {2024})\BibitemShut {NoStop}%
\bibitem [{\citenamefont {Plenio}\ and\ \citenamefont
  {Vitelli}(2001)}]{plenio2001physics}%
  \BibitemOpen
  \bibfield  {author} {\bibinfo {author} {\bibfnamefont {M.~B.}\ \bibnamefont
  {Plenio}}\ and\ \bibinfo {author} {\bibfnamefont {V.}~\bibnamefont
  {Vitelli}},\ }\bibfield  {title} {\bibinfo {title} {The physics of
  forgetting: Landauer's erasure principle and information theory},\ }\href
  {https://doi.org/10.1080/00107510010018916} {\bibfield  {journal} {\bibinfo
  {journal} {Contemp. Phys.}\ }\textbf {\bibinfo {volume} {42}},\ \bibinfo
  {pages} {25} (\bibinfo {year} {2001})}\BibitemShut {NoStop}%
\bibitem [{\citenamefont {Maruyama}\ \emph {et~al.}(2009)\citenamefont
  {Maruyama}, \citenamefont {Nori},\ and\ \citenamefont
  {Vedral}}]{maruyama2009colloquium}%
  \BibitemOpen
  \bibfield  {author} {\bibinfo {author} {\bibfnamefont {K.}~\bibnamefont
  {Maruyama}}, \bibinfo {author} {\bibfnamefont {F.}~\bibnamefont {Nori}},\
  and\ \bibinfo {author} {\bibfnamefont {V.}~\bibnamefont {Vedral}},\
  }\bibfield  {title} {\bibinfo {title} {{Colloquium: The physics of
  Maxwell’s demon and information}},\ }\href
  {https://doi.org/10.1103/RevModPhys.81.1} {\bibfield  {journal} {\bibinfo
  {journal} {Rev. Mod. Phys.}\ }\textbf {\bibinfo {volume} {81}},\ \bibinfo
  {pages} {1} (\bibinfo {year} {2009})}\BibitemShut {NoStop}%
\bibitem [{\citenamefont {Szilard}(1929)}]{szilard1929}%
  \BibitemOpen
  \bibfield  {author} {\bibinfo {author} {\bibfnamefont {L.}~\bibnamefont
  {Szilard}},\ }\bibfield  {title} {\bibinfo {title} {{\"Uber die
  Entropieverminderung in einem thermodynamischen System bei Eingriffen
  intelligenter Wesen}},\ }\href {https://doi.org/10.1007/BF01341281}
  {\bibfield  {journal} {\bibinfo  {journal} {Z. Physik}\ }\textbf {\bibinfo
  {volume} {53}},\ \bibinfo {pages} {840} (\bibinfo {year} {1929})}\BibitemShut
  {NoStop}%
\bibitem [{\citenamefont {Landauer}(1957)}]{landauer1957}%
  \BibitemOpen
  \bibfield  {author} {\bibinfo {author} {\bibfnamefont {R.}~\bibnamefont
  {Landauer}},\ }\bibfield  {title} {\bibinfo {title} {Spatial variation of
  currents and fields due to localized scatterers in metallic conduction},\
  }\href {https://doi.org/10.1147/rd.13.0223} {\bibfield  {journal} {\bibinfo
  {journal} {IBM J. Res. Dev.}\ }\textbf {\bibinfo {volume} {1}},\ \bibinfo
  {pages} {223} (\bibinfo {year} {1957})}\BibitemShut {NoStop}%
\bibitem [{\citenamefont {Bennett}(1982)}]{bennett1982}%
  \BibitemOpen
  \bibfield  {author} {\bibinfo {author} {\bibfnamefont {C.~H.}\ \bibnamefont
  {Bennett}},\ }\bibfield  {title} {\bibinfo {title} {The thermodynamics of
  computation—a review},\ }\href {https://doi.org/10.1007/BF0208415}
  {\bibfield  {journal} {\bibinfo  {journal} {Int. J. Theor. Phys.}\ }\textbf
  {\bibinfo {volume} {21}},\ \bibinfo {pages} {905} (\bibinfo {year}
  {1982})}\BibitemShut {NoStop}%
\bibitem [{\citenamefont {Landauer}(1961)}]{landauer1961irreversibility}%
  \BibitemOpen
  \bibfield  {author} {\bibinfo {author} {\bibfnamefont {R.}~\bibnamefont
  {Landauer}},\ }\bibfield  {title} {\bibinfo {title} {Irreversibility and heat
  generation in the computing process},\ }\href
  {https://doi.org/10.1147/rd.53.0183} {\bibfield  {journal} {\bibinfo
  {journal} {IBM J. Res. Dev.}\ }\textbf {\bibinfo {volume} {5}},\ \bibinfo
  {pages} {183} (\bibinfo {year} {1961})}\BibitemShut {NoStop}%
\bibitem [{\citenamefont {B{\'e}rut}\ \emph {et~al.}(2012)\citenamefont
  {B{\'e}rut}, \citenamefont {Arakelyan}, \citenamefont {Petrosyan},
  \citenamefont {Ciliberto}, \citenamefont {Dillenschneider},\ and\
  \citenamefont {Lutz}}]{berut2012experimental}%
  \BibitemOpen
  \bibfield  {author} {\bibinfo {author} {\bibfnamefont {A.}~\bibnamefont
  {B{\'e}rut}}, \bibinfo {author} {\bibfnamefont {A.}~\bibnamefont
  {Arakelyan}}, \bibinfo {author} {\bibfnamefont {A.}~\bibnamefont
  {Petrosyan}}, \bibinfo {author} {\bibfnamefont {S.}~\bibnamefont
  {Ciliberto}}, \bibinfo {author} {\bibfnamefont {R.}~\bibnamefont
  {Dillenschneider}},\ and\ \bibinfo {author} {\bibfnamefont {E.}~\bibnamefont
  {Lutz}},\ }\bibfield  {title} {\bibinfo {title} {{Experimental verification
  of Landauer’s principle linking information and thermodynamics}},\ }\href
  {https://doi.org/10.1038/nature10872} {\bibfield  {journal} {\bibinfo
  {journal} {Nature}\ }\textbf {\bibinfo {volume} {483}},\ \bibinfo {pages}
  {187} (\bibinfo {year} {2012})}\BibitemShut {NoStop}%
\bibitem [{\citenamefont {B{\'e}rut}\ \emph {et~al.}(2015)\citenamefont
  {B{\'e}rut}, \citenamefont {Petrosyan},\ and\ \citenamefont
  {Ciliberto}}]{berut2015information}%
  \BibitemOpen
  \bibfield  {author} {\bibinfo {author} {\bibfnamefont {A.}~\bibnamefont
  {B{\'e}rut}}, \bibinfo {author} {\bibfnamefont {A.}~\bibnamefont
  {Petrosyan}},\ and\ \bibinfo {author} {\bibfnamefont {S.}~\bibnamefont
  {Ciliberto}},\ }\bibfield  {title} {\bibinfo {title} {Information and
  thermodynamics: experimental verification of landauer's erasure principle},\
  }\href {https://doi.org/10.1088/1742-5468/2015/06/P06015} {\bibfield
  {journal} {\bibinfo  {journal} {J. Stat. Mech.}\ }\textbf {\bibinfo {volume}
  {2015}},\ \bibinfo {pages} {P06015} (\bibinfo {year} {2015})}\BibitemShut
  {NoStop}%
\bibitem [{\citenamefont {Jun}\ \emph {et~al.}(2014)\citenamefont {Jun},
  \citenamefont {Gavrilov},\ and\ \citenamefont {Bechhoefer}}]{jun2014high}%
  \BibitemOpen
  \bibfield  {author} {\bibinfo {author} {\bibfnamefont {Y.}~\bibnamefont
  {Jun}}, \bibinfo {author} {\bibfnamefont {M.}~\bibnamefont {Gavrilov}},\ and\
  \bibinfo {author} {\bibfnamefont {J.}~\bibnamefont {Bechhoefer}},\ }\bibfield
   {title} {\bibinfo {title} {{High-precision test of Landauer’s principle in
  a feedback trap}},\ }\href {https://doi.org/10.1103/PhysRevLett.113.190601}
  {\bibfield  {journal} {\bibinfo  {journal} {Phys. Rev. Lett.}\ }\textbf
  {\bibinfo {volume} {113}},\ \bibinfo {pages} {190601} (\bibinfo {year}
  {2014})}\BibitemShut {NoStop}%
\bibitem [{\citenamefont {Gavrilov}(2017)}]{gavrilov2017erasure}%
  \BibitemOpen
  \bibfield  {author} {\bibinfo {author} {\bibfnamefont {M.}~\bibnamefont
  {Gavrilov}},\ }\href {https://doi.org/10.1007/978-3-319-63694-8} {\emph
  {\bibinfo {title} {Erasure without work in an asymmetric, double-well
  potential}}}\ (\bibinfo  {publisher} {Springer},\ \bibinfo {address}
  {Berlin},\ \bibinfo {year} {2017})\ p.~\bibinfo {pages} {83}\BibitemShut
  {NoStop}%
\bibitem [{\citenamefont {Hong}\ \emph {et~al.}(2016)\citenamefont {Hong},
  \citenamefont {Lambson}, \citenamefont {Dhuey},\ and\ \citenamefont
  {Bokor}}]{hong2016experimental}%
  \BibitemOpen
  \bibfield  {author} {\bibinfo {author} {\bibfnamefont {J.}~\bibnamefont
  {Hong}}, \bibinfo {author} {\bibfnamefont {B.}~\bibnamefont {Lambson}},
  \bibinfo {author} {\bibfnamefont {S.}~\bibnamefont {Dhuey}},\ and\ \bibinfo
  {author} {\bibfnamefont {J.}~\bibnamefont {Bokor}},\ }\bibfield  {title}
  {\bibinfo {title} {{Experimental test of Landauer’s principle in single-bit
  operations on nanomagnetic memory bits}},\ }\href
  {https://doi.org/10.1126/sciadv.1501492} {\bibfield  {journal} {\bibinfo
  {journal} {Science Adv.}\ }\textbf {\bibinfo {volume} {2}},\ \bibinfo {pages}
  {e1501492} (\bibinfo {year} {2016})}\BibitemShut {NoStop}%
\bibitem [{\citenamefont {Martini}\ \emph {et~al.}(2016)\citenamefont
  {Martini}, \citenamefont {Pancaldi}, \citenamefont {Madami}, \citenamefont
  {Vavassori}, \citenamefont {Gubbiotti}, \citenamefont {Tacchi}, \citenamefont
  {Hartmann}, \citenamefont {Emmerling}, \citenamefont {H{\"o}fling},
  \citenamefont {Worschech} \emph {et~al.}}]{martini2016experimental}%
  \BibitemOpen
  \bibfield  {author} {\bibinfo {author} {\bibfnamefont {L.}~\bibnamefont
  {Martini}}, \bibinfo {author} {\bibfnamefont {M.}~\bibnamefont {Pancaldi}},
  \bibinfo {author} {\bibfnamefont {M.}~\bibnamefont {Madami}}, \bibinfo
  {author} {\bibfnamefont {P.}~\bibnamefont {Vavassori}}, \bibinfo {author}
  {\bibfnamefont {G.}~\bibnamefont {Gubbiotti}}, \bibinfo {author}
  {\bibfnamefont {S.}~\bibnamefont {Tacchi}}, \bibinfo {author} {\bibfnamefont
  {F.}~\bibnamefont {Hartmann}}, \bibinfo {author} {\bibfnamefont
  {M.}~\bibnamefont {Emmerling}}, \bibinfo {author} {\bibfnamefont
  {S.}~\bibnamefont {H{\"o}fling}}, \bibinfo {author} {\bibfnamefont
  {L.}~\bibnamefont {Worschech}}, \emph {et~al.},\ }\bibfield  {title}
  {\bibinfo {title} {{Experimental and theoretical analysis of Landauer erasure
  in nano-magnetic switches of different sizes}},\ }\href
  {https://doi.org/10.1016/j.nanoen.2015.10.028} {\bibfield  {journal}
  {\bibinfo  {journal} {Nano Energy}\ }\textbf {\bibinfo {volume} {19}},\
  \bibinfo {pages} {108} (\bibinfo {year} {2016})}\BibitemShut {NoStop}%
\bibitem [{\citenamefont {Gaudenzi}\ \emph {et~al.}(2018)\citenamefont
  {Gaudenzi}, \citenamefont {Burzur{\'\i}}, \citenamefont {Maegawa},
  \citenamefont {Van Der~Zant},\ and\ \citenamefont
  {Luis}}]{gaudenzi2018quantum}%
  \BibitemOpen
  \bibfield  {author} {\bibinfo {author} {\bibfnamefont {R.}~\bibnamefont
  {Gaudenzi}}, \bibinfo {author} {\bibfnamefont {E.}~\bibnamefont
  {Burzur{\'\i}}}, \bibinfo {author} {\bibfnamefont {S.}~\bibnamefont
  {Maegawa}}, \bibinfo {author} {\bibfnamefont {H.}~\bibnamefont {Van
  Der~Zant}},\ and\ \bibinfo {author} {\bibfnamefont {F.}~\bibnamefont
  {Luis}},\ }\bibfield  {title} {\bibinfo {title} {{Quantum Landauer erasure
  with a molecular nanomagnet}},\ }\href
  {https://doi.org/10.1038/s41567-018-0070-7} {\bibfield  {journal} {\bibinfo
  {journal} {Nat. Phys.}\ }\textbf {\bibinfo {volume} {14}},\ \bibinfo {pages}
  {565} (\bibinfo {year} {2018})}\BibitemShut {NoStop}%
\bibitem [{\citenamefont {Saira}\ \emph {et~al.}(2020)\citenamefont {Saira},
  \citenamefont {Matheny}, \citenamefont {Katti}, \citenamefont {Fon},
  \citenamefont {Wimsatt}, \citenamefont {Crutchfield}, \citenamefont {Han},\
  and\ \citenamefont {Roukes}}]{saira2020nonequilibrium}%
  \BibitemOpen
  \bibfield  {author} {\bibinfo {author} {\bibfnamefont {O.-P.}\ \bibnamefont
  {Saira}}, \bibinfo {author} {\bibfnamefont {M.~H.}\ \bibnamefont {Matheny}},
  \bibinfo {author} {\bibfnamefont {R.}~\bibnamefont {Katti}}, \bibinfo
  {author} {\bibfnamefont {W.}~\bibnamefont {Fon}}, \bibinfo {author}
  {\bibfnamefont {G.}~\bibnamefont {Wimsatt}}, \bibinfo {author} {\bibfnamefont
  {J.~P.}\ \bibnamefont {Crutchfield}}, \bibinfo {author} {\bibfnamefont
  {S.}~\bibnamefont {Han}},\ and\ \bibinfo {author} {\bibfnamefont {M.~L.}\
  \bibnamefont {Roukes}},\ }\bibfield  {title} {\bibinfo {title}
  {Nonequilibrium thermodynamics of erasure with superconducting flux logic},\
  }\href {https://doi.org/10.1103/PhysRevResearch.2.013249} {\bibfield
  {journal} {\bibinfo  {journal} {Phys. Rev. Res.}\ }\textbf {\bibinfo {volume}
  {2}},\ \bibinfo {pages} {013249} (\bibinfo {year} {2020})}\BibitemShut
  {NoStop}%
\bibitem [{\citenamefont {Dago}\ \emph {et~al.}(2021)\citenamefont {Dago},
  \citenamefont {Pereda}, \citenamefont {Barros}, \citenamefont {Ciliberto},\
  and\ \citenamefont {Bellon}}]{dago2021information}%
  \BibitemOpen
  \bibfield  {author} {\bibinfo {author} {\bibfnamefont {S.}~\bibnamefont
  {Dago}}, \bibinfo {author} {\bibfnamefont {J.}~\bibnamefont {Pereda}},
  \bibinfo {author} {\bibfnamefont {N.}~\bibnamefont {Barros}}, \bibinfo
  {author} {\bibfnamefont {S.}~\bibnamefont {Ciliberto}},\ and\ \bibinfo
  {author} {\bibfnamefont {L.}~\bibnamefont {Bellon}},\ }\bibfield  {title}
  {\bibinfo {title} {{Information and thermodynamics: fast and precise approach
  to Landauer’s bound in an underdamped micromechanical oscillator}},\ }\href
  {https://doi.org/10.1103/PhysRevLett.126.170601} {\bibfield  {journal}
  {\bibinfo  {journal} {Phys. Rev. Lett.}\ }\textbf {\bibinfo {volume} {126}},\
  \bibinfo {pages} {170601} (\bibinfo {year} {2021})}\BibitemShut {NoStop}%
\bibitem [{\citenamefont {Dago}\ and\ \citenamefont
  {Bellon}(2022)}]{dago2022dynamics}%
  \BibitemOpen
  \bibfield  {author} {\bibinfo {author} {\bibfnamefont {S.}~\bibnamefont
  {Dago}}\ and\ \bibinfo {author} {\bibfnamefont {L.}~\bibnamefont {Bellon}},\
  }\bibfield  {title} {\bibinfo {title} {{Dynamics of information erasure and
  extension of Landauer’s bound to fast processes}},\ }\href
  {https://doi.org/10.1103/PhysRevLett.126.170601} {\bibfield  {journal}
  {\bibinfo  {journal} {Phys. Rev. Lett.}\ }\textbf {\bibinfo {volume} {128}},\
  \bibinfo {pages} {070604} (\bibinfo {year} {2022})}\BibitemShut {NoStop}%
\bibitem [{\citenamefont {Dubi}\ and\ \citenamefont
  {Di~Ventra}(2011)}]{dubi2011colloquium}%
  \BibitemOpen
  \bibfield  {author} {\bibinfo {author} {\bibfnamefont {Y.}~\bibnamefont
  {Dubi}}\ and\ \bibinfo {author} {\bibfnamefont {M.}~\bibnamefont
  {Di~Ventra}},\ }\bibfield  {title} {\bibinfo {title} {Colloquium: Heat flow
  and thermoelectricity in atomic and molecular junctions},\ }\href
  {https://doi.org/10.1103/RevModPhys.83.131} {\bibfield  {journal} {\bibinfo
  {journal} {Rev. Mod. Phys.}\ }\textbf {\bibinfo {volume} {83}},\ \bibinfo
  {pages} {131} (\bibinfo {year} {2011})}\BibitemShut {NoStop}%
\bibitem [{\citenamefont {Blickle}\ and\ \citenamefont
  {Bechinger}(2012)}]{blickle2012realization}%
  \BibitemOpen
  \bibfield  {author} {\bibinfo {author} {\bibfnamefont {V.}~\bibnamefont
  {Blickle}}\ and\ \bibinfo {author} {\bibfnamefont {C.}~\bibnamefont
  {Bechinger}},\ }\bibfield  {title} {\bibinfo {title} {Realization of a
  micrometre-sized stochastic heat engine},\ }\href
  {https://doi.org/10.1038/nphys2163} {\bibfield  {journal} {\bibinfo
  {journal} {Nat. Phys.}\ }\textbf {\bibinfo {volume} {8}},\ \bibinfo {pages}
  {143} (\bibinfo {year} {2012})}\BibitemShut {NoStop}%
\bibitem [{\citenamefont {Mart{\'\i}nez}\ \emph {et~al.}(2016)\citenamefont
  {Mart{\'\i}nez}, \citenamefont {Rold{\'a}n}, \citenamefont {Dinis},
  \citenamefont {Petrov}, \citenamefont {Parrondo},\ and\ \citenamefont
  {Rica}}]{martinez2016brownian}%
  \BibitemOpen
  \bibfield  {author} {\bibinfo {author} {\bibfnamefont {I.~A.}\ \bibnamefont
  {Mart{\'\i}nez}}, \bibinfo {author} {\bibfnamefont {{\'E}.}~\bibnamefont
  {Rold{\'a}n}}, \bibinfo {author} {\bibfnamefont {L.}~\bibnamefont {Dinis}},
  \bibinfo {author} {\bibfnamefont {D.}~\bibnamefont {Petrov}}, \bibinfo
  {author} {\bibfnamefont {J.~M.}\ \bibnamefont {Parrondo}},\ and\ \bibinfo
  {author} {\bibfnamefont {R.~A.}\ \bibnamefont {Rica}},\ }\bibfield  {title}
  {\bibinfo {title} {Brownian carnot engine},\ }\href
  {https://doi.org/10.1038/nphys3518} {\bibfield  {journal} {\bibinfo
  {journal} {Nat. Phys.}\ }\textbf {\bibinfo {volume} {12}},\ \bibinfo {pages}
  {67} (\bibinfo {year} {2016})}\BibitemShut {NoStop}%
\bibitem [{\citenamefont {Ro{\ss}nagel}\ \emph {et~al.}(2016)\citenamefont
  {Ro{\ss}nagel}, \citenamefont {Dawkins}, \citenamefont {Tolazzi},
  \citenamefont {Abah}, \citenamefont {Lutz}, \citenamefont {Schmidt-Kaler},\
  and\ \citenamefont {Singer}}]{rossnagel2016}%
  \BibitemOpen
  \bibfield  {author} {\bibinfo {author} {\bibfnamefont {J.}~\bibnamefont
  {Ro{\ss}nagel}}, \bibinfo {author} {\bibfnamefont {S.~T.}\ \bibnamefont
  {Dawkins}}, \bibinfo {author} {\bibfnamefont {K.~N.}\ \bibnamefont
  {Tolazzi}}, \bibinfo {author} {\bibfnamefont {O.}~\bibnamefont {Abah}},
  \bibinfo {author} {\bibfnamefont {E.}~\bibnamefont {Lutz}}, \bibinfo {author}
  {\bibfnamefont {F.}~\bibnamefont {Schmidt-Kaler}},\ and\ \bibinfo {author}
  {\bibfnamefont {K.}~\bibnamefont {Singer}},\ }\bibfield  {title} {\bibinfo
  {title} {A single-atom heat engine},\ }\href
  {https://doi.org/10.1126/science.aad6320} {\bibfield  {journal} {\bibinfo
  {journal} {Science}\ }\textbf {\bibinfo {volume} {352}},\ \bibinfo {pages}
  {325} (\bibinfo {year} {2016})}\BibitemShut {NoStop}%
\bibitem [{\citenamefont {Maslennikov}\ \emph {et~al.}(2019)\citenamefont
  {Maslennikov}, \citenamefont {Ding}, \citenamefont {Habl{\"a}tzel},
  \citenamefont {Gan}, \citenamefont {Roulet}, \citenamefont {Nimmrichter},
  \citenamefont {Dai}, \citenamefont {Scarani},\ and\ \citenamefont
  {Matsukevich}}]{maslennikov2019quantum}%
  \BibitemOpen
  \bibfield  {author} {\bibinfo {author} {\bibfnamefont {G.}~\bibnamefont
  {Maslennikov}}, \bibinfo {author} {\bibfnamefont {S.}~\bibnamefont {Ding}},
  \bibinfo {author} {\bibfnamefont {R.}~\bibnamefont {Habl{\"a}tzel}}, \bibinfo
  {author} {\bibfnamefont {J.}~\bibnamefont {Gan}}, \bibinfo {author}
  {\bibfnamefont {A.}~\bibnamefont {Roulet}}, \bibinfo {author} {\bibfnamefont
  {S.}~\bibnamefont {Nimmrichter}}, \bibinfo {author} {\bibfnamefont
  {J.}~\bibnamefont {Dai}}, \bibinfo {author} {\bibfnamefont {V.}~\bibnamefont
  {Scarani}},\ and\ \bibinfo {author} {\bibfnamefont {D.}~\bibnamefont
  {Matsukevich}},\ }\bibfield  {title} {\bibinfo {title} {Quantum absorption
  refrigerator with trapped ions},\ }\href
  {https://doi.org/10.1038/s41467-018-08090-0} {\bibfield  {journal} {\bibinfo
  {journal} {Nat. Commun.}\ }\textbf {\bibinfo {volume} {10}},\ \bibinfo
  {pages} {202} (\bibinfo {year} {2019})}\BibitemShut {NoStop}%
\bibitem [{\citenamefont {Peterson}\ \emph {et~al.}(2019)\citenamefont
  {Peterson}, \citenamefont {Batalh\~ao}, \citenamefont {Herrera},
  \citenamefont {Souza}, \citenamefont {Sarthour}, \citenamefont {Oliveira},\
  and\ \citenamefont {Serra}}]{peterson2019experimental}%
  \BibitemOpen
  \bibfield  {author} {\bibinfo {author} {\bibfnamefont {J.~P.~S.}\
  \bibnamefont {Peterson}}, \bibinfo {author} {\bibfnamefont {T.~B.}\
  \bibnamefont {Batalh\~ao}}, \bibinfo {author} {\bibfnamefont
  {M.}~\bibnamefont {Herrera}}, \bibinfo {author} {\bibfnamefont {A.~M.}\
  \bibnamefont {Souza}}, \bibinfo {author} {\bibfnamefont {R.~S.}\ \bibnamefont
  {Sarthour}}, \bibinfo {author} {\bibfnamefont {I.~S.}\ \bibnamefont
  {Oliveira}},\ and\ \bibinfo {author} {\bibfnamefont {R.~M.}\ \bibnamefont
  {Serra}},\ }\bibfield  {title} {\bibinfo {title} {Experimental
  characterization of a spin quantum heat engine},\ }\href
  {https://doi.org/10.1103/PhysRevLett.123.240601} {\bibfield  {journal}
  {\bibinfo  {journal} {Phys. Rev. Lett.}\ }\textbf {\bibinfo {volume} {123}},\
  \bibinfo {pages} {240601} (\bibinfo {year} {2019})}\BibitemShut {NoStop}%
\bibitem [{\citenamefont {von Lindenfels}\ \emph {et~al.}(2019)\citenamefont
  {von Lindenfels}, \citenamefont {Gr\"ab}, \citenamefont {Schmiegelow},
  \citenamefont {Kaushal}, \citenamefont {Schulz}, \citenamefont {Mitchison},
  \citenamefont {Goold}, \citenamefont {Schmidt-Kaler},\ and\ \citenamefont
  {Poschinger}}]{lindenfels2019}%
  \BibitemOpen
  \bibfield  {author} {\bibinfo {author} {\bibfnamefont {D.}~\bibnamefont {von
  Lindenfels}}, \bibinfo {author} {\bibfnamefont {O.}~\bibnamefont {Gr\"ab}},
  \bibinfo {author} {\bibfnamefont {C.~T.}\ \bibnamefont {Schmiegelow}},
  \bibinfo {author} {\bibfnamefont {V.}~\bibnamefont {Kaushal}}, \bibinfo
  {author} {\bibfnamefont {J.}~\bibnamefont {Schulz}}, \bibinfo {author}
  {\bibfnamefont {M.~T.}\ \bibnamefont {Mitchison}}, \bibinfo {author}
  {\bibfnamefont {J.}~\bibnamefont {Goold}}, \bibinfo {author} {\bibfnamefont
  {F.}~\bibnamefont {Schmidt-Kaler}},\ and\ \bibinfo {author} {\bibfnamefont
  {U.~G.}\ \bibnamefont {Poschinger}},\ }\bibfield  {title} {\bibinfo {title}
  {Spin heat engine coupled to a harmonic-oscillator flywheel},\ }\href
  {https://doi.org/10.1103/PhysRevLett.123.080602} {\bibfield  {journal}
  {\bibinfo  {journal} {Phys. Rev. Lett.}\ }\textbf {\bibinfo {volume} {123}},\
  \bibinfo {pages} {080602} (\bibinfo {year} {2019})}\BibitemShut {NoStop}%
\bibitem [{\citenamefont {Onishchenko}\ \emph {et~al.}(2022)\citenamefont
  {Onishchenko}, \citenamefont {Guarnieri}, \citenamefont {Rosillo-Rodes},
  \citenamefont {Pijn}, \citenamefont {Hilder}, \citenamefont {Poschinger},
  \citenamefont {Perarnau-Llobet}, \citenamefont {Eisert},\ and\ \citenamefont
  {Schmidt-Kaler}}]{onishchenko2022probing}%
  \BibitemOpen
  \bibfield  {author} {\bibinfo {author} {\bibfnamefont {O.}~\bibnamefont
  {Onishchenko}}, \bibinfo {author} {\bibfnamefont {G.}~\bibnamefont
  {Guarnieri}}, \bibinfo {author} {\bibfnamefont {P.}~\bibnamefont
  {Rosillo-Rodes}}, \bibinfo {author} {\bibfnamefont {D.}~\bibnamefont {Pijn}},
  \bibinfo {author} {\bibfnamefont {J.}~\bibnamefont {Hilder}}, \bibinfo
  {author} {\bibfnamefont {U.~G.}\ \bibnamefont {Poschinger}}, \bibinfo
  {author} {\bibfnamefont {M.}~\bibnamefont {Perarnau-Llobet}}, \bibinfo
  {author} {\bibfnamefont {J.}~\bibnamefont {Eisert}},\ and\ \bibinfo {author}
  {\bibfnamefont {F.}~\bibnamefont {Schmidt-Kaler}},\ }\href@noop {} {\bibinfo
  {title} {Probing coherent quantum thermodynamics using a trapped ion}}
  (\bibinfo {year} {2022}),\ \Eprint {https://arxiv.org/abs/2207.14325}
  {arXiv:2207.14325} \BibitemShut {NoStop}%
\bibitem [{\citenamefont {Erdman}\ \emph
  {et~al.}(2019{\natexlab{a}})\citenamefont {Erdman}, \citenamefont {Peltonen},
  \citenamefont {Bhandari}, \citenamefont {Dutta}, \citenamefont {Courtois},
  \citenamefont {Fazio}, \citenamefont {Taddei},\ and\ \citenamefont
  {Pekola}}]{erdman2019_prb}%
  \BibitemOpen
  \bibfield  {author} {\bibinfo {author} {\bibfnamefont {P.~A.}\ \bibnamefont
  {Erdman}}, \bibinfo {author} {\bibfnamefont {J.~T.}\ \bibnamefont
  {Peltonen}}, \bibinfo {author} {\bibfnamefont {B.}~\bibnamefont {Bhandari}},
  \bibinfo {author} {\bibfnamefont {B.}~\bibnamefont {Dutta}}, \bibinfo
  {author} {\bibfnamefont {H.}~\bibnamefont {Courtois}}, \bibinfo {author}
  {\bibfnamefont {R.}~\bibnamefont {Fazio}}, \bibinfo {author} {\bibfnamefont
  {F.}~\bibnamefont {Taddei}},\ and\ \bibinfo {author} {\bibfnamefont {J.~P.}\
  \bibnamefont {Pekola}},\ }\bibfield  {title} {\bibinfo {title} {Nonlinear
  thermovoltage in a single-electron transistor},\ }\href
  {https://doi.org/10.1103/PhysRevB.104.075442} {\bibfield  {journal} {\bibinfo
   {journal} {Phys. Rev. B}\ }\textbf {\bibinfo {volume} {99}},\ \bibinfo
  {pages} {165405} (\bibinfo {year} {2019}{\natexlab{a}})}\BibitemShut
  {NoStop}%
\bibitem [{\citenamefont {Bhandari}\ and\ \citenamefont
  {Jordan}(2021)}]{bhandari2021minimal}%
  \BibitemOpen
  \bibfield  {author} {\bibinfo {author} {\bibfnamefont {B.}~\bibnamefont
  {Bhandari}}\ and\ \bibinfo {author} {\bibfnamefont {A.~N.}\ \bibnamefont
  {Jordan}},\ }\bibfield  {title} {\bibinfo {title} {Minimal two-body quantum
  absorption refrigerator},\ }\href
  {https://doi.org/10.1103/PhysRevB.104.075442} {\bibfield  {journal} {\bibinfo
   {journal} {Phys. Rev. B}\ }\textbf {\bibinfo {volume} {104}},\ \bibinfo
  {pages} {075442} (\bibinfo {year} {2021})}\BibitemShut {NoStop}%
\bibitem [{\citenamefont {Elouard}\ \emph {et~al.}(2017)\citenamefont
  {Elouard}, \citenamefont {Herrera-Mart\'{\i}}, \citenamefont {Huard},\ and\
  \citenamefont {Auff\`eves}}]{elouard2017extracting}%
  \BibitemOpen
  \bibfield  {author} {\bibinfo {author} {\bibfnamefont {C.}~\bibnamefont
  {Elouard}}, \bibinfo {author} {\bibfnamefont {D.}~\bibnamefont
  {Herrera-Mart\'{\i}}}, \bibinfo {author} {\bibfnamefont {B.}~\bibnamefont
  {Huard}},\ and\ \bibinfo {author} {\bibfnamefont {A.}~\bibnamefont
  {Auff\`eves}},\ }\bibfield  {title} {\bibinfo {title} {Extracting work from
  quantum measurement in maxwell's demon engines},\ }\href
  {https://doi.org/10.1103/PhysRevLett.118.260603} {\bibfield  {journal}
  {\bibinfo  {journal} {Phys. Rev. Lett.}\ }\textbf {\bibinfo {volume} {118}},\
  \bibinfo {pages} {260603} (\bibinfo {year} {2017})}\BibitemShut {NoStop}%
\bibitem [{\citenamefont {Ding}\ \emph {et~al.}(2018)\citenamefont {Ding},
  \citenamefont {Yi}, \citenamefont {Kim},\ and\ \citenamefont
  {Talkner}}]{ding2018measurement}%
  \BibitemOpen
  \bibfield  {author} {\bibinfo {author} {\bibfnamefont {X.}~\bibnamefont
  {Ding}}, \bibinfo {author} {\bibfnamefont {J.}~\bibnamefont {Yi}}, \bibinfo
  {author} {\bibfnamefont {Y.~W.}\ \bibnamefont {Kim}},\ and\ \bibinfo {author}
  {\bibfnamefont {P.}~\bibnamefont {Talkner}},\ }\bibfield  {title} {\bibinfo
  {title} {Measurement-driven single temperature engine},\ }\href
  {https://doi.org/10.1103/PhysRevE.98.042122} {\bibfield  {journal} {\bibinfo
  {journal} {Phys. Rev. E}\ }\textbf {\bibinfo {volume} {98}},\ \bibinfo
  {pages} {042122} (\bibinfo {year} {2018})}\BibitemShut {NoStop}%
\bibitem [{\citenamefont {Elouard}\ and\ \citenamefont
  {Jordan}(2018)}]{elouard2018efficient}%
  \BibitemOpen
  \bibfield  {author} {\bibinfo {author} {\bibfnamefont {C.}~\bibnamefont
  {Elouard}}\ and\ \bibinfo {author} {\bibfnamefont {A.~N.}\ \bibnamefont
  {Jordan}},\ }\bibfield  {title} {\bibinfo {title} {Efficient quantum
  measurement engines},\ }\href
  {https://doi.org/10.1103/PhysRevLett.120.260601} {\bibfield  {journal}
  {\bibinfo  {journal} {Phys. Rev. Lett.}\ }\textbf {\bibinfo {volume} {120}},\
  \bibinfo {pages} {260601} (\bibinfo {year} {2018})}\BibitemShut {NoStop}%
\bibitem [{\citenamefont {Bresque}\ \emph {et~al.}(2021)\citenamefont
  {Bresque}, \citenamefont {Camati}, \citenamefont {Rogers}, \citenamefont
  {Murch}, \citenamefont {Jordan},\ and\ \citenamefont
  {Auff{\`e}ves}}]{bresque2021two}%
  \BibitemOpen
  \bibfield  {author} {\bibinfo {author} {\bibfnamefont {L.}~\bibnamefont
  {Bresque}}, \bibinfo {author} {\bibfnamefont {P.~A.}\ \bibnamefont {Camati}},
  \bibinfo {author} {\bibfnamefont {S.}~\bibnamefont {Rogers}}, \bibinfo
  {author} {\bibfnamefont {K.}~\bibnamefont {Murch}}, \bibinfo {author}
  {\bibfnamefont {A.~N.}\ \bibnamefont {Jordan}},\ and\ \bibinfo {author}
  {\bibfnamefont {A.}~\bibnamefont {Auff{\`e}ves}},\ }\bibfield  {title}
  {\bibinfo {title} {Two-qubit engine fueled by entanglement and local
  measurements},\ }\href {https://doi.org/10.1103/PhysRevLett.126.120605}
  {\bibfield  {journal} {\bibinfo  {journal} {Phys. Rev. Lett.}\ }\textbf
  {\bibinfo {volume} {126}},\ \bibinfo {pages} {120605} (\bibinfo {year}
  {2021})}\BibitemShut {NoStop}%
\bibitem [{\citenamefont {Buffoni}\ \emph {et~al.}(2019)\citenamefont
  {Buffoni}, \citenamefont {Solfanelli}, \citenamefont {Verrucchi},
  \citenamefont {Cuccoli},\ and\ \citenamefont {Campisi}}]{buffoni2019quantum}%
  \BibitemOpen
  \bibfield  {author} {\bibinfo {author} {\bibfnamefont {L.}~\bibnamefont
  {Buffoni}}, \bibinfo {author} {\bibfnamefont {A.}~\bibnamefont {Solfanelli}},
  \bibinfo {author} {\bibfnamefont {P.}~\bibnamefont {Verrucchi}}, \bibinfo
  {author} {\bibfnamefont {A.}~\bibnamefont {Cuccoli}},\ and\ \bibinfo {author}
  {\bibfnamefont {M.}~\bibnamefont {Campisi}},\ }\bibfield  {title} {\bibinfo
  {title} {Quantum measurement cooling},\ }\href
  {https://doi.org/10.1103/PhysRevLett.122.070603} {\bibfield  {journal}
  {\bibinfo  {journal} {Phys. Rev. Lett.}\ }\textbf {\bibinfo {volume} {122}},\
  \bibinfo {pages} {070603} (\bibinfo {year} {2019})}\BibitemShut {NoStop}%
\bibitem [{\citenamefont {Landi}\ \emph {et~al.}(2022)\citenamefont {Landi},
  \citenamefont {Paternostro},\ and\ \citenamefont
  {Belenchia}}]{landi2022informational}%
  \BibitemOpen
  \bibfield  {author} {\bibinfo {author} {\bibfnamefont {G.~T.}\ \bibnamefont
  {Landi}}, \bibinfo {author} {\bibfnamefont {M.}~\bibnamefont {Paternostro}},\
  and\ \bibinfo {author} {\bibfnamefont {A.}~\bibnamefont {Belenchia}},\
  }\bibfield  {title} {\bibinfo {title} {Informational steady states and
  conditional entropy production in continuously monitored systems},\ }\href
  {https://doi.org/10.1103/PRXQuantum.3.010303} {\bibfield  {journal} {\bibinfo
   {journal} {PRX Quantum}\ }\textbf {\bibinfo {volume} {3}},\ \bibinfo {pages}
  {010303} (\bibinfo {year} {2022})}\BibitemShut {NoStop}%
\bibitem [{\citenamefont {Belenchia}\ \emph {et~al.}(2022)\citenamefont
  {Belenchia}, \citenamefont {Paternostro},\ and\ \citenamefont
  {Landi}}]{belenchia2022informational}%
  \BibitemOpen
  \bibfield  {author} {\bibinfo {author} {\bibfnamefont {A.}~\bibnamefont
  {Belenchia}}, \bibinfo {author} {\bibfnamefont {M.}~\bibnamefont
  {Paternostro}},\ and\ \bibinfo {author} {\bibfnamefont {G.~T.}\ \bibnamefont
  {Landi}},\ }\bibfield  {title} {\bibinfo {title} {{Informational steady
  states and conditional entropy production in continuously monitored systems:
  The case of Gaussian systems}},\ }\href
  {https://doi.org/10.1103/PhysRevA.105.022213} {\bibfield  {journal} {\bibinfo
   {journal} {Phys. Rev. A}\ }\textbf {\bibinfo {volume} {105}},\ \bibinfo
  {pages} {022213} (\bibinfo {year} {2022})}\BibitemShut {NoStop}%
\bibitem [{\citenamefont {Gluza}\ \emph {et~al.}(2021)\citenamefont {Gluza},
  \citenamefont {Sabino}, \citenamefont {Ng}, \citenamefont {Vitagliano},
  \citenamefont {Pezzutto}, \citenamefont {Omar}, \citenamefont {Mazets},
  \citenamefont {Huber}, \citenamefont {Schmiedmayer},\ and\ \citenamefont
  {Eisert}}]{PRXQuantum.2.030310}%
  \BibitemOpen
  \bibfield  {author} {\bibinfo {author} {\bibfnamefont {M.}~\bibnamefont
  {Gluza}}, \bibinfo {author} {\bibfnamefont {J.}~\bibnamefont {Sabino}},
  \bibinfo {author} {\bibfnamefont {N.~H.}\ \bibnamefont {Ng}}, \bibinfo
  {author} {\bibfnamefont {G.}~\bibnamefont {Vitagliano}}, \bibinfo {author}
  {\bibfnamefont {M.}~\bibnamefont {Pezzutto}}, \bibinfo {author}
  {\bibfnamefont {Y.}~\bibnamefont {Omar}}, \bibinfo {author} {\bibfnamefont
  {I.}~\bibnamefont {Mazets}}, \bibinfo {author} {\bibfnamefont
  {M.}~\bibnamefont {Huber}}, \bibinfo {author} {\bibfnamefont
  {J.}~\bibnamefont {Schmiedmayer}},\ and\ \bibinfo {author} {\bibfnamefont
  {J.}~\bibnamefont {Eisert}},\ }\bibfield  {title} {\bibinfo {title} {Quantum
  field thermal machines},\ }\href
  {https://doi.org/10.1103/PRXQuantum.2.030310} {\bibfield  {journal} {\bibinfo
   {journal} {PRX Quantum}\ }\textbf {\bibinfo {volume} {2}},\ \bibinfo {pages}
  {030310} (\bibinfo {year} {2021})}\BibitemShut {NoStop}%
\bibitem [{\citenamefont {Barato}\ and\ \citenamefont
  {Seifert}(2015)}]{barato2015}%
  \BibitemOpen
  \bibfield  {author} {\bibinfo {author} {\bibfnamefont {A.~C.}\ \bibnamefont
  {Barato}}\ and\ \bibinfo {author} {\bibfnamefont {U.}~\bibnamefont
  {Seifert}},\ }\bibfield  {title} {\bibinfo {title} {Thermodynamic uncertainty
  relation for biomolecular processes},\ }\href
  {https://doi.org/10.1103/PhysRevLett.114.158101} {\bibfield  {journal}
  {\bibinfo  {journal} {Phys. Rev. Lett.}\ }\textbf {\bibinfo {volume} {114}},\
  \bibinfo {pages} {158101} (\bibinfo {year} {2015})}\BibitemShut {NoStop}%
\bibitem [{\citenamefont {Gingrich}\ \emph {et~al.}(2016)\citenamefont
  {Gingrich}, \citenamefont {Horowitz}, \citenamefont {Perunov},\ and\
  \citenamefont {England}}]{gingrich2016dissipation}%
  \BibitemOpen
  \bibfield  {author} {\bibinfo {author} {\bibfnamefont {T.~R.}\ \bibnamefont
  {Gingrich}}, \bibinfo {author} {\bibfnamefont {J.~M.}\ \bibnamefont
  {Horowitz}}, \bibinfo {author} {\bibfnamefont {N.}~\bibnamefont {Perunov}},\
  and\ \bibinfo {author} {\bibfnamefont {J.~L.}\ \bibnamefont {England}},\
  }\bibfield  {title} {\bibinfo {title} {Dissipation bounds all steady-state
  current fluctuations},\ }\href
  {https://doi.org/10.1103/PhysRevLett.116.120601} {\bibfield  {journal}
  {\bibinfo  {journal} {Phys. Rev. Lett.}\ }\textbf {\bibinfo {volume} {116}},\
  \bibinfo {pages} {120601} (\bibinfo {year} {2016})}\BibitemShut {NoStop}%
\bibitem [{\citenamefont {Aamir}\ \emph {et~al.}(2023)\citenamefont {Aamir},
  \citenamefont {Suria}, \citenamefont {Guzm{\'a}n}, \citenamefont
  {Castillo-Moreno}, \citenamefont {Epstein}, \citenamefont {Halpern},\ and\
  \citenamefont {Gasparinetti}}]{aamir2023}%
  \BibitemOpen
  \bibfield  {author} {\bibinfo {author} {\bibfnamefont {M.~A.}\ \bibnamefont
  {Aamir}}, \bibinfo {author} {\bibfnamefont {P.~J.}\ \bibnamefont {Suria}},
  \bibinfo {author} {\bibfnamefont {J.~A.~M.}\ \bibnamefont {Guzm{\'a}n}},
  \bibinfo {author} {\bibfnamefont {C.}~\bibnamefont {Castillo-Moreno}},
  \bibinfo {author} {\bibfnamefont {J.~M.}\ \bibnamefont {Epstein}}, \bibinfo
  {author} {\bibfnamefont {N.~Y.}\ \bibnamefont {Halpern}},\ and\ \bibinfo
  {author} {\bibfnamefont {S.}~\bibnamefont {Gasparinetti}},\ }\bibfield
  {title} {\bibinfo {title} {Thermally driven quantum refrigerator autonomously
  resets superconducting qubit},\ }\Eprint
  {https://arxiv.org/abs/arXiv:2305.16710} {arXiv:2305.16710}  (\bibinfo {year}
  {2023})\BibitemShut {NoStop}%
\bibitem [{\citenamefont {Pietzonka}\ \emph {et~al.}(2017)\citenamefont
  {Pietzonka}, \citenamefont {Ritort},\ and\ \citenamefont
  {Seifert}}]{Pietzonka2017a}%
  \BibitemOpen
  \bibfield  {author} {\bibinfo {author} {\bibfnamefont {P.}~\bibnamefont
  {Pietzonka}}, \bibinfo {author} {\bibfnamefont {F.}~\bibnamefont {Ritort}},\
  and\ \bibinfo {author} {\bibfnamefont {U.}~\bibnamefont {Seifert}},\
  }\bibfield  {title} {\bibinfo {title} {{Finite-time generalization of the
  thermodynamic uncertainty relation}},\ }\href
  {https://doi.org/10.1103/PhysRevE.96.012101} {\bibfield  {journal} {\bibinfo
  {journal} {Phys. Rev. E}\ }\textbf {\bibinfo {volume} {96}},\ \bibinfo
  {pages} {012101} (\bibinfo {year} {2017})}\BibitemShut {NoStop}%
\bibitem [{\citenamefont {Dechant}(2018)}]{Dechant2018}%
  \BibitemOpen
  \bibfield  {author} {\bibinfo {author} {\bibfnamefont {A.}~\bibnamefont
  {Dechant}},\ }\bibfield  {title} {\bibinfo {title} {{Multidimensional
  thermodynamic uncertainty relations}},\ }\href
  {https://doi.org/10.1088/1751-8121/aaf3ff} {\bibfield  {journal} {\bibinfo
  {journal} {J. Phys. A}\ }\textbf {\bibinfo {volume} {52}},\ \bibinfo {pages}
  {035001} (\bibinfo {year} {2018})}\BibitemShut {NoStop}%
\bibitem [{\citenamefont {Horowitz}\ and\ \citenamefont
  {Gingrich}(2017)}]{horowitz2017proof}%
  \BibitemOpen
  \bibfield  {author} {\bibinfo {author} {\bibfnamefont {J.}~\bibnamefont
  {Horowitz}}\ and\ \bibinfo {author} {\bibfnamefont {T.}~\bibnamefont
  {Gingrich}},\ }\bibfield  {title} {\bibinfo {title} {Proof of the finite-time
  thermodynamic uncertainty relation for steady-state currents},\ }\href
  {https://doi.org/10.1103/PhysRevE.96.020103} {\bibfield  {journal} {\bibinfo
  {journal} {Phys. Rev. E}\ }\textbf {\bibinfo {volume} {96}},\ \bibinfo
  {pages} {020103} (\bibinfo {year} {2017})}\BibitemShut {NoStop}%
\bibitem [{\citenamefont {Barato}\ \emph {et~al.}(2018)\citenamefont {Barato},
  \citenamefont {Chetrite}, \citenamefont {Faggionato},\ and\ \citenamefont
  {Gabrielli}}]{BaratoNJP2018}%
  \BibitemOpen
  \bibfield  {author} {\bibinfo {author} {\bibfnamefont {A.}~\bibnamefont
  {Barato}}, \bibinfo {author} {\bibfnamefont {R.}~\bibnamefont {Chetrite}},
  \bibinfo {author} {\bibfnamefont {A.}~\bibnamefont {Faggionato}},\ and\
  \bibinfo {author} {\bibfnamefont {D.}~\bibnamefont {Gabrielli}},\ }\bibfield
  {title} {\bibinfo {title} {Bounds on current fluctuations in periodically
  driven systems},\ }\href {https://doi.org/10.1088/1367-2630/aae512}
  {\bibfield  {journal} {\bibinfo  {journal} {New J. Phys.}\ }\textbf {\bibinfo
  {volume} {20}},\ \bibinfo {pages} {103023} (\bibinfo {year}
  {2018})}\BibitemShut {NoStop}%
\bibitem [{\citenamefont {Holubec}\ and\ \citenamefont
  {Ryabov}(2018)}]{Holubec2018PRL}%
  \BibitemOpen
  \bibfield  {author} {\bibinfo {author} {\bibfnamefont {V.}~\bibnamefont
  {Holubec}}\ and\ \bibinfo {author} {\bibfnamefont {A.}~\bibnamefont
  {Ryabov}},\ }\bibfield  {title} {\bibinfo {title} {Cycling tames power
  fluctuations near optimum efficiency},\ }\href
  {https://doi.org/10.1103/PhysRevLett.121.120601} {\bibfield  {journal}
  {\bibinfo  {journal} {Phys. Rev. Lett.}\ }\textbf {\bibinfo {volume} {121}},\
  \bibinfo {pages} {120601} (\bibinfo {year} {2018})}\BibitemShut {NoStop}%
\bibitem [{\citenamefont {Proesmans}\ and\ \citenamefont {Van~den
  Broeck}(2017)}]{proesmans2017discrete}%
  \BibitemOpen
  \bibfield  {author} {\bibinfo {author} {\bibfnamefont {K.}~\bibnamefont
  {Proesmans}}\ and\ \bibinfo {author} {\bibfnamefont {C.}~\bibnamefont
  {Van~den Broeck}},\ }\bibfield  {title} {\bibinfo {title} {Discrete-time
  thermodynamic uncertainty relation},\ }\href
  {https://doi.org/10.1209/0295-5075/119/20001} {\bibfield  {journal} {\bibinfo
   {journal} {Europhys. Lett.}\ }\textbf {\bibinfo {volume} {119}},\ \bibinfo
  {pages} {20001} (\bibinfo {year} {2017})}\BibitemShut {NoStop}%
\bibitem [{\citenamefont {Van~Vu}\ and\ \citenamefont
  {Hasegawa}(2020)}]{VanVu2020}%
  \BibitemOpen
  \bibfield  {author} {\bibinfo {author} {\bibfnamefont {T.}~\bibnamefont
  {Van~Vu}}\ and\ \bibinfo {author} {\bibfnamefont {Y.}~\bibnamefont
  {Hasegawa}},\ }\bibfield  {title} {\bibinfo {title} {Thermodynamic
  uncertainty relations under arbitrary control protocols},\ }\href
  {https://doi.org/10.1103/PhysRevResearch.2.013060} {\bibfield  {journal}
  {\bibinfo  {journal} {Phys. Rev. Res.}\ }\textbf {\bibinfo {volume} {2}},\
  \bibinfo {pages} {013060} (\bibinfo {year} {2020})}\BibitemShut {NoStop}%
\bibitem [{\citenamefont {Miller}\ \emph
  {et~al.}(2021{\natexlab{a}})\citenamefont {Miller}, \citenamefont
  {Mohammady}, \citenamefont {Perarnau-Llobet},\ and\ \citenamefont
  {Guarnieri}}]{Guarnieri2021PRL}%
  \BibitemOpen
  \bibfield  {author} {\bibinfo {author} {\bibfnamefont {H.~J.~D.}\
  \bibnamefont {Miller}}, \bibinfo {author} {\bibfnamefont {M.~H.}\
  \bibnamefont {Mohammady}}, \bibinfo {author} {\bibfnamefont {M.}~\bibnamefont
  {Perarnau-Llobet}},\ and\ \bibinfo {author} {\bibfnamefont {G.}~\bibnamefont
  {Guarnieri}},\ }\bibfield  {title} {\bibinfo {title} {Thermodynamic
  uncertainty relation in slowly driven quantum heat engines},\ }\href
  {https://doi.org/10.1103/PhysRevE.103.052138} {\bibfield  {journal} {\bibinfo
   {journal} {Phys. Rev. Lett.}\ }\textbf {\bibinfo {volume} {126}},\ \bibinfo
  {pages} {210603} (\bibinfo {year} {2021}{\natexlab{a}})}\BibitemShut
  {NoStop}%
\bibitem [{\citenamefont {Miller}\ \emph
  {et~al.}(2021{\natexlab{b}})\citenamefont {Miller}, \citenamefont
  {Mohammady}, \citenamefont {Perarnau-Llobet},\ and\ \citenamefont
  {Guarnieri}}]{Guarnieri2021PRE}%
  \BibitemOpen
  \bibfield  {author} {\bibinfo {author} {\bibfnamefont {H.}~\bibnamefont
  {Miller}}, \bibinfo {author} {\bibfnamefont {M.}~\bibnamefont {Mohammady}},
  \bibinfo {author} {\bibfnamefont {M.}~\bibnamefont {Perarnau-Llobet}},\ and\
  \bibinfo {author} {\bibfnamefont {G.}~\bibnamefont {Guarnieri}},\ }\bibfield
  {title} {\bibinfo {title} {Joint statistics of work and entropy production
  along quantum trajectories},\ }\href
  {https://doi.org/10.1103/PhysRevE.103.052138} {\bibfield  {journal} {\bibinfo
   {journal} {Phys. Rev. E}\ }\textbf {\bibinfo {volume} {103}},\ \bibinfo
  {pages} {052138} (\bibinfo {year} {2021}{\natexlab{b}})}\BibitemShut
  {NoStop}%
\bibitem [{\citenamefont {Falasco}\ \emph {et~al.}(2020)\citenamefont
  {Falasco}, \citenamefont {Esposito},\ and\ \citenamefont
  {Delvenne}}]{falasco2020unifying}%
  \BibitemOpen
  \bibfield  {author} {\bibinfo {author} {\bibfnamefont {G.}~\bibnamefont
  {Falasco}}, \bibinfo {author} {\bibfnamefont {M.}~\bibnamefont {Esposito}},\
  and\ \bibinfo {author} {\bibfnamefont {J.-C.}\ \bibnamefont {Delvenne}},\
  }\bibfield  {title} {\bibinfo {title} {Unifying thermodynamic uncertainty
  relations},\ }\href {https://doi.org/10.1088/1367-2630/ab8679} {\bibfield
  {journal} {\bibinfo  {journal} {New J. Phys.}\ }\textbf {\bibinfo {volume}
  {22}},\ \bibinfo {pages} {053046} (\bibinfo {year} {2020})}\BibitemShut
  {NoStop}%
\bibitem [{\citenamefont {Hasegawa}(2023)}]{Hasegawa2023}%
  \BibitemOpen
  \bibfield  {author} {\bibinfo {author} {\bibfnamefont {Y.}~\bibnamefont
  {Hasegawa}},\ }\bibfield  {title} {\bibinfo {title} {Unifying speed limit,
  thermodynamic uncertainty relation and {H}eisenberg principle via
  bulk-boundary correspondence},\ }\href
  {https://doi.org/10.1038/s41467-023-38074-8} {\bibfield  {journal} {\bibinfo
  {journal} {Nat. Commun.}\ }\textbf {\bibinfo {volume} {14}},\ \bibinfo
  {pages} {2828} (\bibinfo {year} {2023})}\BibitemShut {NoStop}%
\bibitem [{\citenamefont {Vu}\ and\ \citenamefont {Saito}(2023)}]{VuSaito2023}%
  \BibitemOpen
  \bibfield  {author} {\bibinfo {author} {\bibfnamefont {T.~V.}\ \bibnamefont
  {Vu}}\ and\ \bibinfo {author} {\bibfnamefont {K.}~\bibnamefont {Saito}},\
  }\bibfield  {title} {\bibinfo {title} {Thermodynamic unification of optimal
  transport: Thermodynamic uncertainty relation, minimum dissipation, and
  thermodynamic speed limits},\ }\href
  {https://doi.org/10.1103/physrevx.13.011013} {\bibfield  {journal} {\bibinfo
  {journal} {Phys. Rev. X}\ }\textbf {\bibinfo {volume} {13}},\ \bibinfo
  {pages} {011013} (\bibinfo {year} {2023})}\BibitemShut {NoStop}%
\bibitem [{\citenamefont {Dechant}\ and\ \citenamefont
  {Sasa}(2020)}]{dechant2020fluctuation}%
  \BibitemOpen
  \bibfield  {author} {\bibinfo {author} {\bibfnamefont {A.}~\bibnamefont
  {Dechant}}\ and\ \bibinfo {author} {\bibfnamefont {S.}~\bibnamefont {Sasa}},\
  }\bibfield  {title} {\bibinfo {title} {Fluctuation--response inequality out
  of equilibrium},\ }\href {https://doi.org/10.1073/pnas.191838611} {\bibfield
  {journal} {\bibinfo  {journal} {Proc. Natl. Acad. Sci. U.S.A.}\ }\textbf
  {\bibinfo {volume} {117}},\ \bibinfo {pages} {6430} (\bibinfo {year}
  {2020})}\BibitemShut {NoStop}%
\bibitem [{\citenamefont {MacIeszczak}\ \emph {et~al.}(2018)\citenamefont
  {MacIeszczak}, \citenamefont {Brandner},\ and\ \citenamefont
  {Garrahan}}]{MacIeszczak2018}%
  \BibitemOpen
  \bibfield  {author} {\bibinfo {author} {\bibfnamefont {K.}~\bibnamefont
  {MacIeszczak}}, \bibinfo {author} {\bibfnamefont {K.}~\bibnamefont
  {Brandner}},\ and\ \bibinfo {author} {\bibfnamefont {J.}~\bibnamefont
  {Garrahan}},\ }\bibfield  {title} {\bibinfo {title} {Unified thermodynamic
  uncertainty relations in linear response},\ }\href
  {https://doi.org/10.1103/PhysRevLett.121.130601} {\bibfield  {journal}
  {\bibinfo  {journal} {Phys. Rev. Lett.}\ }\textbf {\bibinfo {volume} {121}},\
  \bibinfo {pages} {130601} (\bibinfo {year} {2018})}\BibitemShut {NoStop}%
\bibitem [{\citenamefont {Guarnieri}\ \emph {et~al.}(2019)\citenamefont
  {Guarnieri}, \citenamefont {Landi}, \citenamefont {Clark},\ and\
  \citenamefont {Goold}}]{Guar19c}%
  \BibitemOpen
  \bibfield  {author} {\bibinfo {author} {\bibfnamefont {G.}~\bibnamefont
  {Guarnieri}}, \bibinfo {author} {\bibfnamefont {G.}~\bibnamefont {Landi}},
  \bibinfo {author} {\bibfnamefont {S.}~\bibnamefont {Clark}},\ and\ \bibinfo
  {author} {\bibfnamefont {J.}~\bibnamefont {Goold}},\ }\bibfield  {title}
  {\bibinfo {title} {Thermodynamics of precision in quantum nonequilibrium
  steady states},\ }\href {https://doi.org/10.1103/PhysRevResearch.1.033021}
  {\bibfield  {journal} {\bibinfo  {journal} {Phys. Rev. Res.}\ }\textbf
  {\bibinfo {volume} {1}},\ \bibinfo {pages} {033021} (\bibinfo {year}
  {2019})}\BibitemShut {NoStop}%
\bibitem [{\citenamefont {Agarwalla}\ and\ \citenamefont
  {Segal}(2018)}]{SegalAgarwalla}%
  \BibitemOpen
  \bibfield  {author} {\bibinfo {author} {\bibfnamefont {B.}~\bibnamefont
  {Agarwalla}}\ and\ \bibinfo {author} {\bibfnamefont {D.}~\bibnamefont
  {Segal}},\ }\bibfield  {title} {\bibinfo {title} {Assessing the validity of
  the thermodynamic uncertainty relation in quantum systems},\ }\href
  {https://doi.org/10.1103/PhysRevB.98.155438} {\bibfield  {journal} {\bibinfo
  {journal} {Phys. Rev. B}\ }\textbf {\bibinfo {volume} {98}},\ \bibinfo
  {pages} {155438} (\bibinfo {year} {2018})}\BibitemShut {NoStop}%
\bibitem [{\citenamefont {Saryal}\ \emph {et~al.}(2021)\citenamefont {Saryal},
  \citenamefont {Sadekar},\ and\ \citenamefont {Agarwalla}}]{GerryAgarwalla}%
  \BibitemOpen
  \bibfield  {author} {\bibinfo {author} {\bibfnamefont {S.}~\bibnamefont
  {Saryal}}, \bibinfo {author} {\bibfnamefont {O.}~\bibnamefont {Sadekar}},\
  and\ \bibinfo {author} {\bibfnamefont {B.}~\bibnamefont {Agarwalla}},\
  }\bibfield  {title} {\bibinfo {title} {Thermodynamic uncertainty relation for
  energy transport in a transient regime: A model study},\ }\href
  {https://doi.org/10.1103/PhysRevE.103.022141} {\bibfield  {journal} {\bibinfo
   {journal} {Phys. Rev. E}\ }\textbf {\bibinfo {volume} {103}},\ \bibinfo
  {pages} {022141} (\bibinfo {year} {2021})}\BibitemShut {NoStop}%
\bibitem [{\citenamefont {Miller}\ \emph {et~al.}(2020)\citenamefont {Miller},
  \citenamefont {Guarnieri}, \citenamefont {Mitchison},\ and\ \citenamefont
  {Goold}}]{Miller2020Landauer}%
  \BibitemOpen
  \bibfield  {author} {\bibinfo {author} {\bibfnamefont {H.}~\bibnamefont
  {Miller}}, \bibinfo {author} {\bibfnamefont {G.}~\bibnamefont {Guarnieri}},
  \bibinfo {author} {\bibfnamefont {M.}~\bibnamefont {Mitchison}},\ and\
  \bibinfo {author} {\bibfnamefont {J.}~\bibnamefont {Goold}},\ }\bibfield
  {title} {\bibinfo {title} {Quantum fluctuations hinder finite-time
  information erasure near the {L}andauer limit},\ }\href
  {https://doi.org/10.1103/physrevlett.125.160602} {\bibfield  {journal}
  {\bibinfo  {journal} {Phys. Rev. Lett.}\ }\textbf {\bibinfo {volume} {125}},\
  \bibinfo {pages} {160602} (\bibinfo {year} {2020})}\BibitemShut {NoStop}%
\bibitem [{\citenamefont {Ray}\ \emph {et~al.}(2023)\citenamefont {Ray},
  \citenamefont {Boyd}, \citenamefont {Guarnieri},\ and\ \citenamefont
  {Crutchfield}}]{Ray2023}%
  \BibitemOpen
  \bibfield  {author} {\bibinfo {author} {\bibfnamefont {K.~J.}\ \bibnamefont
  {Ray}}, \bibinfo {author} {\bibfnamefont {A.~B.}\ \bibnamefont {Boyd}},
  \bibinfo {author} {\bibfnamefont {G.}~\bibnamefont {Guarnieri}},\ and\
  \bibinfo {author} {\bibfnamefont {J.~P.}\ \bibnamefont {Crutchfield}},\
  }\bibfield  {title} {\bibinfo {title} {Thermodynamic uncertainty theorem},\
  }\href {https://doi.org/10.1103/PhysRevE.108.054126} {\bibfield  {journal}
  {\bibinfo  {journal} {Phys. Rev. E}\ }\textbf {\bibinfo {volume} {108}},\
  \bibinfo {pages} {054126} (\bibinfo {year} {2023})}\BibitemShut {NoStop}%
\bibitem [{\citenamefont {Jordan}\ and\ \citenamefont
  {Korotkov}(2006)}]{jordan2006}%
  \BibitemOpen
  \bibfield  {author} {\bibinfo {author} {\bibfnamefont {A.~N.}\ \bibnamefont
  {Jordan}}\ and\ \bibinfo {author} {\bibfnamefont {A.~N.}\ \bibnamefont
  {Korotkov}},\ }\bibfield  {title} {\bibinfo {title} {{Qubit feedback and
  control with kicked quantum nondemolition measurements: A quantum Bayesian
  analysis}},\ }\href {https://doi.org/10.1103/PhysRevB.74.085307} {\bibfield
  {journal} {\bibinfo  {journal} {Phys. Rev. B}\ }\textbf {\bibinfo {volume}
  {74}},\ \bibinfo {pages} {085307} (\bibinfo {year} {2006})}\BibitemShut
  {NoStop}%
\bibitem [{\citenamefont {Munson}\ \emph {et~al.}(2024)\citenamefont {Munson},
  \citenamefont {Kothakonda}, \citenamefont {Haferkamp}, \citenamefont
  {Halpern}, \citenamefont {Eisert},\ and\ \citenamefont {Faist}}]{Complexity}%
  \BibitemOpen
  \bibfield  {author} {\bibinfo {author} {\bibfnamefont {A.}~\bibnamefont
  {Munson}}, \bibinfo {author} {\bibfnamefont {N.~B.~T.}\ \bibnamefont
  {Kothakonda}}, \bibinfo {author} {\bibfnamefont {J.}~\bibnamefont
  {Haferkamp}}, \bibinfo {author} {\bibfnamefont {N.~Y.}\ \bibnamefont
  {Halpern}}, \bibinfo {author} {\bibfnamefont {J.}~\bibnamefont {Eisert}},\
  and\ \bibinfo {author} {\bibfnamefont {P.}~\bibnamefont {Faist}},\ }\bibfield
   {title} {\bibinfo {title} {Complexity-constrained quantum thermodynamics},\
  }\Eprint {https://arxiv.org/abs/arXiv:2403.04828} {arXiv:2403.04828}
  (\bibinfo {year} {2024})\BibitemShut {NoStop}%
\bibitem [{\citenamefont {Auff{\`e}ves}(2022)}]{auffeves2022quantum}%
  \BibitemOpen
  \bibfield  {author} {\bibinfo {author} {\bibfnamefont {A.}~\bibnamefont
  {Auff{\`e}ves}},\ }\bibfield  {title} {\bibinfo {title} {Quantum technologies
  need a quantum energy initiative},\ }\href
  {https://doi.org/10.1103/PRXQuantum.3.020101} {\bibfield  {journal} {\bibinfo
   {journal} {PRX Quantum}\ }\textbf {\bibinfo {volume} {3}},\ \bibinfo {pages}
  {020101} (\bibinfo {year} {2022})}\BibitemShut {NoStop}%
\bibitem [{\citenamefont {Bhandari}\ and\ \citenamefont
  {Jordan}(2022)}]{bhandari2022con}%
  \BibitemOpen
  \bibfield  {author} {\bibinfo {author} {\bibfnamefont {B.}~\bibnamefont
  {Bhandari}}\ and\ \bibinfo {author} {\bibfnamefont {A.~N.}\ \bibnamefont
  {Jordan}},\ }\bibfield  {title} {\bibinfo {title} {Continuous measurement
  boosted adiabatic quantum thermal machines},\ }\href
  {https://doi.org/10.1103/PhysRevResearch.4.033103} {\bibfield  {journal}
  {\bibinfo  {journal} {Phys. Rev. Res.}\ }\textbf {\bibinfo {volume} {4}},\
  \bibinfo {pages} {033103} (\bibinfo {year} {2022})}\BibitemShut {NoStop}%
\bibitem [{\citenamefont {Bhandari}\ \emph {et~al.}(2023)\citenamefont
  {Bhandari}, \citenamefont {Czupryniak}, \citenamefont {Erdman},\ and\
  \citenamefont {Jordan}}]{bhandari2023meas}%
  \BibitemOpen
  \bibfield  {author} {\bibinfo {author} {\bibfnamefont {B.}~\bibnamefont
  {Bhandari}}, \bibinfo {author} {\bibfnamefont {R.}~\bibnamefont
  {Czupryniak}}, \bibinfo {author} {\bibfnamefont {P.~A.}\ \bibnamefont
  {Erdman}},\ and\ \bibinfo {author} {\bibfnamefont {A.~N.}\ \bibnamefont
  {Jordan}},\ }\bibfield  {title} {\bibinfo {title} {Measurement-based quantum
  thermal machines with feedback control},\ }\href
  {https://www.mdpi.com/1099-4300/25/2/204} {\bibfield  {journal} {\bibinfo
  {journal} {Entropy}\ }\textbf {\bibinfo {volume} {25}},\ \bibinfo {pages}
  {204} (\bibinfo {year} {2023})}\BibitemShut {NoStop}%
\bibitem [{\citenamefont {Lloyd}(1997)}]{lloyd1997}%
  \BibitemOpen
  \bibfield  {author} {\bibinfo {author} {\bibfnamefont {S.}~\bibnamefont
  {Lloyd}},\ }\bibfield  {title} {\bibinfo {title} {Quantum-mechanical
  maxwell's demon},\ }\href {https://doi.org/10.1103/PhysRevA.56.3374}
  {\bibfield  {journal} {\bibinfo  {journal} {Phys. Rev. A}\ }\textbf {\bibinfo
  {volume} {56}},\ \bibinfo {pages} {3374} (\bibinfo {year}
  {1997})}\BibitemShut {NoStop}%
\bibitem [{\citenamefont {Barato}\ and\ \citenamefont
  {Seifert}(2013)}]{barato2013}%
  \BibitemOpen
  \bibfield  {author} {\bibinfo {author} {\bibfnamefont {A.~C.}\ \bibnamefont
  {Barato}}\ and\ \bibinfo {author} {\bibfnamefont {U.}~\bibnamefont
  {Seifert}},\ }\bibfield  {title} {\bibinfo {title} {An autonomous and
  reversible maxwell's demon},\ }\href
  {https://doi.org/10.1209/0295-5075/101/60001} {\bibfield  {journal} {\bibinfo
   {journal} {EPL}\ }\textbf {\bibinfo {volume} {101}},\ \bibinfo {pages}
  {60001} (\bibinfo {year} {2013})}\BibitemShut {NoStop}%
\bibitem [{\citenamefont {Koski}\ \emph {et~al.}(2014)\citenamefont {Koski},
  \citenamefont {Maisi}, \citenamefont {Pekola},\ and\ \citenamefont
  {Averin}}]{koski2014_pnas}%
  \BibitemOpen
  \bibfield  {author} {\bibinfo {author} {\bibfnamefont {J.~V.}\ \bibnamefont
  {Koski}}, \bibinfo {author} {\bibfnamefont {V.~F.}\ \bibnamefont {Maisi}},
  \bibinfo {author} {\bibfnamefont {J.~P.}\ \bibnamefont {Pekola}},\ and\
  \bibinfo {author} {\bibfnamefont {D.~V.}\ \bibnamefont {Averin}},\ }\bibfield
   {title} {\bibinfo {title} {Experimental realization of a szilard engine with
  a single electron},\ }\href {https://doi.org/10.1073/pnas.1406966111}
  {\bibfield  {journal} {\bibinfo  {journal} {Proc. Natl. Acad. Sci. U.S.A.}\
  }\textbf {\bibinfo {volume} {111}},\ \bibinfo {pages} {13786} (\bibinfo
  {year} {2014})}\BibitemShut {NoStop}%
\bibitem [{\citenamefont {Koski}\ \emph {et~al.}(2015)\citenamefont {Koski},
  \citenamefont {Kutvonen}, \citenamefont {Khaymovich}, \citenamefont
  {Ala-Nissila},\ and\ \citenamefont {Pekola}}]{koski2015}%
  \BibitemOpen
  \bibfield  {author} {\bibinfo {author} {\bibfnamefont {J.~V.}\ \bibnamefont
  {Koski}}, \bibinfo {author} {\bibfnamefont {A.}~\bibnamefont {Kutvonen}},
  \bibinfo {author} {\bibfnamefont {I.~M.}\ \bibnamefont {Khaymovich}},
  \bibinfo {author} {\bibfnamefont {T.}~\bibnamefont {Ala-Nissila}},\ and\
  \bibinfo {author} {\bibfnamefont {J.~P.}\ \bibnamefont {Pekola}},\ }\bibfield
   {title} {\bibinfo {title} {On-chip maxwell's demon as an information-powered
  refrigerator},\ }\href {https://doi.org/10.1103/PhysRevLett.115.260602}
  {\bibfield  {journal} {\bibinfo  {journal} {Phys. Rev. Lett.}\ }\textbf
  {\bibinfo {volume} {115}},\ \bibinfo {pages} {260602} (\bibinfo {year}
  {2015})}\BibitemShut {NoStop}%
\bibitem [{\citenamefont {Camati}\ \emph {et~al.}(2016)\citenamefont {Camati},
  \citenamefont {Peterson}, \citenamefont {Batalh\~ao}, \citenamefont
  {Micadei}, \citenamefont {Souza}, \citenamefont {Sarthour}, \citenamefont
  {Oliveira},\ and\ \citenamefont {Serra}}]{camati2016}%
  \BibitemOpen
  \bibfield  {author} {\bibinfo {author} {\bibfnamefont {P.~A.}\ \bibnamefont
  {Camati}}, \bibinfo {author} {\bibfnamefont {J.~P.~S.}\ \bibnamefont
  {Peterson}}, \bibinfo {author} {\bibfnamefont {T.~B.}\ \bibnamefont
  {Batalh\~ao}}, \bibinfo {author} {\bibfnamefont {K.}~\bibnamefont {Micadei}},
  \bibinfo {author} {\bibfnamefont {A.~M.}\ \bibnamefont {Souza}}, \bibinfo
  {author} {\bibfnamefont {R.~S.}\ \bibnamefont {Sarthour}}, \bibinfo {author}
  {\bibfnamefont {I.~S.}\ \bibnamefont {Oliveira}},\ and\ \bibinfo {author}
  {\bibfnamefont {R.~M.}\ \bibnamefont {Serra}},\ }\bibfield  {title} {\bibinfo
  {title} {Experimental rectification of entropy production by maxwell's demon
  in a quantum system},\ }\href
  {https://doi.org/10.1103/PhysRevLett.117.240502} {\bibfield  {journal}
  {\bibinfo  {journal} {Phys. Rev. Lett.}\ }\textbf {\bibinfo {volume} {117}},\
  \bibinfo {pages} {240502} (\bibinfo {year} {2016})}\BibitemShut {NoStop}%
\bibitem [{\citenamefont {Chida}\ \emph {et~al.}(2017)\citenamefont {Chida},
  \citenamefont {Desai}, \citenamefont {Nishiguchi},\ and\ \citenamefont
  {Fujiwara}}]{chida2017}%
  \BibitemOpen
  \bibfield  {author} {\bibinfo {author} {\bibfnamefont {K.}~\bibnamefont
  {Chida}}, \bibinfo {author} {\bibfnamefont {S.}~\bibnamefont {Desai}},
  \bibinfo {author} {\bibfnamefont {K.}~\bibnamefont {Nishiguchi}},\ and\
  \bibinfo {author} {\bibfnamefont {A.}~\bibnamefont {Fujiwara}},\ }\bibfield
  {title} {\bibinfo {title} {Power generator driven by maxwell's demon},\
  }\href {https://doi.org/10.1038/ncomms15301} {\bibfield  {journal} {\bibinfo
  {journal} {Nat. Commun.}\ }\textbf {\bibinfo {volume} {8}} (\bibinfo {year}
  {2017})}\BibitemShut {NoStop}%
\bibitem [{\citenamefont {Campisi}\ \emph {et~al.}(2017)\citenamefont
  {Campisi}, \citenamefont {Pekola},\ and\ \citenamefont
  {Fazio}}]{campisi2017}%
  \BibitemOpen
  \bibfield  {author} {\bibinfo {author} {\bibfnamefont {M.}~\bibnamefont
  {Campisi}}, \bibinfo {author} {\bibfnamefont {J.}~\bibnamefont {Pekola}},\
  and\ \bibinfo {author} {\bibfnamefont {R.}~\bibnamefont {Fazio}},\ }\bibfield
   {title} {\bibinfo {title} {Feedback-controlled heat transport in quantum
  devices: theory and solid-state experimental proposal},\ }\href
  {https://doi.org/10.1088/1367-2630/aa6acb} {\bibfield  {journal} {\bibinfo
  {journal} {New J. Phys.}\ }\textbf {\bibinfo {volume} {19}},\ \bibinfo
  {pages} {053027} (\bibinfo {year} {2017})}\BibitemShut {NoStop}%
\bibitem [{\citenamefont {Cottet}\ \emph {et~al.}(2017)\citenamefont {Cottet},
  \citenamefont {Jezouin}, \citenamefont {Bretheau}, \citenamefont
  {Campagne-Ibarcq}, \citenamefont {Ficheux}, \citenamefont {Anders},
  \citenamefont {Auffèves}, \citenamefont {Azouit}, \citenamefont {Rouchon},\
  and\ \citenamefont {Huard}}]{cottet2017}%
  \BibitemOpen
  \bibfield  {author} {\bibinfo {author} {\bibfnamefont {N.}~\bibnamefont
  {Cottet}}, \bibinfo {author} {\bibfnamefont {S.}~\bibnamefont {Jezouin}},
  \bibinfo {author} {\bibfnamefont {L.}~\bibnamefont {Bretheau}}, \bibinfo
  {author} {\bibfnamefont {P.}~\bibnamefont {Campagne-Ibarcq}}, \bibinfo
  {author} {\bibfnamefont {Q.}~\bibnamefont {Ficheux}}, \bibinfo {author}
  {\bibfnamefont {J.}~\bibnamefont {Anders}}, \bibinfo {author} {\bibfnamefont
  {A.}~\bibnamefont {Auffèves}}, \bibinfo {author} {\bibfnamefont
  {R.}~\bibnamefont {Azouit}}, \bibinfo {author} {\bibfnamefont
  {P.}~\bibnamefont {Rouchon}},\ and\ \bibinfo {author} {\bibfnamefont
  {B.}~\bibnamefont {Huard}},\ }\bibfield  {title} {\bibinfo {title} {Observing
  a quantum maxwell demon at work},\ }\href
  {https://doi.org/10.1073/pnas.1704827114} {\bibfield  {journal} {\bibinfo
  {journal} {Proc. Natl. Acad. Sci. U.S.A.}\ }\textbf {\bibinfo {volume}
  {114}},\ \bibinfo {pages} {7561} (\bibinfo {year} {2017})}\BibitemShut
  {NoStop}%
\bibitem [{\citenamefont {Potts}\ and\ \citenamefont
  {Samuelsson}(2019)}]{potts2019thermodynamics}%
  \BibitemOpen
  \bibfield  {author} {\bibinfo {author} {\bibfnamefont {P.~P.}\ \bibnamefont
  {Potts}}\ and\ \bibinfo {author} {\bibfnamefont {P.}~\bibnamefont
  {Samuelsson}},\ }\bibfield  {title} {\bibinfo {title} {Thermodynamic
  uncertainty relations including measurement and feedback},\ }\href
  {https://doi.org/10.1103/PhysRevE.100.052137} {\bibfield  {journal} {\bibinfo
   {journal} {Phys. Rev. E}\ }\textbf {\bibinfo {volume} {100}},\ \bibinfo
  {pages} {052137} (\bibinfo {year} {2019})}\BibitemShut {NoStop}%
\bibitem [{\citenamefont {Annby-Andersson}\ \emph {et~al.}(2022)\citenamefont
  {Annby-Andersson}, \citenamefont {Bakhshinezhad}, \citenamefont
  {Bhattacharyya}, \citenamefont {De~Sousa}, \citenamefont {Jarzynski},
  \citenamefont {Samuelsson},\ and\ \citenamefont {Potts}}]{andersson2022}%
  \BibitemOpen
  \bibfield  {author} {\bibinfo {author} {\bibfnamefont {B.}~\bibnamefont
  {Annby-Andersson}}, \bibinfo {author} {\bibfnamefont {F.}~\bibnamefont
  {Bakhshinezhad}}, \bibinfo {author} {\bibfnamefont {D.}~\bibnamefont
  {Bhattacharyya}}, \bibinfo {author} {\bibfnamefont {G.}~\bibnamefont
  {De~Sousa}}, \bibinfo {author} {\bibfnamefont {C.}~\bibnamefont {Jarzynski}},
  \bibinfo {author} {\bibfnamefont {P.}~\bibnamefont {Samuelsson}},\ and\
  \bibinfo {author} {\bibfnamefont {P.~P.}\ \bibnamefont {Potts}},\ }\bibfield
  {title} {\bibinfo {title} {Quantum fokker-planck master equation for
  continuous feedback control},\ }\href
  {https://doi.org/10.1103/PhysRevLett.129.050401} {\bibfield  {journal}
  {\bibinfo  {journal} {Phys. Rev. Lett.}\ }\textbf {\bibinfo {volume} {129}},\
  \bibinfo {pages} {050401} (\bibinfo {year} {2022})}\BibitemShut {NoStop}%
\bibitem [{\citenamefont {Esposito}\ \emph {et~al.}(2010)\citenamefont
  {Esposito}, \citenamefont {Kawai}, \citenamefont {Lindenberg},\ and\
  \citenamefont {Van~den Broeck}}]{espositodot2010}%
  \BibitemOpen
  \bibfield  {author} {\bibinfo {author} {\bibfnamefont {M.}~\bibnamefont
  {Esposito}}, \bibinfo {author} {\bibfnamefont {R.}~\bibnamefont {Kawai}},
  \bibinfo {author} {\bibfnamefont {K.}~\bibnamefont {Lindenberg}},\ and\
  \bibinfo {author} {\bibfnamefont {C.}~\bibnamefont {Van~den Broeck}},\
  }\bibfield  {title} {\bibinfo {title} {{Quantum-dot Carnot engine at maximum
  power}},\ }\href {https://doi.org/10.1103/PhysRevE.81.041106} {\bibfield
  {journal} {\bibinfo  {journal} {Phys. Rev. E}\ }\textbf {\bibinfo {volume}
  {81}},\ \bibinfo {pages} {041106} (\bibinfo {year} {2010})}\BibitemShut
  {NoStop}%
\bibitem [{\citenamefont {Abah}\ \emph {et~al.}(2012)\citenamefont {Abah},
  \citenamefont {Ro\ss{}nagel}, \citenamefont {Jacob}, \citenamefont {Deffner},
  \citenamefont {Schmidt-Kaler}, \citenamefont {Singer},\ and\ \citenamefont
  {Lutz}}]{abah2012single}%
  \BibitemOpen
  \bibfield  {author} {\bibinfo {author} {\bibfnamefont {O.}~\bibnamefont
  {Abah}}, \bibinfo {author} {\bibfnamefont {J.}~\bibnamefont {Ro\ss{}nagel}},
  \bibinfo {author} {\bibfnamefont {G.}~\bibnamefont {Jacob}}, \bibinfo
  {author} {\bibfnamefont {S.}~\bibnamefont {Deffner}}, \bibinfo {author}
  {\bibfnamefont {F.}~\bibnamefont {Schmidt-Kaler}}, \bibinfo {author}
  {\bibfnamefont {K.}~\bibnamefont {Singer}},\ and\ \bibinfo {author}
  {\bibfnamefont {E.}~\bibnamefont {Lutz}},\ }\bibfield  {title} {\bibinfo
  {title} {Single-ion heat engine at maximum power},\ }\href
  {https://doi.org/10.1103/PhysRevLett.109.203006} {\bibfield  {journal}
  {\bibinfo  {journal} {Phys. Rev. Lett.}\ }\textbf {\bibinfo {volume} {109}},\
  \bibinfo {pages} {203006} (\bibinfo {year} {2012})}\BibitemShut {NoStop}%
\bibitem [{\citenamefont {Juergens}\ \emph {et~al.}(2013)\citenamefont
  {Juergens}, \citenamefont {Haupt}, \citenamefont {Moskalets},\ and\
  \citenamefont {Splettstoesser}}]{janine2013dot}%
  \BibitemOpen
  \bibfield  {author} {\bibinfo {author} {\bibfnamefont {S.}~\bibnamefont
  {Juergens}}, \bibinfo {author} {\bibfnamefont {F.}~\bibnamefont {Haupt}},
  \bibinfo {author} {\bibfnamefont {M.}~\bibnamefont {Moskalets}},\ and\
  \bibinfo {author} {\bibfnamefont {J.}~\bibnamefont {Splettstoesser}},\
  }\bibfield  {title} {\bibinfo {title} {Thermoelectric performance of a driven
  double quantum dot},\ }\href {https://doi.org/10.1103/PhysRevB.87.245423}
  {\bibfield  {journal} {\bibinfo  {journal} {Phys. Rev. B}\ }\textbf {\bibinfo
  {volume} {87}},\ \bibinfo {pages} {245423} (\bibinfo {year}
  {2013})}\BibitemShut {NoStop}%
\bibitem [{\citenamefont {del Campo}\ \emph {et~al.}(2014)\citenamefont {del
  Campo}, \citenamefont {Goold},\ and\ \citenamefont
  {Paternostro}}]{campo2014}%
  \BibitemOpen
  \bibfield  {author} {\bibinfo {author} {\bibfnamefont {A.}~\bibnamefont {del
  Campo}}, \bibinfo {author} {\bibfnamefont {J.}~\bibnamefont {Goold}},\ and\
  \bibinfo {author} {\bibfnamefont {M.}~\bibnamefont {Paternostro}},\
  }\bibfield  {title} {\bibinfo {title} {More bang for your buck:
  Super-adiabatic quantum engines},\ }\href {https://doi.org/10.1038/srep06208}
  {\bibfield  {journal} {\bibinfo  {journal} {Sci. Rep.}\ }\textbf {\bibinfo
  {volume} {4}},\ \bibinfo {pages} {6208} (\bibinfo {year} {2014})}\BibitemShut
  {NoStop}%
\bibitem [{\citenamefont {Campisi}\ \emph {et~al.}(2015)\citenamefont
  {Campisi}, \citenamefont {Pekola},\ and\ \citenamefont
  {Fazio}}]{campisi2015nonequilibrium}%
  \BibitemOpen
  \bibfield  {author} {\bibinfo {author} {\bibfnamefont {M.}~\bibnamefont
  {Campisi}}, \bibinfo {author} {\bibfnamefont {J.}~\bibnamefont {Pekola}},\
  and\ \bibinfo {author} {\bibfnamefont {R.}~\bibnamefont {Fazio}},\ }\bibfield
   {title} {\bibinfo {title} {Nonequilibrium fluctuations in quantum heat
  engines: theory, example, and possible solid state experiments},\ }\href
  {https://doi.org/10.1088/1367-2630/17/3/035012} {\bibfield  {journal}
  {\bibinfo  {journal} {New J. Phys.}\ }\textbf {\bibinfo {volume} {17}},\
  \bibinfo {pages} {035012} (\bibinfo {year} {2015})}\BibitemShut {NoStop}%
\bibitem [{\citenamefont {Campisi}\ and\ \citenamefont
  {Fazio}(2016)}]{campisi2016}%
  \BibitemOpen
  \bibfield  {author} {\bibinfo {author} {\bibfnamefont {M.}~\bibnamefont
  {Campisi}}\ and\ \bibinfo {author} {\bibfnamefont {R.}~\bibnamefont
  {Fazio}},\ }\bibfield  {title} {\bibinfo {title} {The power of a critical
  heat engine},\ }\href {https://doi.org/10.1038/ncomms11895} {\bibfield
  {journal} {\bibinfo  {journal} {Nat. Commun.}\ }\textbf {\bibinfo {volume}
  {7}},\ \bibinfo {pages} {11895} (\bibinfo {year} {2016})}\BibitemShut
  {NoStop}%
\bibitem [{\citenamefont {Villazon}\ \emph {et~al.}(2019)\citenamefont
  {Villazon}, \citenamefont {Polkovnikov},\ and\ \citenamefont
  {Chandran}}]{villazon2019}%
  \BibitemOpen
  \bibfield  {author} {\bibinfo {author} {\bibfnamefont {T.}~\bibnamefont
  {Villazon}}, \bibinfo {author} {\bibfnamefont {A.}~\bibnamefont
  {Polkovnikov}},\ and\ \bibinfo {author} {\bibfnamefont {A.}~\bibnamefont
  {Chandran}},\ }\bibfield  {title} {\bibinfo {title} {Swift heat transfer by
  fast-forward driving in open quantum systems},\ }\href
  {https://doi.org/10.1103/PhysRevA.100.012126} {\bibfield  {journal} {\bibinfo
   {journal} {Phys. Rev. A}\ }\textbf {\bibinfo {volume} {100}},\ \bibinfo
  {pages} {012126} (\bibinfo {year} {2019})}\BibitemShut {NoStop}%
\bibitem [{\citenamefont {Miller}\ \emph {et~al.}(2019)\citenamefont {Miller},
  \citenamefont {Scandi}, \citenamefont {Anders},\ and\ \citenamefont
  {Perarnau-Llobet}}]{miller2019work}%
  \BibitemOpen
  \bibfield  {author} {\bibinfo {author} {\bibfnamefont {H.~J.}\ \bibnamefont
  {Miller}}, \bibinfo {author} {\bibfnamefont {M.}~\bibnamefont {Scandi}},
  \bibinfo {author} {\bibfnamefont {J.}~\bibnamefont {Anders}},\ and\ \bibinfo
  {author} {\bibfnamefont {M.}~\bibnamefont {Perarnau-Llobet}},\ }\bibfield
  {title} {\bibinfo {title} {Work fluctuations in slow processes: quantum
  signatures and optimal control},\ }\href
  {https://doi.org/10.1103/PhysRevLett.123.230603} {\bibfield  {journal}
  {\bibinfo  {journal} {Phys. Rev. Lett.}\ }\textbf {\bibinfo {volume} {123}},\
  \bibinfo {pages} {230603} (\bibinfo {year} {2019})}\BibitemShut {NoStop}%
\bibitem [{\citenamefont {Tobalina}\ \emph {et~al.}(2019)\citenamefont
  {Tobalina}, \citenamefont {Lizuain},\ and\ \citenamefont
  {Muga}}]{tobalina2019vanishing}%
  \BibitemOpen
  \bibfield  {author} {\bibinfo {author} {\bibfnamefont {A.}~\bibnamefont
  {Tobalina}}, \bibinfo {author} {\bibfnamefont {I.}~\bibnamefont {Lizuain}},\
  and\ \bibinfo {author} {\bibfnamefont {J.~G.}\ \bibnamefont {Muga}},\
  }\bibfield  {title} {\bibinfo {title} {{Vanishing efficiency of a speeded-up
  ion-in-Paul-trap Otto engine (a)}},\ }\href
  {https://doi.org/10.1209/0295-5075/127/20005} {\bibfield  {journal} {\bibinfo
   {journal} {Europhys. Lett.}\ }\textbf {\bibinfo {volume} {127}},\ \bibinfo
  {pages} {20005} (\bibinfo {year} {2019})}\BibitemShut {NoStop}%
\bibitem [{\citenamefont {Erdman}\ \emph
  {et~al.}(2019{\natexlab{b}})\citenamefont {Erdman}, \citenamefont {Cavina},
  \citenamefont {Fazio}, \citenamefont {Taddei},\ and\ \citenamefont
  {Giovannetti}}]{erdman2019maximum}%
  \BibitemOpen
  \bibfield  {author} {\bibinfo {author} {\bibfnamefont {P.~A.}\ \bibnamefont
  {Erdman}}, \bibinfo {author} {\bibfnamefont {V.}~\bibnamefont {Cavina}},
  \bibinfo {author} {\bibfnamefont {R.}~\bibnamefont {Fazio}}, \bibinfo
  {author} {\bibfnamefont {F.}~\bibnamefont {Taddei}},\ and\ \bibinfo {author}
  {\bibfnamefont {V.}~\bibnamefont {Giovannetti}},\ }\bibfield  {title}
  {\bibinfo {title} {Maximum power and corresponding efficiency for two-level
  heat engines and refrigerators: optimality of fast cycles},\ }\href
  {https://doi.org/10.1088/1367-2630/ab4dca} {\bibfield  {journal} {\bibinfo
  {journal} {New J. Phys.}\ }\textbf {\bibinfo {volume} {21}},\ \bibinfo
  {pages} {103049} (\bibinfo {year} {2019}{\natexlab{b}})}\BibitemShut
  {NoStop}%
\bibitem [{\citenamefont {Abiuso}\ and\ \citenamefont
  {Giovannetti}(2019)}]{abiuso2019nonmarkov}%
  \BibitemOpen
  \bibfield  {author} {\bibinfo {author} {\bibfnamefont {P.}~\bibnamefont
  {Abiuso}}\ and\ \bibinfo {author} {\bibfnamefont {V.}~\bibnamefont
  {Giovannetti}},\ }\bibfield  {title} {\bibinfo {title} {{Non-Markov
  enhancement of maximum power for quantum thermal machines}},\ }\href
  {https://doi.org/10.1103/PhysRevA.99.052106} {\bibfield  {journal} {\bibinfo
  {journal} {Phys. Rev. A}\ }\textbf {\bibinfo {volume} {99}},\ \bibinfo
  {pages} {052106} (\bibinfo {year} {2019})}\BibitemShut {NoStop}%
\bibitem [{\citenamefont {Abiuso}\ and\ \citenamefont
  {Perarnau-Llobet}(2020)}]{abiuso2020_prl}%
  \BibitemOpen
  \bibfield  {author} {\bibinfo {author} {\bibfnamefont {P.}~\bibnamefont
  {Abiuso}}\ and\ \bibinfo {author} {\bibfnamefont {M.}~\bibnamefont
  {Perarnau-Llobet}},\ }\bibfield  {title} {\bibinfo {title} {Optimal cycles
  for low-dissipation heat engines},\ }\href
  {https://doi.org/10.1103/PhysRevLett.124.110606} {\bibfield  {journal}
  {\bibinfo  {journal} {Phys. Rev. Lett.}\ }\textbf {\bibinfo {volume} {124}},\
  \bibinfo {pages} {110606} (\bibinfo {year} {2020})}\BibitemShut {NoStop}%
\bibitem [{\citenamefont {Bhandari}\ \emph {et~al.}(2020)\citenamefont
  {Bhandari}, \citenamefont {Alonso}, \citenamefont {Taddei}, \citenamefont
  {von Oppen}, \citenamefont {Fazio},\ and\ \citenamefont
  {Arrachea}}]{bhandari2020geo}%
  \BibitemOpen
  \bibfield  {author} {\bibinfo {author} {\bibfnamefont {B.}~\bibnamefont
  {Bhandari}}, \bibinfo {author} {\bibfnamefont {P.~T.}\ \bibnamefont
  {Alonso}}, \bibinfo {author} {\bibfnamefont {F.}~\bibnamefont {Taddei}},
  \bibinfo {author} {\bibfnamefont {F.}~\bibnamefont {von Oppen}}, \bibinfo
  {author} {\bibfnamefont {R.}~\bibnamefont {Fazio}},\ and\ \bibinfo {author}
  {\bibfnamefont {L.}~\bibnamefont {Arrachea}},\ }\bibfield  {title} {\bibinfo
  {title} {Geometric properties of adiabatic quantum thermal machines},\ }\href
  {https://doi.org/10.1103/PhysRevB.102.155407} {\bibfield  {journal} {\bibinfo
   {journal} {Phys. Rev. B}\ }\textbf {\bibinfo {volume} {102}},\ \bibinfo
  {pages} {155407} (\bibinfo {year} {2020})}\BibitemShut {NoStop}%
\bibitem [{\citenamefont {Abiuso}\ \emph {et~al.}(2020)\citenamefont {Abiuso},
  \citenamefont {Miller}, \citenamefont {Perarnau-Llobet},\ and\ \citenamefont
  {Scandi}}]{abiuso2020geometric}%
  \BibitemOpen
  \bibfield  {author} {\bibinfo {author} {\bibfnamefont {P.}~\bibnamefont
  {Abiuso}}, \bibinfo {author} {\bibfnamefont {H.~J.}\ \bibnamefont {Miller}},
  \bibinfo {author} {\bibfnamefont {M.}~\bibnamefont {Perarnau-Llobet}},\ and\
  \bibinfo {author} {\bibfnamefont {M.}~\bibnamefont {Scandi}},\ }\bibfield
  {title} {\bibinfo {title} {Geometric optimisation of quantum thermodynamic
  processes},\ }\href {https://doi.org/10.3390/e22101076} {\bibfield  {journal}
  {\bibinfo  {journal} {Entropy}\ }\textbf {\bibinfo {volume} {22}},\ \bibinfo
  {pages} {1076} (\bibinfo {year} {2020})}\BibitemShut {NoStop}%
\bibitem [{\citenamefont {Pancotti}\ \emph {et~al.}(2020)\citenamefont
  {Pancotti}, \citenamefont {Scandi}, \citenamefont {Mitchison},\ and\
  \citenamefont {Perarnau-Llobet}}]{pancotti2020speed}%
  \BibitemOpen
  \bibfield  {author} {\bibinfo {author} {\bibfnamefont {N.}~\bibnamefont
  {Pancotti}}, \bibinfo {author} {\bibfnamefont {M.}~\bibnamefont {Scandi}},
  \bibinfo {author} {\bibfnamefont {M.~T.}\ \bibnamefont {Mitchison}},\ and\
  \bibinfo {author} {\bibfnamefont {M.}~\bibnamefont {Perarnau-Llobet}},\
  }\bibfield  {title} {\bibinfo {title} {Speed-ups to isothermality: Enhanced
  quantum thermal machines through control of the system-bath coupling},\
  }\href {https://doi.org/10.1103/PhysRevX.10.031015} {\bibfield  {journal}
  {\bibinfo  {journal} {Phys. Rev. X}\ }\textbf {\bibinfo {volume} {10}},\
  \bibinfo {pages} {031015} (\bibinfo {year} {2020})}\BibitemShut {NoStop}%
\bibitem [{\citenamefont {Dann}\ and\ \citenamefont
  {Kosloff}(2020)}]{dann2020quantum}%
  \BibitemOpen
  \bibfield  {author} {\bibinfo {author} {\bibfnamefont {R.}~\bibnamefont
  {Dann}}\ and\ \bibinfo {author} {\bibfnamefont {R.}~\bibnamefont {Kosloff}},\
  }\bibfield  {title} {\bibinfo {title} {Quantum signatures in the quantum
  carnot cycle},\ }\href {https://doi.org/10.1088/1367-2630/ab6876} {\bibfield
  {journal} {\bibinfo  {journal} {New J. Phys.}\ }\textbf {\bibinfo {volume}
  {22}},\ \bibinfo {pages} {013055} (\bibinfo {year} {2020})}\BibitemShut
  {NoStop}%
\bibitem [{\citenamefont {Wang}\ \emph {et~al.}(2011)\citenamefont {Wang},
  \citenamefont {He},\ and\ \citenamefont {He}}]{wang2011performance}%
  \BibitemOpen
  \bibfield  {author} {\bibinfo {author} {\bibfnamefont {J.}~\bibnamefont
  {Wang}}, \bibinfo {author} {\bibfnamefont {J.}~\bibnamefont {He}},\ and\
  \bibinfo {author} {\bibfnamefont {X.}~\bibnamefont {He}},\ }\bibfield
  {title} {\bibinfo {title} {Performance analysis of a two-state quantum heat
  engine working with a single-mode radiation field in a cavity},\ }\href
  {https://doi.org/10.1103/PhysRevE.84.041127} {\bibfield  {journal} {\bibinfo
  {journal} {Phys. Rev. E}\ }\textbf {\bibinfo {volume} {84}},\ \bibinfo
  {pages} {041127} (\bibinfo {year} {2011})}\BibitemShut {NoStop}%
\bibitem [{\citenamefont {Cavina}\ \emph {et~al.}(2021)\citenamefont {Cavina},
  \citenamefont {Erdman}, \citenamefont {Abiuso}, \citenamefont {Tolomeo},\
  and\ \citenamefont {Giovannetti}}]{cavina2021}%
  \BibitemOpen
  \bibfield  {author} {\bibinfo {author} {\bibfnamefont {V.}~\bibnamefont
  {Cavina}}, \bibinfo {author} {\bibfnamefont {P.~A.}\ \bibnamefont {Erdman}},
  \bibinfo {author} {\bibfnamefont {P.}~\bibnamefont {Abiuso}}, \bibinfo
  {author} {\bibfnamefont {L.}~\bibnamefont {Tolomeo}},\ and\ \bibinfo {author}
  {\bibfnamefont {V.}~\bibnamefont {Giovannetti}},\ }\bibfield  {title}
  {\bibinfo {title} {Maximum-power heat engines and refrigerators in the
  fast-driving regime},\ }\href {https://doi.org/10.1103/PhysRevA.104.032226}
  {\bibfield  {journal} {\bibinfo  {journal} {Phys. Rev. A}\ }\textbf {\bibinfo
  {volume} {104}},\ \bibinfo {pages} {032226} (\bibinfo {year}
  {2021})}\BibitemShut {NoStop}%
\bibitem [{\citenamefont {Terr\'en~Alonso}\ \emph {et~al.}(2022)\citenamefont
  {Terr\'en~Alonso}, \citenamefont {Abiuso}, \citenamefont {Perarnau-Llobet},\
  and\ \citenamefont {Arrachea}}]{pablo2022geo}%
  \BibitemOpen
  \bibfield  {author} {\bibinfo {author} {\bibfnamefont {P.}~\bibnamefont
  {Terr\'en~Alonso}}, \bibinfo {author} {\bibfnamefont {P.}~\bibnamefont
  {Abiuso}}, \bibinfo {author} {\bibfnamefont {M.}~\bibnamefont
  {Perarnau-Llobet}},\ and\ \bibinfo {author} {\bibfnamefont {L.}~\bibnamefont
  {Arrachea}},\ }\bibfield  {title} {\bibinfo {title} {Geometric optimization
  of nonequilibrium adiabatic thermal machines and implementation in a qubit
  system},\ }\href {https://doi.org/10.1103/PRXQuantum.3.010326} {\bibfield
  {journal} {\bibinfo  {journal} {PRX Quantum}\ }\textbf {\bibinfo {volume}
  {3}},\ \bibinfo {pages} {010326} (\bibinfo {year} {2022})}\BibitemShut
  {NoStop}%
\bibitem [{\citenamefont {Nettersheim}\ \emph {et~al.}(2022)\citenamefont
  {Nettersheim}, \citenamefont {Burgardt}, \citenamefont {Bouton},
  \citenamefont {Adam}, \citenamefont {Lutz},\ and\ \citenamefont
  {Widera}}]{nettersheim2022power}%
  \BibitemOpen
  \bibfield  {author} {\bibinfo {author} {\bibfnamefont {J.}~\bibnamefont
  {Nettersheim}}, \bibinfo {author} {\bibfnamefont {S.}~\bibnamefont
  {Burgardt}}, \bibinfo {author} {\bibfnamefont {Q.}~\bibnamefont {Bouton}},
  \bibinfo {author} {\bibfnamefont {D.}~\bibnamefont {Adam}}, \bibinfo {author}
  {\bibfnamefont {E.}~\bibnamefont {Lutz}},\ and\ \bibinfo {author}
  {\bibfnamefont {A.}~\bibnamefont {Widera}},\ }\bibfield  {title} {\bibinfo
  {title} {{Power of a quasi-spin quantum Otto engine at negative effective
  temperature}},\ }\Eprint {https://arxiv.org/abs/arXiv:2207.09272}
  {arXiv:2207.09272}  (\bibinfo {year} {2022})\BibitemShut {NoStop}%
\bibitem [{\citenamefont {Das}\ \emph {et~al.}(2023)\citenamefont {Das},
  \citenamefont {Thomas},\ and\ \citenamefont {Jordan}}]{das2023}%
  \BibitemOpen
  \bibfield  {author} {\bibinfo {author} {\bibfnamefont {D.}~\bibnamefont
  {Das}}, \bibinfo {author} {\bibfnamefont {G.}~\bibnamefont {Thomas}},\ and\
  \bibinfo {author} {\bibfnamefont {A.~N.}\ \bibnamefont {Jordan}},\ }\bibfield
   {title} {\bibinfo {title} {Quantum stirling heat engine operating in finite
  time},\ }\href {https://doi.org/10.1103/PhysRevA.108.012220} {\bibfield
  {journal} {\bibinfo  {journal} {Phys. Rev. A}\ }\textbf {\bibinfo {volume}
  {108}},\ \bibinfo {pages} {012220} (\bibinfo {year} {2023})}\BibitemShut
  {NoStop}%
\bibitem [{\citenamefont {Carrega}\ \emph {et~al.}(2024)\citenamefont
  {Carrega}, \citenamefont {Razzoli}, \citenamefont {Erdman}, \citenamefont
  {Cavaliere}, \citenamefont {Benenti},\ and\ \citenamefont
  {Sassetti}}]{carrega2024}%
  \BibitemOpen
  \bibfield  {author} {\bibinfo {author} {\bibfnamefont {M.}~\bibnamefont
  {Carrega}}, \bibinfo {author} {\bibfnamefont {L.}~\bibnamefont {Razzoli}},
  \bibinfo {author} {\bibfnamefont {P.~A.}\ \bibnamefont {Erdman}}, \bibinfo
  {author} {\bibfnamefont {F.}~\bibnamefont {Cavaliere}}, \bibinfo {author}
  {\bibfnamefont {G.}~\bibnamefont {Benenti}},\ and\ \bibinfo {author}
  {\bibfnamefont {M.}~\bibnamefont {Sassetti}},\ }\bibfield  {title} {\bibinfo
  {title} {Dissipation-induced collective advantage of a quantum thermal
  machine},\ }\href {https://doi.org/10.1116/5.0190340} {\bibfield  {journal}
  {\bibinfo  {journal} {AVS Quantum Sci.}\ }\textbf {\bibinfo {volume} {6}},\
  \bibinfo {pages} {025001} (\bibinfo {year} {2024})}\BibitemShut {NoStop}%
\bibitem [{\citenamefont {Razzoli}\ \emph {et~al.}(2024)\citenamefont
  {Razzoli}, \citenamefont {Cavaliere}, \citenamefont {Carrega}, \citenamefont
  {Sassetti},\ and\ \citenamefont {Benenti}}]{razzoli2024}%
  \BibitemOpen
  \bibfield  {author} {\bibinfo {author} {\bibfnamefont {L.}~\bibnamefont
  {Razzoli}}, \bibinfo {author} {\bibfnamefont {F.}~\bibnamefont {Cavaliere}},
  \bibinfo {author} {\bibfnamefont {M.}~\bibnamefont {Carrega}}, \bibinfo
  {author} {\bibfnamefont {M.}~\bibnamefont {Sassetti}},\ and\ \bibinfo
  {author} {\bibfnamefont {G.}~\bibnamefont {Benenti}},\ }\bibfield  {title}
  {\bibinfo {title} {Efficiency and thermodynamic uncertainty relations of a
  dynamical quantum heat engine},\ }\href
  {https://doi.org/10.1140/epjs/s11734-023-00949-8} {\bibfield  {journal}
  {\bibinfo  {journal} {Eur. Phys. J. Spec. Top.}\ }\textbf {\bibinfo {volume}
  {233}},\ \bibinfo {pages} {1263} (\bibinfo {year} {2024})}\BibitemShut
  {NoStop}%
\bibitem [{\citenamefont {Jordan}\ and\ \citenamefont
  {Mitchell}(2015)}]{jordan2015}%
  \BibitemOpen
  \bibfield  {author} {\bibinfo {author} {\bibfnamefont {M.~I.}\ \bibnamefont
  {Jordan}}\ and\ \bibinfo {author} {\bibfnamefont {T.~M.}\ \bibnamefont
  {Mitchell}},\ }\bibfield  {title} {\bibinfo {title} {Machine learning:
  Trends, perspectives, and prospects},\ }\href
  {https://doi.org/10.1126/science.aaa8415} {\bibfield  {journal} {\bibinfo
  {journal} {Science}\ }\textbf {\bibinfo {volume} {349}},\ \bibinfo {pages}
  {255} (\bibinfo {year} {2015})}\BibitemShut {NoStop}%
\bibitem [{\citenamefont {Carrasquilla}(2020)}]{carrasquilla2020}%
  \BibitemOpen
  \bibfield  {author} {\bibinfo {author} {\bibfnamefont {J.}~\bibnamefont
  {Carrasquilla}},\ }\bibfield  {title} {\bibinfo {title} {Machine learning for
  quantum matter},\ }\href {https://doi.org/10.1080/23746149.2020.1797528}
  {\bibfield  {journal} {\bibinfo  {journal} {Adv. Phys.: X}\ }\textbf
  {\bibinfo {volume} {5}},\ \bibinfo {pages} {1797528} (\bibinfo {year}
  {2020})}\BibitemShut {NoStop}%
\bibitem [{\citenamefont {Krenn}\ \emph {et~al.}(2023)\citenamefont {Krenn},
  \citenamefont {Landgraf}, \citenamefont {Foesel},\ and\ \citenamefont
  {Marquardt}}]{krenn2023}%
  \BibitemOpen
  \bibfield  {author} {\bibinfo {author} {\bibfnamefont {M.}~\bibnamefont
  {Krenn}}, \bibinfo {author} {\bibfnamefont {J.}~\bibnamefont {Landgraf}},
  \bibinfo {author} {\bibfnamefont {T.}~\bibnamefont {Foesel}},\ and\ \bibinfo
  {author} {\bibfnamefont {F.}~\bibnamefont {Marquardt}},\ }\bibfield  {title}
  {\bibinfo {title} {Artificial intelligence and machine learning for quantum
  technologies},\ }\href {https://doi.org/10.1103/PhysRevA.107.010101}
  {\bibfield  {journal} {\bibinfo  {journal} {Phys. Rev. A}\ }\textbf {\bibinfo
  {volume} {107}},\ \bibinfo {pages} {010101} (\bibinfo {year}
  {2023})}\BibitemShut {NoStop}%
\bibitem [{\citenamefont {Dawid}\ \emph {et~al.}(2023)\citenamefont {Dawid},
  \citenamefont {Arnold}, \citenamefont {Requena}, \citenamefont {Gresch},
  \citenamefont {Płodzień}, \citenamefont {Donatella}, \citenamefont
  {Nicoli}, \citenamefont {Stornati}, \citenamefont {Koch}, \citenamefont
  {Büttner}, \citenamefont {Okuła}, \citenamefont {Muñoz-Gil}, \citenamefont
  {Vargas-Hernández}, \citenamefont {Cervera-Lierta}, \citenamefont
  {Carrasquilla}, \citenamefont {Dunjko}, \citenamefont {Gabrié},
  \citenamefont {Huembeli}, \citenamefont {van Nieuwenburg}, \citenamefont
  {Vicentini}, \citenamefont {Wang}, \citenamefont {Wetzel}, \citenamefont
  {Carleo}, \citenamefont {Greplová}, \citenamefont {Krems}, \citenamefont
  {Marquardt}, \citenamefont {Tomza}, \citenamefont {Lewenstein},\ and\
  \citenamefont {Dauphin}}]{dawid2023}%
  \BibitemOpen
  \bibfield  {author} {\bibinfo {author} {\bibfnamefont {A.}~\bibnamefont
  {Dawid}}, \bibinfo {author} {\bibfnamefont {J.}~\bibnamefont {Arnold}},
  \bibinfo {author} {\bibfnamefont {B.}~\bibnamefont {Requena}}, \bibinfo
  {author} {\bibfnamefont {A.}~\bibnamefont {Gresch}}, \bibinfo {author}
  {\bibfnamefont {M.}~\bibnamefont {Płodzień}}, \bibinfo {author}
  {\bibfnamefont {K.}~\bibnamefont {Donatella}}, \bibinfo {author}
  {\bibfnamefont {K.~A.}\ \bibnamefont {Nicoli}}, \bibinfo {author}
  {\bibfnamefont {P.}~\bibnamefont {Stornati}}, \bibinfo {author}
  {\bibfnamefont {R.}~\bibnamefont {Koch}}, \bibinfo {author} {\bibfnamefont
  {M.}~\bibnamefont {Büttner}}, \bibinfo {author} {\bibfnamefont
  {R.}~\bibnamefont {Okuła}}, \bibinfo {author} {\bibfnamefont
  {G.}~\bibnamefont {Muñoz-Gil}}, \bibinfo {author} {\bibfnamefont {R.~A.}\
  \bibnamefont {Vargas-Hernández}}, \bibinfo {author} {\bibfnamefont
  {A.}~\bibnamefont {Cervera-Lierta}}, \bibinfo {author} {\bibfnamefont
  {J.}~\bibnamefont {Carrasquilla}}, \bibinfo {author} {\bibfnamefont
  {V.}~\bibnamefont {Dunjko}}, \bibinfo {author} {\bibfnamefont
  {M.}~\bibnamefont {Gabrié}}, \bibinfo {author} {\bibfnamefont
  {P.}~\bibnamefont {Huembeli}}, \bibinfo {author} {\bibfnamefont
  {E.}~\bibnamefont {van Nieuwenburg}}, \bibinfo {author} {\bibfnamefont
  {F.}~\bibnamefont {Vicentini}}, \bibinfo {author} {\bibfnamefont
  {L.}~\bibnamefont {Wang}}, \bibinfo {author} {\bibfnamefont {S.~J.}\
  \bibnamefont {Wetzel}}, \bibinfo {author} {\bibfnamefont {G.}~\bibnamefont
  {Carleo}}, \bibinfo {author} {\bibfnamefont {E.}~\bibnamefont {Greplová}},
  \bibinfo {author} {\bibfnamefont {R.}~\bibnamefont {Krems}}, \bibinfo
  {author} {\bibfnamefont {F.}~\bibnamefont {Marquardt}}, \bibinfo {author}
  {\bibfnamefont {M.}~\bibnamefont {Tomza}}, \bibinfo {author} {\bibfnamefont
  {M.}~\bibnamefont {Lewenstein}},\ and\ \bibinfo {author} {\bibfnamefont
  {A.}~\bibnamefont {Dauphin}},\ }\bibfield  {title} {\bibinfo {title} {Modern
  applications of machine learning in quantum sciences},\ }\Eprint
  {https://arxiv.org/abs/2204.04198} {arXiv:2204.04198}  (\bibinfo {year}
  {2023})\BibitemShut {NoStop}%
\bibitem [{\citenamefont {Bukov}\ \emph {et~al.}(2018)\citenamefont {Bukov},
  \citenamefont {Day}, \citenamefont {Sels}, \citenamefont {Weinberg},
  \citenamefont {Polkovnikov},\ and\ \citenamefont {Mehta}}]{bukov2018}%
  \BibitemOpen
  \bibfield  {author} {\bibinfo {author} {\bibfnamefont {M.}~\bibnamefont
  {Bukov}}, \bibinfo {author} {\bibfnamefont {A.~G.~R.}\ \bibnamefont {Day}},
  \bibinfo {author} {\bibfnamefont {D.}~\bibnamefont {Sels}}, \bibinfo {author}
  {\bibfnamefont {P.}~\bibnamefont {Weinberg}}, \bibinfo {author}
  {\bibfnamefont {A.}~\bibnamefont {Polkovnikov}},\ and\ \bibinfo {author}
  {\bibfnamefont {P.}~\bibnamefont {Mehta}},\ }\bibfield  {title} {\bibinfo
  {title} {Reinforcement learning in different phases of quantum control},\
  }\href {https://doi.org/10.1103/PhysRevX.8.031086} {\bibfield  {journal}
  {\bibinfo  {journal} {Phys. Rev. X}\ }\textbf {\bibinfo {volume} {8}},\
  \bibinfo {pages} {031086} (\bibinfo {year} {2018})}\BibitemShut {NoStop}%
\bibitem [{\citenamefont {Zhang}\ \emph {et~al.}(2019)\citenamefont {Zhang},
  \citenamefont {Wei}, \citenamefont {Asad}, \citenamefont {Yang},\ and\
  \citenamefont {Wang}}]{zhang2019}%
  \BibitemOpen
  \bibfield  {author} {\bibinfo {author} {\bibfnamefont {X.-M.}\ \bibnamefont
  {Zhang}}, \bibinfo {author} {\bibfnamefont {Z.}~\bibnamefont {Wei}}, \bibinfo
  {author} {\bibfnamefont {R.}~\bibnamefont {Asad}}, \bibinfo {author}
  {\bibfnamefont {X.-C.}\ \bibnamefont {Yang}},\ and\ \bibinfo {author}
  {\bibfnamefont {X.}~\bibnamefont {Wang}},\ }\bibfield  {title} {\bibinfo
  {title} {When does reinforcement learning stand out in quantum control? a
  comparative study on state preparation},\ }\href
  {https://doi.org/10.1038/s41534-019-0201-8} {\bibfield  {journal} {\bibinfo
  {journal} {npj Quantum Inf.}\ }\textbf {\bibinfo {volume} {5}},\ \bibinfo
  {pages} {85} (\bibinfo {year} {2019})}\BibitemShut {NoStop}%
\bibitem [{\citenamefont {Dalgaard}\ \emph {et~al.}(2020)\citenamefont
  {Dalgaard}, \citenamefont {Motzoi}, \citenamefont {S{\o}rensen},\ and\
  \citenamefont {Sherson}}]{dalgaard2020}%
  \BibitemOpen
  \bibfield  {author} {\bibinfo {author} {\bibfnamefont {M.}~\bibnamefont
  {Dalgaard}}, \bibinfo {author} {\bibfnamefont {F.}~\bibnamefont {Motzoi}},
  \bibinfo {author} {\bibfnamefont {J.~J.}\ \bibnamefont {S{\o}rensen}},\ and\
  \bibinfo {author} {\bibfnamefont {J.}~\bibnamefont {Sherson}},\ }\bibfield
  {title} {\bibinfo {title} {Global optimization of quantum dynamics with
  alphazero deep exploration},\ }\href
  {https://doi.org/10.1038/s41534-019-0241-0} {\bibfield  {journal} {\bibinfo
  {journal} {npj Quantum Inf.}\ }\textbf {\bibinfo {volume} {6}},\ \bibinfo
  {pages} {6} (\bibinfo {year} {2020})}\BibitemShut {NoStop}%
\bibitem [{\citenamefont {Mackeprang}\ \emph {et~al.}(2020)\citenamefont
  {Mackeprang}, \citenamefont {Dasari},\ and\ \citenamefont
  {Wrachtrup}}]{mackeprang2020}%
  \BibitemOpen
  \bibfield  {author} {\bibinfo {author} {\bibfnamefont {J.}~\bibnamefont
  {Mackeprang}}, \bibinfo {author} {\bibfnamefont {D.~B.~R.}\ \bibnamefont
  {Dasari}},\ and\ \bibinfo {author} {\bibfnamefont {J.}~\bibnamefont
  {Wrachtrup}},\ }\bibfield  {title} {\bibinfo {title} {A reinforcement
  learning approach for quantum state engineering},\ }\href
  {https://doi.org/10.1007/s42484-020-00016-8} {\bibfield  {journal} {\bibinfo
  {journal} {Quantum Mach. Intell.}\ }\textbf {\bibinfo {volume} {2}},\
  \bibinfo {pages} {5} (\bibinfo {year} {2020})}\BibitemShut {NoStop}%
\bibitem [{\citenamefont {Brown}\ \emph {et~al.}(2021)\citenamefont {Brown},
  \citenamefont {Sgroi}, \citenamefont {Giannelli}, \citenamefont {Paraoanu},
  \citenamefont {Paladino}, \citenamefont {Falci}, \citenamefont
  {Paternostro},\ and\ \citenamefont {Ferraro}}]{brown2021}%
  \BibitemOpen
  \bibfield  {author} {\bibinfo {author} {\bibfnamefont {J.}~\bibnamefont
  {Brown}}, \bibinfo {author} {\bibfnamefont {S.}~\bibnamefont {Sgroi}},
  \bibinfo {author} {\bibfnamefont {L.}~\bibnamefont {Giannelli}}, \bibinfo
  {author} {\bibfnamefont {G.~S.}\ \bibnamefont {Paraoanu}}, \bibinfo {author}
  {\bibfnamefont {E.}~\bibnamefont {Paladino}}, \bibinfo {author}
  {\bibfnamefont {G.}~\bibnamefont {Falci}}, \bibinfo {author} {\bibfnamefont
  {M.}~\bibnamefont {Paternostro}},\ and\ \bibinfo {author} {\bibfnamefont
  {A.}~\bibnamefont {Ferraro}},\ }\bibfield  {title} {\bibinfo {title}
  {Reinforcement learning-enhanced protocols for coherent population-transfer
  in three-level quantum systems},\ }\href
  {https://doi.org/10.1088/1367-2630/ac2393} {\bibfield  {journal} {\bibinfo
  {journal} {New J. Phys.}\ }\textbf {\bibinfo {volume} {23}},\ \bibinfo
  {pages} {093035} (\bibinfo {year} {2021})}\BibitemShut {NoStop}%
\bibitem [{\citenamefont {Porotti}\ \emph {et~al.}(2022)\citenamefont
  {Porotti}, \citenamefont {Essig}, \citenamefont {Huard},\ and\ \citenamefont
  {Marquardt}}]{porotti2022}%
  \BibitemOpen
  \bibfield  {author} {\bibinfo {author} {\bibfnamefont {R.}~\bibnamefont
  {Porotti}}, \bibinfo {author} {\bibfnamefont {A.}~\bibnamefont {Essig}},
  \bibinfo {author} {\bibfnamefont {B.}~\bibnamefont {Huard}},\ and\ \bibinfo
  {author} {\bibfnamefont {F.}~\bibnamefont {Marquardt}},\ }\bibfield  {title}
  {\bibinfo {title} {Deep reinforcement learning for quantum state preparation
  with weak nonlinear measurements},\ }\href
  {https://doi.org/10.22331/q-2022-06-28-747} {\bibfield  {journal} {\bibinfo
  {journal} {Quantum}\ }\textbf {\bibinfo {volume} {6}},\ \bibinfo {pages}
  {747} (\bibinfo {year} {2022})}\BibitemShut {NoStop}%
\bibitem [{\citenamefont {Metz}\ and\ \citenamefont {Bukov}(2023)}]{metz2023}%
  \BibitemOpen
  \bibfield  {author} {\bibinfo {author} {\bibfnamefont {F.}~\bibnamefont
  {Metz}}\ and\ \bibinfo {author} {\bibfnamefont {M.}~\bibnamefont {Bukov}},\
  }\bibfield  {title} {\bibinfo {title} {Self-correcting quantum many-body
  control using reinforcement learning with tensor networks},\ }\href
  {https://doi.org/10.1038/s42256-023-00687-5} {\bibfield  {journal} {\bibinfo
  {journal} {Nature Mach. Intell.}\ }\textbf {\bibinfo {volume} {5}},\ \bibinfo
  {pages} {780} (\bibinfo {year} {2023})}\BibitemShut {NoStop}%
\bibitem [{\citenamefont {F{\"o}sel}\ \emph {et~al.}(2018)\citenamefont
  {F{\"o}sel}, \citenamefont {Tighineanu}, \citenamefont {Weiss},\ and\
  \citenamefont {Marquardt}}]{fosel2018}%
  \BibitemOpen
  \bibfield  {author} {\bibinfo {author} {\bibfnamefont {T.}~\bibnamefont
  {F{\"o}sel}}, \bibinfo {author} {\bibfnamefont {P.}~\bibnamefont
  {Tighineanu}}, \bibinfo {author} {\bibfnamefont {T.}~\bibnamefont {Weiss}},\
  and\ \bibinfo {author} {\bibfnamefont {F.}~\bibnamefont {Marquardt}},\
  }\bibfield  {title} {\bibinfo {title} {Reinforcement learning with neural
  networks for quantum feedback},\ }\href
  {https://doi.org/10.1103/PhysRevX.8.031084} {\bibfield  {journal} {\bibinfo
  {journal} {Phys. Rev. X}\ }\textbf {\bibinfo {volume} {8}},\ \bibinfo {pages}
  {031084} (\bibinfo {year} {2018})}\BibitemShut {NoStop}%
\bibitem [{\citenamefont {Reuer}\ \emph {et~al.}(2023)\citenamefont {Reuer},
  \citenamefont {Landgraf}, \citenamefont {F{\"o}sel}, \citenamefont
  {O'Sullivan}, \citenamefont {Belt{\'a}n}, \citenamefont {Akin}, \citenamefont
  {Norris}, \citenamefont {Remm}, \citenamefont {Kerschbaum}, \citenamefont
  {Besse}, \citenamefont {Marquardt}, \citenamefont {Wallraff},\ and\
  \citenamefont {Eichler}}]{reuer2023}%
  \BibitemOpen
  \bibfield  {author} {\bibinfo {author} {\bibfnamefont {K.}~\bibnamefont
  {Reuer}}, \bibinfo {author} {\bibfnamefont {J.}~\bibnamefont {Landgraf}},
  \bibinfo {author} {\bibfnamefont {T.}~\bibnamefont {F{\"o}sel}}, \bibinfo
  {author} {\bibfnamefont {J.}~\bibnamefont {O'Sullivan}}, \bibinfo {author}
  {\bibfnamefont {L.}~\bibnamefont {Belt{\'a}n}}, \bibinfo {author}
  {\bibfnamefont {A.}~\bibnamefont {Akin}}, \bibinfo {author} {\bibfnamefont
  {G.~J.}\ \bibnamefont {Norris}}, \bibinfo {author} {\bibfnamefont
  {A.}~\bibnamefont {Remm}}, \bibinfo {author} {\bibfnamefont {M.}~\bibnamefont
  {Kerschbaum}}, \bibinfo {author} {\bibfnamefont {J.-C.}\ \bibnamefont
  {Besse}}, \bibinfo {author} {\bibfnamefont {F.}~\bibnamefont {Marquardt}},
  \bibinfo {author} {\bibfnamefont {A.}~\bibnamefont {Wallraff}},\ and\
  \bibinfo {author} {\bibfnamefont {C.}~\bibnamefont {Eichler}},\ }\bibfield
  {title} {\bibinfo {title} {Realizing a deep reinforcement learning agent for
  real-time quantum feedback},\ }\href
  {https://doi.org/10.1038/s41467-023-42901-3} {\bibfield  {journal} {\bibinfo
  {journal} {Nat. Commun.}\ }\textbf {\bibinfo {volume} {14}},\ \bibinfo
  {pages} {7138} (\bibinfo {year} {2023})}\BibitemShut {NoStop}%
\bibitem [{\citenamefont {An}\ and\ \citenamefont {Zhou}(2019)}]{an2019}%
  \BibitemOpen
  \bibfield  {author} {\bibinfo {author} {\bibfnamefont {Z.}~\bibnamefont
  {An}}\ and\ \bibinfo {author} {\bibfnamefont {D.}~\bibnamefont {Zhou}},\
  }\bibfield  {title} {\bibinfo {title} {Deep reinforcement learning for
  quantum gate control},\ }\href {https://doi.org/10.1209/0295-5075/126/60002}
  {\bibfield  {journal} {\bibinfo  {journal} {EPL}\ }\textbf {\bibinfo {volume}
  {126}},\ \bibinfo {pages} {60002} (\bibinfo {year} {2019})}\BibitemShut
  {NoStop}%
\bibitem [{\citenamefont {Niu}\ \emph {et~al.}(2019)\citenamefont {Niu},
  \citenamefont {Boixo}, \citenamefont {Smelyanskiy},\ and\ \citenamefont
  {Neven}}]{niu2019}%
  \BibitemOpen
  \bibfield  {author} {\bibinfo {author} {\bibfnamefont {M.~Y.}\ \bibnamefont
  {Niu}}, \bibinfo {author} {\bibfnamefont {S.}~\bibnamefont {Boixo}}, \bibinfo
  {author} {\bibfnamefont {V.~N.}\ \bibnamefont {Smelyanskiy}},\ and\ \bibinfo
  {author} {\bibfnamefont {H.}~\bibnamefont {Neven}},\ }\bibfield  {title}
  {\bibinfo {title} {Universal quantum control through deep reinforcement
  learning},\ }\href {https://doi.org/10.1038/s41534-019-0141-3} {\bibfield
  {journal} {\bibinfo  {journal} {npj Quantum Inf.}\ }\textbf {\bibinfo
  {volume} {5}},\ \bibinfo {pages} {33} (\bibinfo {year} {2019})}\BibitemShut
  {NoStop}%
\bibitem [{\citenamefont {Sweke}\ \emph {et~al.}(2020)\citenamefont {Sweke},
  \citenamefont {Kesselring}, \citenamefont {van Nieuwenburg},\ and\
  \citenamefont {Eisert}}]{sweke2020}%
  \BibitemOpen
  \bibfield  {author} {\bibinfo {author} {\bibfnamefont {R.}~\bibnamefont
  {Sweke}}, \bibinfo {author} {\bibfnamefont {M.~S.}\ \bibnamefont
  {Kesselring}}, \bibinfo {author} {\bibfnamefont {E.~P.~L.}\ \bibnamefont {van
  Nieuwenburg}},\ and\ \bibinfo {author} {\bibfnamefont {J.}~\bibnamefont
  {Eisert}},\ }\bibfield  {title} {\bibinfo {title} {Reinforcement learning
  decoders for fault-tolerant quantum computation},\ }\href
  {https://doi.org/10.1088/1367-2630/ab4dca} {\bibfield  {journal} {\bibinfo
  {journal} {Mach. Learn. Sci. Technol.}\ }\textbf {\bibinfo {volume} {2}},\
  \bibinfo {pages} {025005} (\bibinfo {year} {2020})}\BibitemShut {NoStop}%
\bibitem [{\citenamefont {Sgroi}\ \emph {et~al.}(2021)\citenamefont {Sgroi},
  \citenamefont {Palma},\ and\ \citenamefont {Paternostro}}]{sgroi2021}%
  \BibitemOpen
  \bibfield  {author} {\bibinfo {author} {\bibfnamefont {S.}~\bibnamefont
  {Sgroi}}, \bibinfo {author} {\bibfnamefont {G.~M.}\ \bibnamefont {Palma}},\
  and\ \bibinfo {author} {\bibfnamefont {M.}~\bibnamefont {Paternostro}},\
  }\bibfield  {title} {\bibinfo {title} {Reinforcement learning approach to
  nonequilibrium quantum thermodynamics},\ }\href
  {https://doi.org/10.1103/PhysRevLett.126.020601} {\bibfield  {journal}
  {\bibinfo  {journal} {Phys. Rev. Lett.}\ }\textbf {\bibinfo {volume} {126}},\
  \bibinfo {pages} {020601} (\bibinfo {year} {2021})}\BibitemShut {NoStop}%
\bibitem [{\citenamefont {Erdman}\ and\ \citenamefont
  {No{\'e}}(2022)}]{erdman2022}%
  \BibitemOpen
  \bibfield  {author} {\bibinfo {author} {\bibfnamefont {P.~A.}\ \bibnamefont
  {Erdman}}\ and\ \bibinfo {author} {\bibfnamefont {F.}~\bibnamefont
  {No{\'e}}},\ }\bibfield  {title} {\bibinfo {title} {Identifying optimal
  cycles in quantum thermal machines with reinforcement-learning},\ }\href
  {https://doi.org/10.1038/s41534-021-00512-0} {\bibfield  {journal} {\bibinfo
  {journal} {npj Quantum Inf.}\ }\textbf {\bibinfo {volume} {8}},\ \bibinfo
  {pages} {1} (\bibinfo {year} {2022})}\BibitemShut {NoStop}%
\bibitem [{\citenamefont {Erdman}\ \emph {et~al.}(2022)\citenamefont {Erdman},
  \citenamefont {Andolina}, \citenamefont {Giovannetti},\ and\ \citenamefont
  {No{\'e}}}]{erdman2022_arxiv}%
  \BibitemOpen
  \bibfield  {author} {\bibinfo {author} {\bibfnamefont {P.~A.}\ \bibnamefont
  {Erdman}}, \bibinfo {author} {\bibfnamefont {G.~M.}\ \bibnamefont
  {Andolina}}, \bibinfo {author} {\bibfnamefont {V.}~\bibnamefont
  {Giovannetti}},\ and\ \bibinfo {author} {\bibfnamefont {F.}~\bibnamefont
  {No{\'e}}},\ }\bibfield  {title} {\bibinfo {title} {{Reinforcement learning
  optimization of the charging of a Dicke quantum battery}},\ }\Eprint
  {https://arxiv.org/abs/arXiv:2212.12397} {arXiv:2212.12397}  (\bibinfo {year}
  {2022})\BibitemShut {NoStop}%
\bibitem [{\citenamefont {Erdman}\ and\ \citenamefont
  {No{\'e}}(2023)}]{erdman2023_pnas}%
  \BibitemOpen
  \bibfield  {author} {\bibinfo {author} {\bibfnamefont {P.~A.}\ \bibnamefont
  {Erdman}}\ and\ \bibinfo {author} {\bibfnamefont {F.}~\bibnamefont
  {No{\'e}}},\ }\bibfield  {title} {\bibinfo {title} {{Model-free optimization
  of power/efficiency tradeoffs in quantum thermal machines using reinforcement
  learning}},\ }\href {https://doi.org/10.1093/pnasnexus/pgad248} {\bibfield
  {journal} {\bibinfo  {journal} {PNAS Nexus}\ }\textbf {\bibinfo {volume}
  {2}},\ \bibinfo {pages} {pgad248} (\bibinfo {year} {2023})}\BibitemShut
  {NoStop}%
\bibitem [{\citenamefont {Erdman}\ \emph {et~al.}(2023)\citenamefont {Erdman},
  \citenamefont {Rolandi}, \citenamefont {Abiuso}, \citenamefont
  {Perarnau-Llobet},\ and\ \citenamefont {No\'e}}]{erdman2023_prr}%
  \BibitemOpen
  \bibfield  {author} {\bibinfo {author} {\bibfnamefont {P.~A.}\ \bibnamefont
  {Erdman}}, \bibinfo {author} {\bibfnamefont {A.}~\bibnamefont {Rolandi}},
  \bibinfo {author} {\bibfnamefont {P.}~\bibnamefont {Abiuso}}, \bibinfo
  {author} {\bibfnamefont {M.}~\bibnamefont {Perarnau-Llobet}},\ and\ \bibinfo
  {author} {\bibfnamefont {F.}~\bibnamefont {No\'e}},\ }\bibfield  {title}
  {\bibinfo {title} {Pareto-optimal cycles for power, efficiency and
  fluctuations of quantum heat engines using reinforcement learning},\ }\href
  {https://doi.org/10.1103/PhysRevResearch.5.L022017} {\bibfield  {journal}
  {\bibinfo  {journal} {Phys. Rev. Res.}\ }\textbf {\bibinfo {volume} {5}},\
  \bibinfo {pages} {L022017} (\bibinfo {year} {2023})}\BibitemShut {NoStop}%
\bibitem [{\citenamefont {Deng}\ \emph {et~al.}(2024)\citenamefont {Deng},
  \citenamefont {Ai}, \citenamefont {Wang}, \citenamefont {Shao}, \citenamefont
  {Liu},\ and\ \citenamefont {Cui}}]{deng2024}%
  \BibitemOpen
  \bibfield  {author} {\bibinfo {author} {\bibfnamefont {G.-x.}\ \bibnamefont
  {Deng}}, \bibinfo {author} {\bibfnamefont {H.}~\bibnamefont {Ai}}, \bibinfo
  {author} {\bibfnamefont {B.}~\bibnamefont {Wang}}, \bibinfo {author}
  {\bibfnamefont {W.}~\bibnamefont {Shao}}, \bibinfo {author} {\bibfnamefont
  {Y.}~\bibnamefont {Liu}},\ and\ \bibinfo {author} {\bibfnamefont
  {Z.}~\bibnamefont {Cui}},\ }\bibfield  {title} {\bibinfo {title} {Exploring
  the optimal cycle for a quantum heat engine using reinforcement learning},\
  }\href {https://doi.org/10.1103/PhysRevA.109.022246} {\bibfield  {journal}
  {\bibinfo  {journal} {Phys. Rev. A}\ }\textbf {\bibinfo {volume} {109}},\
  \bibinfo {pages} {022246} (\bibinfo {year} {2024})}\BibitemShut {NoStop}%
\bibitem [{\citenamefont {Gogolin}\ and\ \citenamefont
  {Eisert}(2016)}]{Thermalization}%
  \BibitemOpen
  \bibfield  {author} {\bibinfo {author} {\bibfnamefont {C.}~\bibnamefont
  {Gogolin}}\ and\ \bibinfo {author} {\bibfnamefont {J.}~\bibnamefont
  {Eisert}},\ }\bibfield  {title} {\bibinfo {title} {Equilibration,
  thermalisation, and the emergence of statistical mechanics in closed quantum
  systems},\ }\href {https://doi.org/10.1088/0034-4885/79/5/056001} {\bibfield
  {journal} {\bibinfo  {journal} {Rep. Prog. Phys.}\ }\textbf {\bibinfo
  {volume} {79}},\ \bibinfo {pages} {056001} (\bibinfo {year}
  {2016})}\BibitemShut {NoStop}%
\bibitem [{\citenamefont {Gorini}\ \emph {et~al.}(1976)\citenamefont {Gorini},
  \citenamefont {Kossakowski},\ and\ \citenamefont {Sudarshan}}]{gorini1976}%
  \BibitemOpen
  \bibfield  {author} {\bibinfo {author} {\bibfnamefont {V.}~\bibnamefont
  {Gorini}}, \bibinfo {author} {\bibfnamefont {A.}~\bibnamefont
  {Kossakowski}},\ and\ \bibinfo {author} {\bibfnamefont {E.~C.~G.}\
  \bibnamefont {Sudarshan}},\ }\bibfield  {title} {\bibinfo {title} {Completely
  positive dynamical semigroups of $n$‐level systems},\ }\href
  {https://doi.org/10.1063/1.522979} {\bibfield  {journal} {\bibinfo  {journal}
  {J. Math. Phys.}\ }\textbf {\bibinfo {volume} {17}},\ \bibinfo {pages} {821}
  (\bibinfo {year} {1976})}\BibitemShut {NoStop}%
\bibitem [{\citenamefont {Lindblad}(1976)}]{lindblad1976}%
  \BibitemOpen
  \bibfield  {author} {\bibinfo {author} {\bibfnamefont {G.}~\bibnamefont
  {Lindblad}},\ }\bibfield  {title} {\bibinfo {title} {On the generators of
  quantum dynamical semigroups},\ }\href {https://doi.org/10.1007/BF01608499}
  {\bibfield  {journal} {\bibinfo  {journal} {Commun. Math. Phys}\ }\textbf
  {\bibinfo {volume} {48}},\ \bibinfo {pages} {119} (\bibinfo {year}
  {1976})}\BibitemShut {NoStop}%
\bibitem [{\citenamefont {Hofer}\ \emph {et~al.}(2017)\citenamefont {Hofer},
  \citenamefont {Brask}, \citenamefont {Perarnau-Llobet},\ and\ \citenamefont
  {Brunner}}]{hofer2017quantum}%
  \BibitemOpen
  \bibfield  {author} {\bibinfo {author} {\bibfnamefont {P.~P.}\ \bibnamefont
  {Hofer}}, \bibinfo {author} {\bibfnamefont {J.~B.}\ \bibnamefont {Brask}},
  \bibinfo {author} {\bibfnamefont {M.}~\bibnamefont {Perarnau-Llobet}},\ and\
  \bibinfo {author} {\bibfnamefont {N.}~\bibnamefont {Brunner}},\ }\bibfield
  {title} {\bibinfo {title} {Quantum thermal machine as a thermometer},\ }\href
  {https://doi.org/10.1103/PhysRevLett.119.090603} {\bibfield  {journal}
  {\bibinfo  {journal} {Phys. Rev. Lett.}\ }\textbf {\bibinfo {volume} {119}},\
  \bibinfo {pages} {090603} (\bibinfo {year} {2017})}\BibitemShut {NoStop}%
\bibitem [{\citenamefont {Zhang}\ \emph {et~al.}(2017)\citenamefont {Zhang},
  \citenamefont {Liu}, \citenamefont {Wu}, \citenamefont {Jacobs},\ and\
  \citenamefont {Nori}}]{zhang2017quantum}%
  \BibitemOpen
  \bibfield  {author} {\bibinfo {author} {\bibfnamefont {J.}~\bibnamefont
  {Zhang}}, \bibinfo {author} {\bibfnamefont {Y.-x.}\ \bibnamefont {Liu}},
  \bibinfo {author} {\bibfnamefont {R.-B.}\ \bibnamefont {Wu}}, \bibinfo
  {author} {\bibfnamefont {K.}~\bibnamefont {Jacobs}},\ and\ \bibinfo {author}
  {\bibfnamefont {F.}~\bibnamefont {Nori}},\ }\bibfield  {title} {\bibinfo
  {title} {Quantum feedback: theory, experiments, and applications},\ }\href
  {https://doi.org/10.1016/j.physrep.2017.02.003} {\bibfield  {journal}
  {\bibinfo  {journal} {Phys. Rep.}\ }\textbf {\bibinfo {volume} {679}},\
  \bibinfo {pages} {1} (\bibinfo {year} {2017})}\BibitemShut {NoStop}%
\bibitem [{\citenamefont {Yanik}\ \emph {et~al.}(2022)\citenamefont {Yanik},
  \citenamefont {Bhandari}, \citenamefont {Manikandan},\ and\ \citenamefont
  {Jordan}}]{yanik2022thermodynamics}%
  \BibitemOpen
  \bibfield  {author} {\bibinfo {author} {\bibfnamefont {K.}~\bibnamefont
  {Yanik}}, \bibinfo {author} {\bibfnamefont {B.}~\bibnamefont {Bhandari}},
  \bibinfo {author} {\bibfnamefont {S.~K.}\ \bibnamefont {Manikandan}},\ and\
  \bibinfo {author} {\bibfnamefont {A.~N.}\ \bibnamefont {Jordan}},\ }\bibfield
   {title} {\bibinfo {title} {{Thermodynamics of quantum measurement and
  Maxwell's demon's arrow of time}},\ }\href
  {https://doi.org/10.1103/PhysRevA.106.042221} {\bibfield  {journal} {\bibinfo
   {journal} {Phys. Rev. A}\ }\textbf {\bibinfo {volume} {106}},\ \bibinfo
  {pages} {042221} (\bibinfo {year} {2022})}\BibitemShut {NoStop}%
\bibitem [{\citenamefont {Jacobs}(2003)}]{jacobs2003}%
  \BibitemOpen
  \bibfield  {author} {\bibinfo {author} {\bibfnamefont {K.}~\bibnamefont
  {Jacobs}},\ }\bibfield  {title} {\bibinfo {title} {How to project qubits
  faster using quantum feedback},\ }\href
  {https://doi.org/10.1103/PhysRevA.67.030301} {\bibfield  {journal} {\bibinfo
  {journal} {Phys. Rev. A}\ }\textbf {\bibinfo {volume} {67}},\ \bibinfo
  {pages} {030301} (\bibinfo {year} {2003})}\BibitemShut {NoStop}%
\bibitem [{\citenamefont {Wiseman}\ and\ \citenamefont
  {Milburn}(2009)}]{wiseman2009}%
  \BibitemOpen
  \bibfield  {author} {\bibinfo {author} {\bibfnamefont {H.~M.}\ \bibnamefont
  {Wiseman}}\ and\ \bibinfo {author} {\bibfnamefont {G.~J.}\ \bibnamefont
  {Milburn}},\ }\href {https://doi.org/10.1017/CBO9780511813948} {\emph
  {\bibinfo {title} {Quantum measurement and control}}}\ (\bibinfo  {publisher}
  {Cambridge University Press},\ \bibinfo {year} {2009})\BibitemShut {NoStop}%
\bibitem [{\citenamefont {Dressel}\ \emph {et~al.}(2017)\citenamefont
  {Dressel}, \citenamefont {Chantasri}, \citenamefont {Jordan},\ and\
  \citenamefont {Korotkov}}]{dressel2017}%
  \BibitemOpen
  \bibfield  {author} {\bibinfo {author} {\bibfnamefont {J.}~\bibnamefont
  {Dressel}}, \bibinfo {author} {\bibfnamefont {A.}~\bibnamefont {Chantasri}},
  \bibinfo {author} {\bibfnamefont {A.~N.}\ \bibnamefont {Jordan}},\ and\
  \bibinfo {author} {\bibfnamefont {A.~N.}\ \bibnamefont {Korotkov}},\
  }\bibfield  {title} {\bibinfo {title} {Arrow of time for continuous quantum
  measurement},\ }\href {https://doi.org/10.1103/PhysRevLett.119.220507}
  {\bibfield  {journal} {\bibinfo  {journal} {Phys. Rev. Lett.}\ }\textbf
  {\bibinfo {volume} {119}},\ \bibinfo {pages} {220507} (\bibinfo {year}
  {2017})}\BibitemShut {NoStop}%
\bibitem [{\citenamefont {Alicki}(1979)}]{alicki1979}%
  \BibitemOpen
  \bibfield  {author} {\bibinfo {author} {\bibfnamefont {R.}~\bibnamefont
  {Alicki}},\ }\bibfield  {title} {\bibinfo {title} {The quantum open system as
  a model of the heat engine},\ }\href
  {https://doi.org/10.1088/0305-4470/12/5/007} {\bibfield  {journal} {\bibinfo
  {journal} {J. Phys. A}\ }\textbf {\bibinfo {volume} {12}},\ \bibinfo {pages}
  {L103} (\bibinfo {year} {1979})}\BibitemShut {NoStop}%
\bibitem [{\citenamefont {Binder}\ \emph {et~al.}(2019)\citenamefont {Binder},
  \citenamefont {Correa}, \citenamefont {Gogolin}, \citenamefont {Anders},\
  and\ \citenamefont {Adesso}}]{binder2019}%
  \BibitemOpen
  \bibfield  {author} {\bibinfo {author} {\bibfnamefont {F.}~\bibnamefont
  {Binder}}, \bibinfo {author} {\bibfnamefont {L.}~\bibnamefont {Correa}},
  \bibinfo {author} {\bibfnamefont {C.}~\bibnamefont {Gogolin}}, \bibinfo
  {author} {\bibfnamefont {J.}~\bibnamefont {Anders}},\ and\ \bibinfo {author}
  {\bibfnamefont {G.}~\bibnamefont {Adesso}},\ }\href@noop {} {\emph {\bibinfo
  {title} {Thermodynamics in the Quantum Regime: Fundamental Aspects and New
  Directions}}},\ Fundamental Theories of Physics\ (\bibinfo  {publisher}
  {Springer International Publishing},\ \bibinfo {year} {2019})\BibitemShut
  {NoStop}%
\bibitem [{\citenamefont {Sutton}\ and\ \citenamefont
  {Barto}(2018)}]{sutton2018}%
  \BibitemOpen
  \bibfield  {author} {\bibinfo {author} {\bibfnamefont {R.~S.}\ \bibnamefont
  {Sutton}}\ and\ \bibinfo {author} {\bibfnamefont {A.~G.}\ \bibnamefont
  {Barto}},\ }\href {https://doi.org/10.1016/S0925-2312(00)00324-6} {\emph
  {\bibinfo {title} {Reinforcement learning: An introduction}}}\ (\bibinfo
  {publisher} {MIT press},\ \bibinfo {year} {2018})\BibitemShut {NoStop}%
\bibitem [{\citenamefont {Haarnoja}\ \emph
  {et~al.}(2018{\natexlab{a}})\citenamefont {Haarnoja}, \citenamefont {Zhou},
  \citenamefont {Abbeel},\ and\ \citenamefont {Levine}}]{haarnoja2018_pmlr}%
  \BibitemOpen
  \bibfield  {author} {\bibinfo {author} {\bibfnamefont {T.}~\bibnamefont
  {Haarnoja}}, \bibinfo {author} {\bibfnamefont {A.}~\bibnamefont {Zhou}},
  \bibinfo {author} {\bibfnamefont {P.}~\bibnamefont {Abbeel}},\ and\ \bibinfo
  {author} {\bibfnamefont {S.}~\bibnamefont {Levine}},\ }\bibfield  {title}
  {\bibinfo {title} {Soft actor-critic: Off-policy maximum entropy deep
  reinforcement learning with a stochastic actor},\ }in\ \href
  {http://proceedings.mlr.press/v80/haarnoja18b.html} {\emph {\bibinfo
  {booktitle} {International Conference on Machine Learning}}},\ Vol.~\bibinfo
  {volume} {80}\ (\bibinfo {organization} {PMLR},\ \bibinfo {year} {2018})\ p.\
  \bibinfo {pages} {1861}\BibitemShut {NoStop}%
\bibitem [{\citenamefont {Haarnoja}\ \emph
  {et~al.}(2018{\natexlab{b}})\citenamefont {Haarnoja}, \citenamefont {Zhou},
  \citenamefont {Hartikainen}, \citenamefont {Tucker}, \citenamefont {Ha},
  \citenamefont {Tan}, \citenamefont {Kumar}, \citenamefont {Zhu},
  \citenamefont {Gupta}, \citenamefont {Abbeel} \emph
  {et~al.}}]{haarnoja2018_arxiv_sac}%
  \BibitemOpen
  \bibfield  {author} {\bibinfo {author} {\bibfnamefont {T.}~\bibnamefont
  {Haarnoja}}, \bibinfo {author} {\bibfnamefont {A.}~\bibnamefont {Zhou}},
  \bibinfo {author} {\bibfnamefont {K.}~\bibnamefont {Hartikainen}}, \bibinfo
  {author} {\bibfnamefont {G.}~\bibnamefont {Tucker}}, \bibinfo {author}
  {\bibfnamefont {S.}~\bibnamefont {Ha}}, \bibinfo {author} {\bibfnamefont
  {J.}~\bibnamefont {Tan}}, \bibinfo {author} {\bibfnamefont {V.}~\bibnamefont
  {Kumar}}, \bibinfo {author} {\bibfnamefont {H.}~\bibnamefont {Zhu}}, \bibinfo
  {author} {\bibfnamefont {A.}~\bibnamefont {Gupta}}, \bibinfo {author}
  {\bibfnamefont {P.}~\bibnamefont {Abbeel}}, \emph {et~al.},\ }\bibfield
  {title} {\bibinfo {title} {Soft actor-critic algorithms and applications},\
  }\Eprint {https://arxiv.org/abs/arXiv:1812.05905} {arXiv:1812.05905}
  (\bibinfo {year} {2018}{\natexlab{b}})\BibitemShut {NoStop}%
\bibitem [{\citenamefont {Haarnoja}\ \emph
  {et~al.}(2018{\natexlab{c}})\citenamefont {Haarnoja}, \citenamefont {Ha},
  \citenamefont {Zhou}, \citenamefont {Tan}, \citenamefont {Tucker},\ and\
  \citenamefont {Levine}}]{haarnoja2018_arxiv_walk}%
  \BibitemOpen
  \bibfield  {author} {\bibinfo {author} {\bibfnamefont {T.}~\bibnamefont
  {Haarnoja}}, \bibinfo {author} {\bibfnamefont {S.}~\bibnamefont {Ha}},
  \bibinfo {author} {\bibfnamefont {A.}~\bibnamefont {Zhou}}, \bibinfo {author}
  {\bibfnamefont {J.}~\bibnamefont {Tan}}, \bibinfo {author} {\bibfnamefont
  {G.}~\bibnamefont {Tucker}},\ and\ \bibinfo {author} {\bibfnamefont
  {S.}~\bibnamefont {Levine}},\ }\bibfield  {title} {\bibinfo {title} {Learning
  to walk via deep reinforcement learning},\ }\Eprint
  {https://arxiv.org/abs/arXiv:1912.11077} {arXiv:1912.11077}  (\bibinfo {year}
  {2018}{\natexlab{c}})\BibitemShut {NoStop}%
\bibitem [{\citenamefont {Paszke}\ \emph {et~al.}(2017)\citenamefont {Paszke},
  \citenamefont {Gross}, \citenamefont {Chintala}, \citenamefont {Chanan},
  \citenamefont {Yang}, \citenamefont {DeVito}, \citenamefont {Lin},
  \citenamefont {Desmaison}, \citenamefont {Antiga},\ and\ \citenamefont
  {Lerer}}]{paszke2017}%
  \BibitemOpen
  \bibfield  {author} {\bibinfo {author} {\bibfnamefont {A.}~\bibnamefont
  {Paszke}}, \bibinfo {author} {\bibfnamefont {S.}~\bibnamefont {Gross}},
  \bibinfo {author} {\bibfnamefont {S.}~\bibnamefont {Chintala}}, \bibinfo
  {author} {\bibfnamefont {G.}~\bibnamefont {Chanan}}, \bibinfo {author}
  {\bibfnamefont {E.}~\bibnamefont {Yang}}, \bibinfo {author} {\bibfnamefont
  {Z.}~\bibnamefont {DeVito}}, \bibinfo {author} {\bibfnamefont
  {Z.}~\bibnamefont {Lin}}, \bibinfo {author} {\bibfnamefont {A.}~\bibnamefont
  {Desmaison}}, \bibinfo {author} {\bibfnamefont {L.}~\bibnamefont {Antiga}},\
  and\ \bibinfo {author} {\bibfnamefont {A.}~\bibnamefont {Lerer}},\ }\bibfield
   {title} {\bibinfo {title} {Automatic differentiation in pytorch},\ }in\
  \href {https://openreview.net/forum?id=BJJsrmfCZ} {\emph {\bibinfo
  {booktitle} {NIPS 2017 Workshop on Autodiff}}}\ (\bibinfo {year}
  {2017})\BibitemShut {NoStop}%
\bibitem [{\citenamefont {Maillet}\ \emph {et~al.}(2019)\citenamefont
  {Maillet}, \citenamefont {Erdman}, \citenamefont {Cavina}, \citenamefont
  {Bhandari}, \citenamefont {Mannila}, \citenamefont {Peltonen}, \citenamefont
  {Mari}, \citenamefont {Taddei}, \citenamefont {Jarzynski}, \citenamefont
  {Giovannetti},\ and\ \citenamefont {Pekola}}]{maillet2019}%
  \BibitemOpen
  \bibfield  {author} {\bibinfo {author} {\bibfnamefont {O.}~\bibnamefont
  {Maillet}}, \bibinfo {author} {\bibfnamefont {P.~A.}\ \bibnamefont {Erdman}},
  \bibinfo {author} {\bibfnamefont {V.}~\bibnamefont {Cavina}}, \bibinfo
  {author} {\bibfnamefont {B.}~\bibnamefont {Bhandari}}, \bibinfo {author}
  {\bibfnamefont {E.~T.}\ \bibnamefont {Mannila}}, \bibinfo {author}
  {\bibfnamefont {J.~T.}\ \bibnamefont {Peltonen}}, \bibinfo {author}
  {\bibfnamefont {A.}~\bibnamefont {Mari}}, \bibinfo {author} {\bibfnamefont
  {F.}~\bibnamefont {Taddei}}, \bibinfo {author} {\bibfnamefont
  {C.}~\bibnamefont {Jarzynski}}, \bibinfo {author} {\bibfnamefont
  {V.}~\bibnamefont {Giovannetti}},\ and\ \bibinfo {author} {\bibfnamefont
  {J.~P.}\ \bibnamefont {Pekola}},\ }\bibfield  {title} {\bibinfo {title}
  {Optimal probabilistic work extraction beyond the free energy difference with
  a single-electron device},\ }\href
  {https://doi.org/10.1103/PhysRevLett.122.150604} {\bibfield  {journal}
  {\bibinfo  {journal} {Phys. Rev. Lett.}\ }\textbf {\bibinfo {volume} {122}},\
  \bibinfo {pages} {150604} (\bibinfo {year} {2019})}\BibitemShut {NoStop}%
\bibitem [{\citenamefont {Scandi}\ \emph {et~al.}(2022)\citenamefont {Scandi},
  \citenamefont {Barker}, \citenamefont {Lehmann}, \citenamefont {Dick},
  \citenamefont {Maisi},\ and\ \citenamefont {Perarnau-Llobet}}]{scandi2022}%
  \BibitemOpen
  \bibfield  {author} {\bibinfo {author} {\bibfnamefont {M.}~\bibnamefont
  {Scandi}}, \bibinfo {author} {\bibfnamefont {D.}~\bibnamefont {Barker}},
  \bibinfo {author} {\bibfnamefont {S.}~\bibnamefont {Lehmann}}, \bibinfo
  {author} {\bibfnamefont {K.~A.}\ \bibnamefont {Dick}}, \bibinfo {author}
  {\bibfnamefont {V.~F.}\ \bibnamefont {Maisi}},\ and\ \bibinfo {author}
  {\bibfnamefont {M.}~\bibnamefont {Perarnau-Llobet}},\ }\bibfield  {title}
  {\bibinfo {title} {Minimally dissipative information erasure in a quantum dot
  via thermodynamic length},\ }\href
  {https://doi.org/10.1103/PhysRevLett.129.270601} {\bibfield  {journal}
  {\bibinfo  {journal} {Phys. Rev. Lett.}\ }\textbf {\bibinfo {volume} {129}},\
  \bibinfo {pages} {270601} (\bibinfo {year} {2022})}\BibitemShut {NoStop}%
\bibitem [{\citenamefont {Gunyh{\'o}}\ \emph {et~al.}(2024)\citenamefont
  {Gunyh{\'o}}, \citenamefont {Kundu}, \citenamefont {Ma}, \citenamefont {Liu},
  \citenamefont {Niemel{\"a}}, \citenamefont {Catto}, \citenamefont {Vadimov},
  \citenamefont {Vesterinen}, \citenamefont {Singh}, \citenamefont {Chen} \emph
  {et~al.}}]{gunyho2024single}%
  \BibitemOpen
  \bibfield  {author} {\bibinfo {author} {\bibfnamefont {A.~M.}\ \bibnamefont
  {Gunyh{\'o}}}, \bibinfo {author} {\bibfnamefont {S.}~\bibnamefont {Kundu}},
  \bibinfo {author} {\bibfnamefont {J.}~\bibnamefont {Ma}}, \bibinfo {author}
  {\bibfnamefont {W.}~\bibnamefont {Liu}}, \bibinfo {author} {\bibfnamefont
  {S.}~\bibnamefont {Niemel{\"a}}}, \bibinfo {author} {\bibfnamefont
  {G.}~\bibnamefont {Catto}}, \bibinfo {author} {\bibfnamefont
  {V.}~\bibnamefont {Vadimov}}, \bibinfo {author} {\bibfnamefont
  {V.}~\bibnamefont {Vesterinen}}, \bibinfo {author} {\bibfnamefont
  {P.}~\bibnamefont {Singh}}, \bibinfo {author} {\bibfnamefont
  {Q.}~\bibnamefont {Chen}}, \emph {et~al.},\ }\bibfield  {title} {\bibinfo
  {title} {Single-shot readout of a superconducting qubit using a thermal
  detector},\ }\href
  {https://doi.org/https://doi.org/10.1038/s41928-024-01147-7} {\bibfield
  {journal} {\bibinfo  {journal} {Nat. Electron.}\ }\textbf {\bibinfo {volume}
  {7}},\ \bibinfo {pages} {288} (\bibinfo {year} {2024})}\BibitemShut {NoStop}%
\bibitem [{\citenamefont {Nguyen}\ \emph {et~al.}(2024)\citenamefont {Nguyen},
  \citenamefont {Kim}, \citenamefont {Hashim}, \citenamefont {Goss},
  \citenamefont {Marinelli}, \citenamefont {Bhandari}, \citenamefont {Das},
  \citenamefont {Naik}, \citenamefont {Kreikebaum}, \citenamefont {Jordan}
  \emph {et~al.}}]{nguyen2024programmable}%
  \BibitemOpen
  \bibfield  {author} {\bibinfo {author} {\bibfnamefont {L.~B.}\ \bibnamefont
  {Nguyen}}, \bibinfo {author} {\bibfnamefont {Y.}~\bibnamefont {Kim}},
  \bibinfo {author} {\bibfnamefont {A.}~\bibnamefont {Hashim}}, \bibinfo
  {author} {\bibfnamefont {N.}~\bibnamefont {Goss}}, \bibinfo {author}
  {\bibfnamefont {B.}~\bibnamefont {Marinelli}}, \bibinfo {author}
  {\bibfnamefont {B.}~\bibnamefont {Bhandari}}, \bibinfo {author}
  {\bibfnamefont {D.}~\bibnamefont {Das}}, \bibinfo {author} {\bibfnamefont
  {R.~K.}\ \bibnamefont {Naik}}, \bibinfo {author} {\bibfnamefont {J.~M.}\
  \bibnamefont {Kreikebaum}}, \bibinfo {author} {\bibfnamefont {A.~N.}\
  \bibnamefont {Jordan}}, \emph {et~al.},\ }\bibfield  {title} {\bibinfo
  {title} {{Programmable Heisenberg interactions between Floquet qubits}},\
  }\href {https://www.nature.com/articles/s41567-023-02326-7} {\bibfield
  {journal} {\bibinfo  {journal} {Nat. Phys.}\ }\textbf {\bibinfo {volume}
  {20}},\ \bibinfo {pages} {240} (\bibinfo {year} {2024})}\BibitemShut
  {NoStop}%
\bibitem [{\citenamefont {Huang}\ \emph {et~al.}(2021)\citenamefont {Huang},
  \citenamefont {Mundada}, \citenamefont {Gyenis}, \citenamefont {Schuster},
  \citenamefont {Houck},\ and\ \citenamefont
  {Koch}}]{PhysRevApplied.15.034065}%
  \BibitemOpen
  \bibfield  {author} {\bibinfo {author} {\bibfnamefont {Z.}~\bibnamefont
  {Huang}}, \bibinfo {author} {\bibfnamefont {P.~S.}\ \bibnamefont {Mundada}},
  \bibinfo {author} {\bibfnamefont {A.}~\bibnamefont {Gyenis}}, \bibinfo
  {author} {\bibfnamefont {D.~I.}\ \bibnamefont {Schuster}}, \bibinfo {author}
  {\bibfnamefont {A.~A.}\ \bibnamefont {Houck}},\ and\ \bibinfo {author}
  {\bibfnamefont {J.}~\bibnamefont {Koch}},\ }\bibfield  {title} {\bibinfo
  {title} {Engineering dynamical sweet spots to protect qubits from $1/f$
  noise},\ }\href {https://doi.org/10.1103/PhysRevApplied.15.034065} {\bibfield
   {journal} {\bibinfo  {journal} {Phys. Rev. Appl.}\ }\textbf {\bibinfo
  {volume} {15}},\ \bibinfo {pages} {034065} (\bibinfo {year}
  {2021})}\BibitemShut {NoStop}%
\bibitem [{\citenamefont {Pietzonka}\ and\ \citenamefont
  {Seifert}(2018)}]{pietzonka2018universal}%
  \BibitemOpen
  \bibfield  {author} {\bibinfo {author} {\bibfnamefont {P.}~\bibnamefont
  {Pietzonka}}\ and\ \bibinfo {author} {\bibfnamefont {U.}~\bibnamefont
  {Seifert}},\ }\bibfield  {title} {\bibinfo {title} {Universal trade-off
  between power, efficiency, and constancy in steady-state heat engines},\
  }\href {https://doi.org/10.1103/PhysRevLett.120.190602} {\bibfield  {journal}
  {\bibinfo  {journal} {Phys. Rev. Lett.}\ }\textbf {\bibinfo {volume} {120}},\
  \bibinfo {pages} {190602} (\bibinfo {year} {2018})}\BibitemShut {NoStop}%
\bibitem [{\citenamefont {Johansson}\ \emph {et~al.}(2013)\citenamefont
  {Johansson}, \citenamefont {Nation},\ and\ \citenamefont
  {Nori}}]{johansson2013}%
  \BibitemOpen
  \bibfield  {author} {\bibinfo {author} {\bibfnamefont {J.~R.}\ \bibnamefont
  {Johansson}}, \bibinfo {author} {\bibfnamefont {P.~D.}\ \bibnamefont
  {Nation}},\ and\ \bibinfo {author} {\bibfnamefont {F.}~\bibnamefont {Nori}},\
  }\bibfield  {title} {\bibinfo {title} {Qutip 2: A python framework for the
  dynamics of open quantum systems},\ }\href
  {https://doi.org/10.1016/j.cpc.2012.11.019} {\bibfield  {journal} {\bibinfo
  {journal} {Comp. Phys. Comm.}\ }\textbf {\bibinfo {volume} {184}},\ \bibinfo
  {pages} {1234} (\bibinfo {year} {2013})}\BibitemShut {NoStop}%
\bibitem [{\citenamefont {Christodoulou}(2019)}]{christodoulou2019}%
  \BibitemOpen
  \bibfield  {author} {\bibinfo {author} {\bibfnamefont {P.}~\bibnamefont
  {Christodoulou}},\ }\bibfield  {title} {\bibinfo {title} {Soft actor-critic
  for discrete action settings},\ }\Eprint
  {https://arxiv.org/abs/arXiv:1910.07207} {arXiv:1910.07207}  (\bibinfo {year}
  {2019})\BibitemShut {NoStop}%
\bibitem [{\citenamefont {Delalleau}\ \emph {et~al.}(2019)\citenamefont
  {Delalleau}, \citenamefont {Peter}, \citenamefont {Alonso},\ and\
  \citenamefont {Logut}}]{delalleau2019}%
  \BibitemOpen
  \bibfield  {author} {\bibinfo {author} {\bibfnamefont {O.}~\bibnamefont
  {Delalleau}}, \bibinfo {author} {\bibfnamefont {M.}~\bibnamefont {Peter}},
  \bibinfo {author} {\bibfnamefont {E.}~\bibnamefont {Alonso}},\ and\ \bibinfo
  {author} {\bibfnamefont {A.}~\bibnamefont {Logut}},\ }\bibfield  {title}
  {\bibinfo {title} {Discrete and continuous action representation for
  practical rl in video games},\ }\Eprint
  {https://arxiv.org/abs/arXiv:1912.11077} {arXiv:1912.11077}  (\bibinfo {year}
  {2019})\BibitemShut {NoStop}%
\bibitem [{\citenamefont {Achiam}(2018)}]{spinningup2018}%
  \BibitemOpen
  \bibfield  {author} {\bibinfo {author} {\bibfnamefont {J.}~\bibnamefont
  {Achiam}},\ }\bibfield  {title} {\bibinfo {title} {{Spinning Up in Deep
  Reinforcement Learning}}} (\bibinfo {year} {2018})\BibitemShut {NoStop}%
\bibitem [{\citenamefont {Kingma}\ and\ \citenamefont {Ba}(2014)}]{kingma2014}%
  \BibitemOpen
  \bibfield  {author} {\bibinfo {author} {\bibfnamefont {D.~P.}\ \bibnamefont
  {Kingma}}\ and\ \bibinfo {author} {\bibfnamefont {J.}~\bibnamefont {Ba}},\
  }\bibfield  {title} {\bibinfo {title} {Adam: A method for stochastic
  optimization},\ }\Eprint {https://arxiv.org/abs/arXiv:1412.6980}
  {arXiv:1412.6980}  (\bibinfo {year} {2014})\BibitemShut {NoStop}%
\end{thebibliography}
\end{document}